# Answering the persona, not the person: Over-determination in the construction of synthetic respondents

**Konrad Suchecki: konrad.suchecki@wp.pl; Laurent Florès, Université Paris-Panthéon-Assas and FlashInsight: Laurent.Flores@assas-universite.fr (correspondence: drlaurentflores@gmail.com)**

## Abstract

Since the release of GPT-4 in 2023 persona-prompted large language models (LLMs) have been increasingly used as synthetic survey respondents and digital twins, and validated by how well they replicate human aggregate responses. That leaves a prior much more important question open: how do these models actually produce a synthetic answer in the first place? To answer that we built each of 14,000 synthetic respondents from the observed attributes of one respondent in the European Social Survey (Round 11, 28 countries) and used that person's own answers to 15 attitude questions as the benchmark. Two models were tested: Gemini 2.5 Flash and GPT-5-mini.

We discovered that synthetic answers sit closer to the typical answer for a profile than real people with that profile do, yet they barely track the person they were built from. They correlate 0.59 (Gemini) and 0.55 (GPT-5-mini) with the estimated human conditional mean, against 0.44 for the matched humans themselves, a value that mirrors human predictability (√0.192). With the matched individual, the LLMs correlation is 0.28 and 0.26. The rest of the evidence describes how such answers are made. Measured in units of human answer variance, the persona accounts for 0.617 (Gemini) and 0.468 (GPT-5-mini) of response variance, against 0.192 for humans: 2.4 to 3.2 times as much. Synthetic answer profiles span about five effective dimensions; human profiles, nearly ten. Repeated generation adds run-to-run variation worth only 12% of the human residual, so re-prompting does not bring the individual back. Personal values carry more than half (Gemini) and about seven tenths (GPT-5-mini) of the predictable component, against roughly one third for humans, and the cue whose meaning is closest to the question tends to dominate. Withholding the three dominant cues reduces about half of the persona-explained variation. The synthetic answers concentrate, and the model re-anchors on a few of the attributes that remain available for LLMl to use. Synthetic respondents answer as if they applied a low-dimensional, substituting rule to a handful of cues. We call the result over-determination: the persona determines synthetic answers far more strongly than it determines the answers of the people it describes.

These findings indicate where synthetic respondents can be used. They can support directional exploration, questionnaire pretesting, and comparison of broad population segments. Our results do not support using synthetic respondents as individual digital twins or as a guide to differences within a segment. Adding persona detail may raise predictability without recovering the missing individual layer. Validating synthetic respondents requires more than aggregate agreement with human benchmarks: it should also examine conditional variance, dimensional structure, and individual heterogeneity.

## 1. Introduction

Synthetic respondents have moved from research curiosity to commercial products far faster than the evidence needed to evaluate them. Persona-prompted language models promise survey answers within minutes, at a fraction of the cost of fieldwork, and for populations that are hard or expensive to reach. For research buyers and providers, the key question is when such respondents can be trusted, and under what conditions. Decisions about products, messages, and policies may increasingly rest on answers that no one gave.

The current debate runs along two lines. One side shows that synthetic samples reproduce rankings, group contrasts, and aggregate patterns, and reads this as evidence of usefulness. The other shows that they fail for individuals and distributions, and reads this as evidence of unreliability. Both sides measure how far synthetic answers deviate from human ones. Neither explains why the deviation takes the shape it does. Without that explanation, the same model can be judged fit or unfit depending on which benchmark happens to be chosen.

Our approach differs from both. We neither defend nor reject synthetic respondents. We try to describe how a synthetic answer is made, and from that, where synthetic data can be used and where they cannot. Our argument is that both lines of evidence follow from one property of persona-prompted models: they answer much closer to what a person with a given profile would be expected to say than real people with that profile actually do. We tend to think of the model as it behaves like a sampled respondent. In fact, models operate as a judge working from a few cues in the persona, and it applies them with a consistent determination unknown to real surveyed people. If this is right, aggregate success and individual failure are two consequences of one process. The validation process will need to establish what aspects of human responding are actually reproduced with synthetic respondents.

## 1.1 Synthetic respondents as a practical and scientific question

Large language models can be prompted to answer survey questions as if they were given a persona describing a specific person. Populations of such answers, often referred to as silicon samples, are now used for opinion estimation, market research, and the pretesting and piloting of questionnaires (Kumagae & Kaneko, 2026). Synthetic data have also entered commercial research practice, prompting Esomar to issue guidance for buyers on evaluating synthetic-data solutions in business and decision-making contexts. (Esomar, 2025).

The academic debate has developed along three lines. One documents aggregate-level fidelity. A second documents failure at the individual and distributional levels. A third, more recent, builds diagnostic frameworks that locate where synthetic data differs from human data. Existing approaches assess accuracy in different ways, but accuracy alone tells not enough about how a synthetic answer is produced. We wanted to ask that more fundamental question: what process generates an LLM's response when it answers as a persona?

## 1.2 Silicon samples reproduce population patterns under some conditions

Argyle et al. (2023) advanced the concept of algorithmic fidelity, demonstrating that persona-conditioned LLM outputs reproduce attitude patterns from US surveys, including data from the American National Election Studies. In marketing research, silicon samples have been proposed primarily for upstream stages of the process, such as qualitative pretesting and pilot studies (Sarstedt et al., 2024). Semantic Similarity Rating (SSR), developed by PyMC Labs in collaboration with Colgate-Palmolive, elicits free-text responses from LLMs and maps them onto Likert distributions through embedding similarity to reference statements. Across 57 personal-care product concept surveys comprising 9,300 human responses, SSR reproduces human purchase-intent rankings at approximately 90% of human test–retest reliability while maintaining realistic response distributions (KS similarity > 0.85) (Maier et al., 2025).

Aggregate fidelity, however, varies unevenly across items and contexts. In ESS Round 11, adding a country label raises the median correlation between simulated and observed country means from −0.03 to 0.52, with item-level recovery showing substantial variation (C. Wang, 2026). A review of 25 studies finds mixed evidence: fidelity is documented for some outcomes, whereas other studies report distortion of groups and cross-cultural deviation (Zhang, 2026).

## 1.3 Individual-level and distributional failure

A second body of work documents where synthetic respondents fail. Synthetic responses exhibit insufficient variance and remain sensitive to prompt wording and timing (Bisbee et al., 2024). Boelaert et al. (2025) describe a "machine bias" characterized by strong directional bias combined with low within-topic variance, where the bias direction varies unpredictably across topics. LLMs also misrepresent and flatten identity groups (Wang, Morgenstern & Dickerson, 2025).

Studies that match synthetic respondents to specific individuals reach similar conclusions. In a mega-study of 19 preregistered studies with digital twins built from rich individual-level data, individual-level correlations average approximately 0.2; twins exhibit reduced variance compared with humans; and richer individual data improve relative heterogeneity without recovering exact individual responses (Peng et al., 2026). Within an inferential framework, digital twins reproduce average human effects while providing little information about which individuals differ from those averages (S. Wang et al., 2026). In ESS Round 11, individual-level recovery is negligible under all conditions and shows no association with aggregate recovery (C. Wang, 2026). In the Global Preference Survey, covering 67,201 matched respondents across 74 countries, individual-level alignment is systematic but modest; LLMs recover most demographic gradients but only about half of the cultural dimensions (Chen et al., 2026).

What these studies share is a measurement. They establish the size and direction of synthetic–human divergence but say little about the mechanism behind it.

## 1.4 Tools for locating divergence

A third line of research provides diagnostic frameworks. The Total Survey Error (TSE) framework serves as the central organizing structure in survey methodology and categorizes survey error into representation components covering coverage, sampling, and nonresponse, measurement components involving validity, measurement error, and processing, while distinguishing systematic bias from variable error (Groves & Lyberg, 2010). TS2E adapts TSE to the simulation lifecycle and attributes error to design stages. In their ANES 2024 vote-choice case study, the strongest and most consistent effect on performance came from including attitudinal variables, such as party identification and ideology, in the persona (Sen et al., 2026).

The respondent coverage gap (RCG) evaluates entire answer profiles rather than responses to individual questions. It measures how far each human respondent lies from the nearest generated profile, relative to distances between humans (Kumagae & Kaneko, 2026). On two US social surveys, all six LLMs tested leave human respondents uncovered. Under the authors' main settings, these models underperform relative to a baseline that samples each answer independently from human marginals; additionally, every LLM exhibits a higher RCG than a Gaussian copula fitted to the human data.

This stream shifts the question from whether synthetic respondents are accurate to where, and in what form, they diverge. These tools identify and describe divergence without specifying the generative process that produces it.

## 1.5 The implicit model: the respondent as a probabilistic draw

Survey methodology treats an answer as a stochastic outcome. Zaller and Feldman (1992) model the answer as an average over considerations sampled at the moment of response, accounting for response instability and context effects. Classical test theory represents an observed score as the sum of a true score and measurement error, conventionally expressed as $X = T + E$ (Lord & Novick, 1968). Survey-response theory further conceptualizes an observed answer as the product of several cognitive stages: comprehension of the question, retrieval of relevant information, judgment formation, and translation of that judgment into an available response category (Tourangeau, Rips & Rasinski, 2000). Collectively, these perspectives imply that an observed survey response is neither perfectly stable nor error-free.

Evaluations of silicon samples implicitly transfer this model to synthetic respondents. In such frameworks, the LLM is treated as a respondent drawn from the conditional distribution of individuals matching the persona, with aggregates of such draws expected to approximate the population. If this assumption holds, two implications should follow. First, a synthetic respondent should exhibit predictability from the persona comparable to that observed in matched human respondents. Second, a population of synthetic respondents should display response dimensionality similar to that of the human population.

There are reasons from the LLM literature to doubt that this assumption holds. Kirk et al. (2024) find that reinforcement learning from human feedback (RLHF), significantly reduces output diversity relative to supervised fine-tuning across multiple measures and tasks. Park, Schoenegger and Zhu (2024) document zero or near-zero variation in repeated GPT-3.5 responses to several nuanced social-science questions, describing this phenomenon as a "correct answer" effect. These findings make low output diversity a plausible LLM-side mechanism. However, they do not establish that alignment procedures or RLHF caused the variance contraction observed in the present models.

## 1.6 Thesis: over-determination

We test this assumption using a matched one-to-one design. Each synthetic respondent is constructed from the observed attributes of a single ESS Round 11 respondent (28 countries, N = 14,000), with human responses serving as the anchor.

Our main claim is that the LLM answers close to the expected answer for the persona (the conditional expectation): the answer a typical person with that profile would give. The model's version of this expected answer departs from the human one. In this paper we would like to show how:

- differences between personas can be exaggerated or muted (Section 7).
- all answers are shifted by a model-specific amount (Section 3.5).
- the answer leans on persona attributes whose meaning is close to the question (Section 3.6).

A real respondent is one of many people who could share the same profile, and each of them would answer somewhat differently. The model gives the answer characteristic of the profile. We call this over-determination: the persona determines the synthetic answer far more strongly than it determines the answer of the matched human.

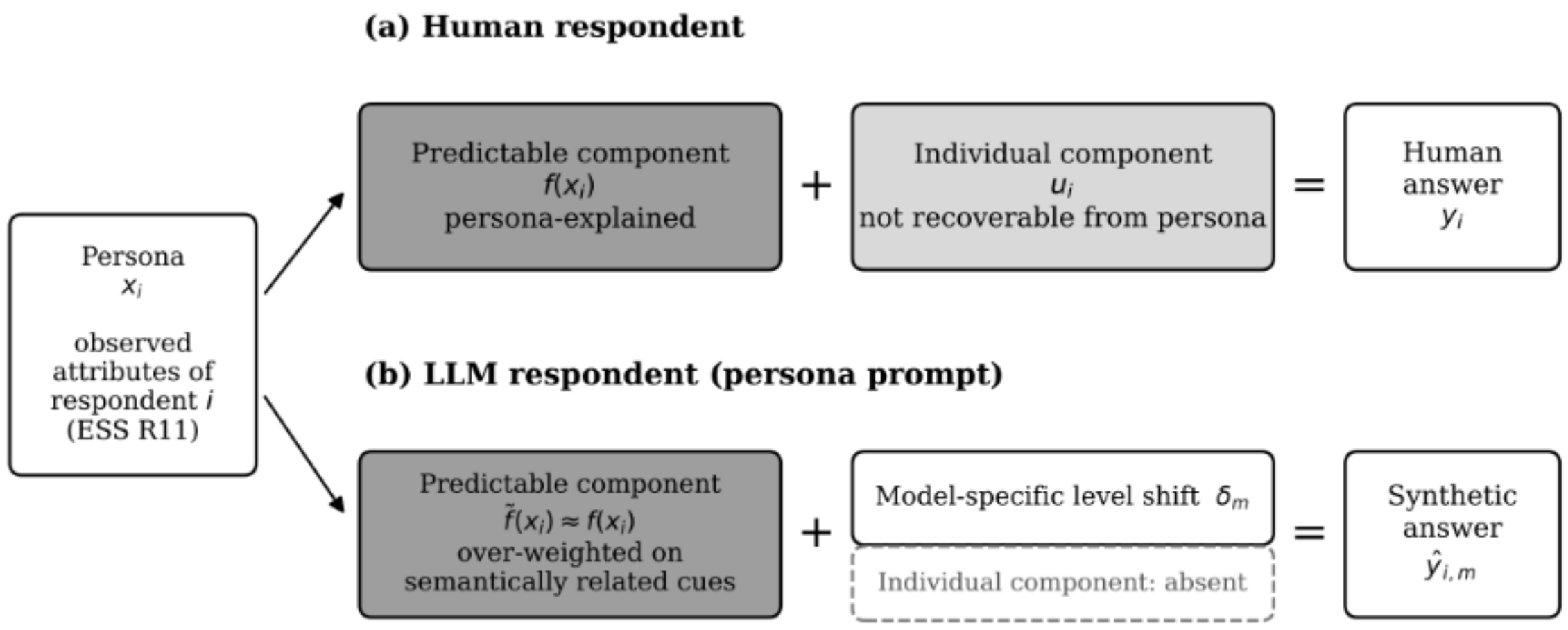

Figure 1. Conceptual diagram.

There is an established vocabulary for this kind of behaviour, and it comes from the study of judgement. Brunswik's lens model describes a judge who infers an unobserved criterion from imperfect cues (Brunswik, 1956). Brunswik’s lens model distinguishes between the predictive value of cues and the weight assigned to them by the judge. Validity describes how strongly each cue relates to the criterion. Cue utilisation captures the weight given to that cue by the judge. The framework examines how closely these two align. Our matched design has this structure. The human data supply the validities: how strongly each persona attribute relates to what real respondents say. The model's answers supply the utilisation: how strongly each attribute shapes what the model says. Read this way, the paper's diagnostics stop being a list of separate statistics. Predictability shows how completely the cues determine the answer. Dimensionality shows how few distinct patterns organise answers across questions. The Shapley decomposition shows which cues are used, and masking shows what happens when the dominant ones are taken away.

A second classic result anticipates the central finding. Dawes (1979) showed that simple linear rules applied consistently to a few cues tend to outperform holistic judgement, even when their weights are imperfect. . Synthetic respondents behave much like such a rule: closer to the human conditional mean than real respondents are, and only weakly correlated with the individual (Section 7). The substitution we observe under masking resembles the cue-by-cue search of fast-and-frugal models of judgement (Gigerenzer & Goldstein, 1996). We do not claim that LLMs implement any of these models. The answers look as if they were produced by a low-dimensional rule that substitutes one cue for another when the first is missing. This is a pattern the vocabulary of judgement describes better than the vocabulary of sampling.

This mechanism fits what all three streams of literature report. Aggregate rankings and group contrasts survive because they depend mainly on the predictable component. Individual answers and distributions fail because the individual-level variation is largely missing. And coverage of human answer profiles is limited, since a lower-dimensional generator confines synthetic responses to a subspace of the human answer space.

The argument has an implication for the field. Benchmarking synthetic respondents against real ones is a necessary stage. The next one is to understand how synthetic answers are generated, because that mechanism should govern how personas are designed and how synthetic data are used.

## 1.7 Research questions

The paper addresses five research questions.

RQ1 (predictability): To what extent can responses be predicted from persona-level attributes for both humans and LLMs?

RQ2 (dimensionality): Do synthetic responses span the same effective dimensionality as human responses?

RQ3 (stochasticity): Do repeated generations for the same persona recover the human individual-level component?

RQ4 (mechanism): Which persona attributes drive synthetic answers relative to human answers, and what occurs when dominant attributes are removed from the prompt?

RQ5 (conditional mean): Are LLM responses closer to the human conditional mean than the responses of matched human respondents?

## 1.8 Contributions

The paper makes six contributions.

First, it introduces a matched one-to-one design using ESS Round 11 data (28 countries, N = 14,000) and two LLMs (Gemini and GPT-5-mini), alongside the human benchmark constructed from the same persona-level information.

Second, it documents a predictability gap. Measured in units of matched human answer variance, the persona accounts for 0.617 (Gemini) and 0.468 (GPT-5-mini) of response variance compared to 0.192 for humans (3.2 and 2.4 times as much). As a share of each source's own variance (out-of-fold $R^2$), the corresponding values are 0.695 and 0.731 against 0.192 (3.6 and 3.8 times as much). The first measure compares how much answer variation the persona produces in each source on a common human scale. The second shows how much of each source's own answer variation the persona accounts for. The gap persists under a reliability sensitivity analysis that corrects GPT-5-mini using its own test–retest reliability and varies human reliability from 1.0 to 0.5.

Third, it separates two properties of synthetic answers: over-determination by the persona and compression of answer variance. These two measures rank models differently, which indicates that the properties are distinct.

Fourth, it documents dimensional flattening: a participation ratio of 9.67 for humans against 5.03 for Gemini and 5.31 for GPT-5-mini (12,442 respondents with valid answers on all 15 outcomes in all three sources).

Fifth, it provides evidence on mechanism. A Shapley decomposition shows predictability shifting from country towards personal values. A test–retest experiment puts run-to-run variation at 12% of the human residual. A masking experiment shows that withholding the three dominant attributes removes about half of the persona-explained variation; the surviving attributes recover only 14–19% of the withheld contribution (Section 6.8), and the model re-anchors on a few of them. When only the leading attribute is withheld, the second-ranked attribute takes over in 13 of 15 outcomes.

Sixth, it offers a reading of these results as cue utilisation. Viewed through the lens model, the findings suggest that the model forms judgments by relying on a narrow set of persona cues and adapting when the most influential cues are unavailable. The reading is interpretive and behavioural; it rests on the same estimates as the rest of the paper.

## 1.9 Structure of the paper

Section 2 describes the data and design. Section 3 reports the predictability gap and its decomposition. Section 4 examines dimensional flattening and addresses the reliability objection. Section 5 reports the test–retest experiment. Section 6 examines the masking of the three dominant attributes. Section 7 compares synthetic and human answers against the human conditional mean. Section 8 reads the results as cue-based judgement and discusses robustness, implications, and limitations.

# 2. Data, Matched Design and Elicitation

## 2.1 Design logic

The design rests on a single principle. Each synthetic respondent is constructed from the observed answers of one real ESS respondent, and that respondent's own survey answers serve as the benchmark. The model receives that person's attributes, and human and model answers are predicted from the same set of predictors. Differences in predictability can therefore be attributed to how the answer is produced rather than to differences between samples.

Section 1 showed that aggregate agreement can mask failure at the individual level. Comparing each synthetic respondent with its human counterpart is therefore the minimum design required to test whether the persona determines the synthetic answer more strongly than it determines the human answer.

The human answer represents one noisy observation of the person's attitude. Matching controls the input conditions but does not convert the human answer into ground truth. Section 4 addresses this concern through a reliability sensitivity analysis.

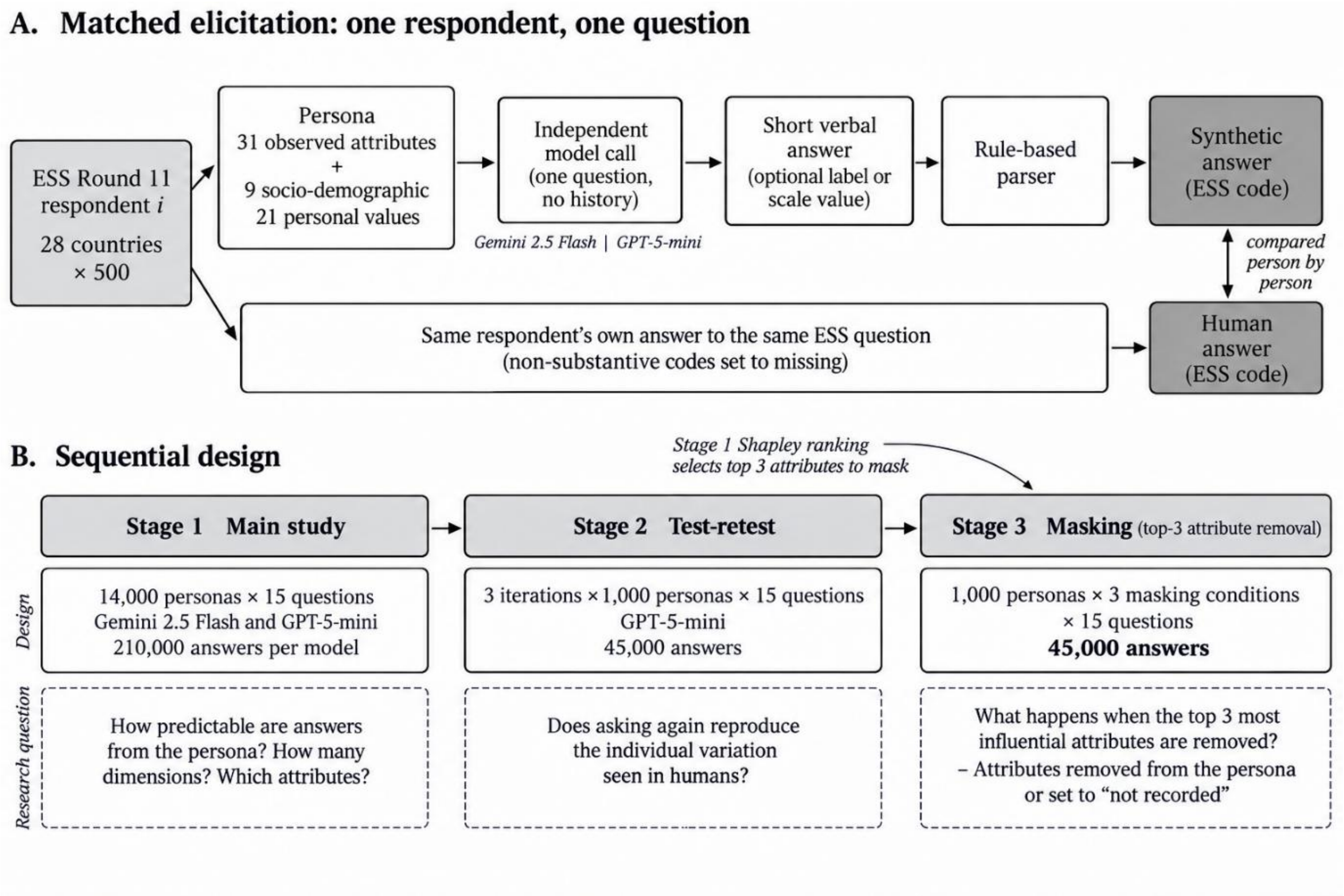

Figure 2. Design scheme.

## 2.2 Data source

The study uses the European Social Survey Round 11 integrated file, edition 3.0 (ESS11e03), released on 2 June 2025. This edition contains data from 28 participating countries and was accessed through the ESS Data Portal hosted by Sikt (ESS ERIC, 2025). The survey draws strict probability samples in every country, with country designs approved by the ESS sampling and weighting expert panel.

The European Social Survey was selected for four reasons. First, it uses probability samples. Second, measurement is standardised across all countries. Third, it includes a validated values instrument called the Human Values Scale, which is based on Schwartz's Portrait Values Questionnaire. Fourth, it covers a broad set of attitudes within a single questionnaire, including trust, politics, immigration, redistribution, climate, religion, authority, and well-being. C. Wang (2026) used the same ESS round for silicon sampling, which provides a direct reference point.

## 2.3 Sample

The sample comprises 14,000 respondents from 28 countries, with exactly 500 per country, randomly drawn from the ESS11 file. The list of selected respondents (country and respondent ID) is released with the replication package.

The sample was designed to study how LLMs use persona information to generate survey responses. Equal allocation assigns each country the same weight and statistical power. This enables cross-country comparisons and yields a broad range of personas across cultural contexts, value profiles, and social positions. Consequently, pooled estimates reflect this design rather than the European population as a whole.

## 2.4 Weighting

The main analysis is unweighted because the quantity of interest is the relation between persona and answer within matched pairs rather than a population estimate. Using human survey weights would assume that humans and models share the same representation errors, which TS2E terms adjustment error (Sen et al., 2026). The population-size weight (anweight) is not used because, with equal country allocation, it would convert the estimand into a European population quantity.

As a sensitivity analysis, each synthetic answer was assigned the post-stratification weight (pspwght) of its matched respondent. These weights were incorporated into model fitting (weighted ridge) and used in $R^2$ and correlation calculations. The participation ratio was computed using the weighted 15-outcome correlation matrix on the same 12,442 complete cases. Out-of-fold $R^2$ values are 0.192 for humans, 0.695 for Gemini, and 0.731 for GPT-5-mini unweighted, versus 0.179, 0.692, and 0.730 weighted. Participation ratios were 9.67, 5.03, and 5.31 unweighted, versus 9.85, 5.14, and 5.38 weighted. Weighting does not alter the reported pattern, and the gap between human and model predictability increases slightly under weighting.

## 2.5 Persona design

Each persona corresponds to one real respondent, described by 31 attributes presented in a consistent order: country; nine socio-demographic and life-situation characteristics; and 21 personal values (Table 1). The socio-demographic block includes age, gender, type of place of residence, years of full-time education, main activity during the last seven days, number of people in the household, whether children have ever lived in the household, perception of household income, and self-rated health.

The values block comprises 21 items from the ESS Human Values Scale, based on Schwartz's Portrait Values Questionnaire. Each item assesses how much a particular goal or way of life matters to the respondent, for instance, being creative, being rich, equal treatment of people, living in safe surroundings, following rules, caring for nature, and following traditions. In the original ESS instrument, respondents indicated how similar they were to a person described by a short verbal portrait, using a six-point scale from "Very much like me" to "Not like me at all". In the persona prompts, we retained each respondent's recorded ESS response and its corresponding verbal scale label while replacing the full portrait wording with a short label, such as "Understanding different people". The prompt labels are therefore abbreviated representations of the ESS Human Values Scale items (ESS ERIC, 2024).

All attribute values correspond to actual respondent answers, rendered with ESS verbal labels (for example "Female", "A big city", "Coping on present income"). No text was generated, imputed, or paraphrased by a model. This is a deliberate design choice; if the model had written the persona description itself, part of the predictability we measure would reflect coherence that the model created itself. The outcome questions were not included in the persona prompt.

The prompt frames the persona as a consumer by using the instruction "You are role-playing as a consumer with these attributes". This framing originates from the market-research use case, whereas the outcome variables concern social and political attitudes.

Attribute order was standardized, with the personal-values block consistently positioned at the end directly preceding the outcome question. LLM-generated responses are sensitive to both the ordering of prompt elements (Guan et al., 2025) and formulation of persona prompts (Lutz et al., 2025), so our observed results may partially reflect this specific prompt architecture. Attribute ordering was not experimentally varied and is acknowledged as a design limitation.

| Attribute as shown to the model | ESS code | Categories in the data (main sample) |
|---|---|---|
| Age | agea | years |
| Gender | gndr | Female, Male |
| Lives in | domicil | A big city; Suburbs or outskirts of big city; Town or small city; Country village; Farm or home in countryside |
| Years of full-time education | eduyrs | years |
| Main activity last 7 days | mnactic | Paid work; Education; Unemployed, looking for job; Unemployed, not looking for job; Permanently sick or disabled; Retired; Community or military service; Housework, looking after children, others; Other |
| People living in household | hhmmb | number |
| Ever had children living in household | chldhhe | Yes; No; Not applicable (routing) |
| Feeling about household income nowadays | hincfel | Living comfortably; Coping; Difficult; Very difficult (on present income) |
| Subjective general health | health | Very good; Good; Fair; Bad; Very bad |
| Values (21 items) | ip…/imp… | Very much like me; Like me; Somewhat like me; A little like me; Not like me; Not like me at all |

table 1. Socio-demographic and value attributes in the persona. Table note: any attribute may also take Don't know, Refusal, No answer or Not available.

## 2.6 Missing information in personas

Persona attributes retain ESS non-substantive codes as verbal labels drawn from the ESS file. These values are not removed or imputed. Two types of missing information are distinguished.

First, item non-response affects 629 of 14,000 personas (4.5%), who have at least one non-substantive persona attribute. Of these, 555 carry at least one Don't Know, Refusal, or No Answer response, while the remaining 74 carry "Not Available" information in at least one attribute. Excluding these cases leaves a sample of n = 13,371 with out-of-fold $R^2$ values of 0.195 (humans), 0.695 (Gemini), and 0.728 (GPT-5-mini), alongside participation ratios of 9.61, 5.05, and 5.41. The overall conclusion remains unchanged.

The second type is "Not Applicable" due to questionnaire routing. This occurs exclusively for the item on whether children have ever lived in the household (4,833 personas, 34.5%). The question is administered only to respondents without children currently living at home; thus, "Not Applicable" identifies respondents who currently reside with children. The prompt displays this label without explaining the routing logic, meaning the model receives a less informative cue than the label would otherwise imply. This is documented as a limitation. In the regression models, non-substantive values are pooled with the reference category, whereas "Not Applicable" forms its own category, so both human and synthetic answers are modelled using identical inputs.

## 2.7 Question wording

The questions were drawn from the ESS Round 11 source questionnaire with three adaptations for self-administered elicitation. First, the country placeholder "[country]" was replaced by "your country". This change affects four items: immigration of people from a different race or ethnic group, whether immigrants make the country a worse or better place to live, influence of the political system, and loyalty towards leaders. Second, interviewer and showcard instructions (for example, "Using this card...") were replaced by the list of answer options and an instruction to select exactly one response. Third, agree–disagree statements are followed by "To what extent do you agree or disagree?", which corresponds to the interviewer prompt in ESS.

The generic country placeholder was used because ESS national questionnaires include the specific country name, whereas in this study the country is provided directly as a persona attribute. This generic wording maintains a single prompt template across all 28 countries, ensuring that question text remains identical regardless of country context. The trade-off is that the model must connect "your country" to the corresponding country attribute within the persona, while ESS respondents were informed explicitly about their own country name. The 15 outcome variables are listed in Table 2.

| Domain | Question (short) | ESS code | Answer format | Coding |
|---|---|---|---|---|
| Social trust | Most people can be trusted, or you can't be too careful | ppltrst | 0–10 | 0–10 |
| | Most people would try to take advantage of you, or try to be fair | pplfair | 0–10 | 0–10 |
| | People mostly try to be helpful, or look out for themselves | pplhlp | 0–10 | 0–10 |
| Politics | How interested in politics | polintr | 4 labels, very … not at all interested | 1–4 |
| | Political system allows people like you to have a say | psppipla | 5 labels, not at all … a great deal | 1–5 |
| | Confident in own ability to participate in politics | cptppola | 5 labels, not at all … completely confident | 1–5 |
| Immigration | Allow people of a different race or ethnic group to come and live here | imdfetn | 4 labels, allow many … allow none | 1–4 |
| | Country made a worse or better place to live by immigrants | imwbcnt | 0–10 | 0–10 |
| Redistribution | Government should reduce differences in income levels | gincdif | 5 labels, agree strongly … disagree strongly | 1–5 |

| Domain | Question (short) | ESS code | Answer format | Coding |
|---|---|---|---|---|
| Authority | Obedience and respect for authority are the most important values children should learn | lrnobed | 5 labels, agree strongly … disagree strongly | 1–5 |
| | What the country needs most is loyalty towards its leaders | loylead | 5 labels, agree strongly … disagree strongly | 1–5 |
| Climate | How worried about climate change | wrclmch | 5 labels, not at all … extremely worried | 1–5 |
| | Personal responsibility to reduce climate change | ccrdprs | 0–10 | 0–10 |
| Religion | How religious are you | rlgdgr | 0–10 | 0–10 |
| Well-being | How satisfied with life as a whole | stflife | 0–10 | 0–10 |

Table 2. Outcomes: domain, short question, ESS code, answer format and coding.

## 2.8 Elicitation

Each prompt comprises a persona description, an instruction to respond from the persona's perspective, the ESS question, the list of answer options with instructions to select exactly one, and an instruction to return only the selected option in the exact wording specified without explanation (Box 1). Each call includes one persona and one question, without any conversation history maintained. Earlier responses are not carried into subsequent calls; therefore, consistency across questions cannot emerge from a shared conversational context. All prompts are written in English, with country information entering as a persona attribute.

The model returns a short verdict, which is either an answer label or a scale value. A rule-based parser converts this verdict into ESS numeric codes through a three-step process: first, exact match to an option; second, match after removing a leading number (for example, "7 – …"); third, unambiguous shortened option of at least 10 characters. Unmatched answers are classified as missing data and are not imputed or re-elicited. Unlike Semantic Similarity Rating (Maier et al., 2025), our parser relies exclusively on the response options provided in the prompt and does not use embeddings or semantic similarity calculations. Coverage in the main study is 209,982 of 210,000 answers for Gemini and 209,999 of 210,000 for GPT-5-mini (Table A2).

You are role-playing as a consumer with these attributes:
- Country: [Poland]
- Age: [47]
- Gender: [Female]
- Lives in: [Town or small city]
- Years of full-time education: [14]
- Main activity last 7 days: [Paid work]
- People living in household: [3]
- Ever had children living in household: [Not applicable]
- Feeling about household income nowadays: [Coping on present income]
- Subjective general health: [Good]

How important these things are to you (from 'Very much like me' to 'Not like me at all'):
- Thinking up new ideas and being creative: [Like me]
- Being rich, having money and expensive things: [A little like me]

- That people are treated equally and have equal opportunities: [Very much like me]
- Showing abilities and being admired: [Somewhat like me]
- Living in secure and safe surroundings: [Very much like me]
- Trying new and different things in life: [Somewhat like me]
- Doing what is told and following rules: [Like me]
- Understanding different people: [Like me]
- Being humble and modest, not drawing attention: [Like me]
- Having a good time: [Somewhat like me]
- Making own decisions and being free: [Very much like me]
- Helping people and caring for others' well-being: [Very much like me]
- Being successful and having achievements recognised: [Somewhat like me]
- That government is strong and ensures safety: [Like me]
- Seeking adventures and having an exciting life: [A little like me]
- Behaving properly: [Like me]
- Getting respect from others: [Somewhat like me]
- Being loyal to friends and devoted to people close: [Very much like me]
- Caring for nature and the environment: [Like me]
- Following traditions and customs: [Somewhat like me]
- Seeking fun and things that give pleasure: [Somewhat like me]

Now answer the following question from the perspective of this consumer profile.
(b) Question block
Question: "How worried are you about climate change? Select exactly one of the following answers:
- Not at all worried
- Not very worried
- Somewhat worried
- Very worried
- Extremely worried
Return only the selected answer, using the exact wording shown above. Do not return an explanation, punctuation, or any additional text."

Box 1. Full prompt for one persona and one question. Answer options were always listed in the same order and were not randomised.

## 2.9 Models

Gemini 2.5 Flash was accessed via Google Vertex AI at temperature = 0 with default settings. GPT-5-mini (version gpt-5-mini-2025-08-07) was accessed through Azure OpenAI using a fixed model snapshot and identical generation parameters for all calls. Temperature was not specified because GPT-5-series reasoning models do not support user-controlled temperature or top_p parameters (Microsoft, 2026; OpenAI, 2025). All other generation parameters remained consistent across API calls.

Sampling differs between the two models and cannot be aligned. Our intention was not comparing randomness across models. We decided to test within-model variability for GPT-5-mini only (Section 5). The Gemini main study was collected from 15 October 2026; the GPT-5-mini main study from 19 October 2026; test–retest data on 22 October 2026; and masking data (FULL, DELETE TOP-3, and NOT RECORDED TOP-3 in one final run) on 27–28 October 2026. Models served through provider APIs may change after this period.

## 2.10 Role of reasoning

GPT-5-mini is a reasoning model, and Gemini 2.5 Flash can also reason internally before generating answers. Both models exhibit the same pattern across different providers and sampling strategies. We do not claim that reasoning doesn't have any significant effect, but rather that the observed pattern does not depend on a single model family or sampling configuration. Related evidence suggests that increasing reasoning effort does not necessarily improve the fidelity of synthetic survey responses. Pew Research Center (2026) found similar person- and question-level match rates across low, medium and high reasoning settings. Testing the effect of varying reasoning effort is left for future work.

## 2.11 Sequential design

The study is organized in three stages (Table 3). The main study establishes the overall pattern, whereas two smaller experiments are designed to test its underlying sources.

Stage 1, designated as the main study, comprises 14,000 personas × 15 questions for both models, yielding 210,000 answers per model. This stage supports analyses of predictability, dimensionality, attribution of explained variance (Shapley decomposition), and comparison with the human conditional mean.

Stage 2, designated as the test–retest experiment, comprises 1,000 personas × 15 questions × 3 repetitions for GPT-5-mini, yielding 45,000 answers. This experiment assesses whether repeated generation of the same persona reproduces individual variation observed in human respondents. Gemini is excluded because repeated calls at temperature = 0 would be expected to return near-identical answers.

Stage 3, designated as the masking experiment, comprises 1,000 personas × 15 questions × 3 conditions for GPT-5-mini, yielding 45,000 answers. For each question, the three persona attributes contributing most to the explained variance of GPT-5-mini answers in Stage 1 are withheld simultaneously by either deleting their rows (DELETE TOP-3) or replacing their values with "not recorded" (NOT RECORDED TOP-3). The FULL condition involves re-elicitation of answers within the same run. This experiment examines what synthetic responses rely upon when their strongest predictive cues are absent. The masking targets were selected from the final Stage 1 Shapley ranking (N = 14,000) and fixed prior to data collection for the masking conditions.

Both 1,000-person samples were randomly drawn from the main study, stratified by country with a distribution of 35–36 personas per country.

| Stage | Model(s) | Personas | Questions | Conditions / repetitions | Answers | Research question |
|---|---|---|---|---|---|---|
| **1. Main study** | Gemini 2.5 Flash; GPT-5-mini | 14,000 (28 countries × 500) | 15 | 1 | 210,000 per model | How predictable are answers from the persona, how many dimensions do they span, and which attributes drive them? |
| **2. Test-retest** | GPT-5-mini | 1,000 (stratified by country, drawn from Stage 1) | 15 | 3 repetitions | 45,000 | Does asking the same persona again reproduce the individual variation seen in humans? |

| Stage | Model(s) | Personas | Questions | Conditions / repetitions | Answers | Research question |
|---|---|---|---|---|---|---|
| **3. Masking** | GPT-5-mini | 1,000 (stratified by country, drawn from Stage 1) | 15 | 3 conditions: full persona; Top-3 rows deleted; Top-3 values marked "not recorded" | 45,000 | What does the synthetic answer rely on when its strongest persona cues are removed? |

Table 3. Sequential design: the three stages of data collection.

# 3. The Predictability Gap

## 3.1 Question and logic

The first research question we wanted to ask is how much of the answer can be predicted from persona-level attributes, for real respondents and for the synthetic respondents built from them. Two accounts of synthetic responding make different predictions. If the model behaves like a respondent drawn from the conditional distribution given the persona, the persona should produce about as much answer variation for the model as for the matched human. If the model answers close to the conditional expectation, the persona should produce far more variation in synthetic answers and leave little unexplained. In cue terms, the first account describes a judge who uses the persona cues about as much as they are worth in human data; the second, a judge who leans on them much harder. Predictors, sample, estimator, and folds are identical across all three sources. Only the answers being predicted differ.

## 3.2 Measures

For each of the 15 outcomes and each source, the answer is predicted from the 31 persona attributes using out-of-fold ridge regression (Hoerl & Kennard, 1970; penalty $\lambda = 1$). All attributes were encoded as categorical indicators, and five folds were constructed using stratification by country. $R^2$ was estimated from pooled out-of-fold predictions for each outcome. The cross-validation design, predictor set, and penalty parameters are identical across humans, Gemini, and GPT-5-mini.

Age, years of education, and household size were included as categorical predictors rather than continuous variables. Age was grouped into seven bands (15–24, 25–34, 35–44, 45–54, 55–64, 65–74, and 75–100 years), whereas years of education were categorized into six bands (0–8, 9–10, 11–12, 13–14, 15–16, and 17–30 years). Household size was treated as categorical and capped at a maximum value of six. Non-substantive persona values (Don't know, Refusal, No answer) were set to missing and, under indicator coding, pooled with the reference category of the attribute; "Not applicable" for chldhhe forms its own category (Table A1). After one-hot encoding with one reference category omitted per attribute, the full-sample design matrix contained 170 columns. All summary statistics represent unweighted means across the 15 outcome variables, and ratios such as "3.2 times" correspond to ratios of these means (0.617 ÷ 0.192 = 3.2).

Two measures of predictability are reported because synthetic responses exhibit differential variance from human answers. On average across outcomes, synthetic response variance is 0.872 for Gemini and 0.615 for GPT-5-mini relative to the matched human variance. Specifically, synthetic response variance refers to the extent of variation in model-generated answers across all personas for a given question, expressed as a ratio to the variation observed in matched human responses to the same question. $R^2$ values alone may therefore appear high not merely because the model explains additional variance, but also because there is limited variance to explain in the first place.

The primary metric is absolute explained variance (AEV). This measures the variance predicted by persona-level attributes, expressed as a ratio to the matched human response variance for the same outcome. The AEV indicates how much answer variation the persona generates and enables comparison across sources whose responses differ in spread:

$$\mathrm{AEV} = \mathrm{R}^2 \times \mathrm{Var}(\text{synthetic answers}) \;/\; \mathrm{Var}(\text{matched human answers})$$

The secondary metric is out-of-fold $R^2$, which represents the proportion of each source's own response variance predicted by persona-level attributes. This measure indicates how strongly the persona constrains the answers of each source, independent of answer spread. Residual variance, calculated as $(1 - R^2) \times$ relative variance, corresponds to the portion of the answer not predicted by the persona, expressed on the same human response scale. AEV and residual variance sum to the total relative variance; for example, in the case of GPT-5-mini, 0.468 + 0.147 = 0.615. On this standardized scale, a value of 1 represents the total variance of human responses to the same question, thereby allowing all three quantities to be interpreted as fractions of the human benchmark.

## 3.3 Main result

The persona generates substantially more response variation in synthetic respondents than in the individuals they describe. Mean AEV across outcomes is 0.192 for humans, 0.617 for Gemini, and 0.468 for GPT-5-mini (corresponding to 3.2 and 2.4 times the human baseline) (Figure 3, Panel A). The secondary metric yields consistent results. Out-of-fold $R^2$ is 0.192 for humans, 0.695 for Gemini, and 0.731 for GPT-5-mini (corresponding to 3.6 and 3.8 times the human baseline) (Figure 3, Panel B).

Just as telling is how little remains unexplained. Residual variance is 0.808 for humans, 0.255 for Gemini, and 0.147 for GPT-5-mini (in human units). Synthetic answers thus keep about one-third of the human unexplained variation for Gemini and less than one-fifth for GPT-5-mini. For humans, the residual holds everything the 31 observed attributes miss: personal history, momentary context, measurement error. Earlier work likewise finds that persona variables account for only a modest share of the variance in human responses on subjective annotation tasks (Hu & Collier, 2024). For the models it is simply the part of the generated answer that the persona attributes do not account for. A rule, in this sense, has almost no residual; a person has a large one.

The two measures rank the models differently. GPT-5-mini is more strongly constrained by the persona, as evidenced by its higher $R^2$ value. Gemini produces substantially more persona-driven variation in absolute terms because GPT-5-mini exhibits greater variance compression (relative variance 0.615 versus 0.872). These findings demonstrate that variance compression and over-determination by the persona are distinct properties of synthetic responses, and that a high $R^2$ cannot be interpreted solely as a consequence of low response variance. The two models were also set up differently (temperature 0 for Gemini; the fixed default of 1 for GPT-5-mini), which could affect total variance, and the AEV/$R^2$ ordering. The direction argues against this explanation: sampling noise at the higher temperature should raise GPT-5-mini's answer variance and lower its $R^2$, whereas we observe the opposite. With only two models, however, decoding cannot be fully separated from other model-specific differences.

Despite differences in provider and sampling configuration (temperature 0 versus the default value of 1) the $R^2$ profiles for both models across the 15 outcomes are highly similar (correlation coefficient = 0.82). Consequently, questions on which one model is most tightly constrained by the persona tend to be the same for the other model.

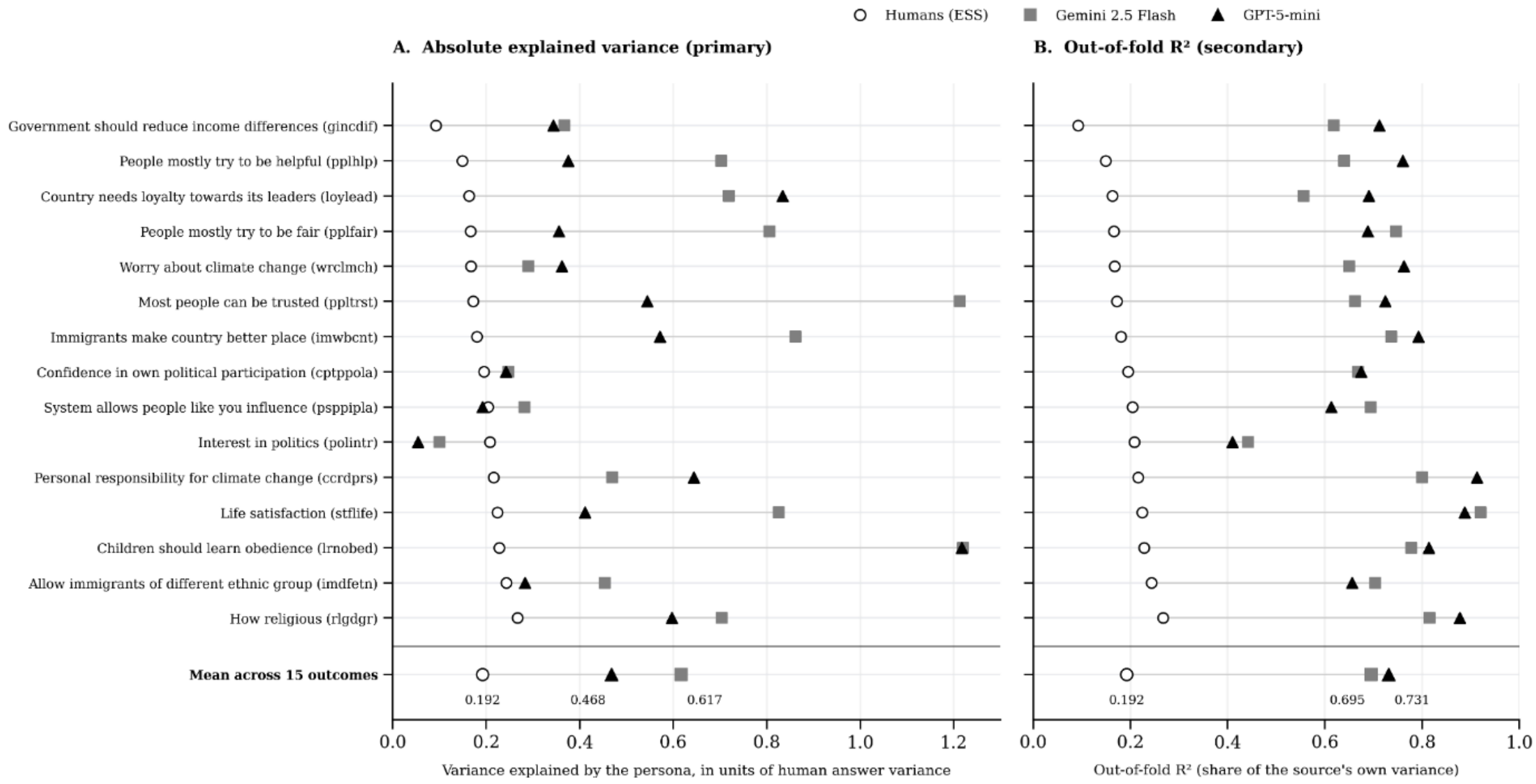

Figure 3. Panel A: absolute explained variance; Panel B: out-of-fold $R^2$, by outcome sorted by human value, with means at the bottom.

## 3.4 Pattern across outcomes

The gap is consistent across outcomes For $R^2$, both models exceed humans on all 15 outcomes (human values range from 0.092 to 0.267; model values range from 0.410 to 0.920). For AEV, Gemini exceeds humans on 14 of 15 outcomes and GPT-5-mini on 13 (Table A3).

These exceptions are informative. Across both models, interest in politics represents the exception (AEV = 0.100 for Gemini and 0.054 for GPT-5-mini, compared to 0.208 for humans). Section 3.5 explains this pattern. For GPT-5-mini only, a second exception is observed: whether the political system allows people like you to have influence (0.192 versus 0.204).

The largest absolute excess appears on items concerning authority and social trust. For obedience as a value for children, AEV equals 1.220 (Gemini) and 1.217 (GPT-5-mini), compared to 0.228 for humans, a ratio of approximately 5.3 times. For loyalty towards leaders, the corresponding values are 0.719 and 0.834 against 0.163, or 4.4 and 5.1 times respectively. For trust in people, Gemini yields 1.213 versus 0.172 for humans, representing a ratio of 7.1 times. Values exceeding 1.0 indicate that persona-driven variation in model-generated responses alone exceeds the total variation observed in human responses to the same question.

On these items, synthetic responses exhibit greater variation than human responses, as evidenced by relative variance exceeding 1. This pattern holds for six of 15 outcomes in Gemini (the three social-trust items, immigrants making the country a better place to live, loyalty towards leaders, and obedience) and for two outcomes in GPT-5-mini (loyalty towards leaders and obedience). Consequently, synthetic responses are not uniformly compressed. Where synthetic responses exhibit greater variation than human responses, this additional variation follows the persona.

## 3.5 Level shift and concentration of answers

Two further properties of synthetic responses provide context for the predictability results. Section 7 examines these properties relative to the human conditional mean.

The first property is a level shift, whereby the mean synthetic response differs from the mean human response. These shifts are expressed in units of human standard deviations and signed according to substantive direction; consequently, a positive value indicates greater endorsement of the attitude named in the question:

$$\text{shift} = (\text{mean synthetic answer} - \text{mean human answer}) \ / \ \text{SD of human answers}$$

Shifts are computed on matched pairs containing valid responses from both sources, with sign reversal applied to items coded against the substantive direction of measurement. The mean absolute shift across outcomes is 0.44 standard deviations for Gemini and 0.50 standard deviations for GPT-5-mini. This measure is at least 0.3 SD on ten outcomes for Gemini and eleven outcomes for GPT-5-mini (Table 4). A shift of 0.5 SD indicates that the average synthetic respondent is positioned half a human standard deviation away from the average real respondent on that particular question.

Both models shift in the same direction on 12 of 15 outcomes. Compared with the matched humans, synthetic respondents (Gemini / GPT-5-mini) are more interested in politics (+0.79 / +0.64); more confident in their own political participation (+0.62 / +0.53) and in their influence on politics (+0.38 / +0.47); more worried about climate change (+0.43 / +0.14) and more likely to feel personal responsibility for it (+0.66 / +0.54); more supportive of income redistribution (+0.48 / +0.47); more open to immigrants of a different ethnic group (+0.19 / +0.56); and more likely to see people as helpful (+0.51 / +1.02) and fair (+0.17 / +1.06). They are also less religious (−0.25 / −0.17) and less in agreement with obedience as a value for children (−0.41 / −0.58) and with loyalty to leaders (−0.81 / −0.09). The models diverge on generalised trust (Gemini −0.56, GPT-5-mini +0.49) and on whether immigrants make the country a better place (−0.27 / +0.73). Life satisfaction shows almost no shift (−0.09 / +0.02).

The shift is partly common to both models and partly model-specific. It corresponds to the level component δ_m in Figure 1.

The direction of mean shifts should be interpreted cautiously. Across outcomes, the common tendency is toward responses associated with greater political engagement and efficacy, stronger pro-environmental attitudes, lower authoritarianism, and a more benevolent view of others. This pattern is compatible with prior evidence that LLMs can shift survey responses toward socially desirable positions when they infer that they are being evaluated (Salecha et al., 2024). It is also broadly consistent with evidence that the opinions expressed by language models can differ systematically from those of human demographic groups, including reported left-leaning tendencies among some human-feedback-tuned models (Santurkar et al., 2023). These studies provide interpretive context rather than a direct explanation of the shifts observed in this study.

The second property is concentration, as synthetic responses are more frequently concentrated in a single response option. The most frequent response option constitutes approximately 46.5% (Gemini) and 45.6% (GPT-5-mini) of all responses, compared to 29.8% for humans. Concentration exceeds that observed in human respondents on 14 of 15 outcomes for each model.

Interest in politics represents the extreme case. The response option "Quite interested" constitutes 79.7% (Gemini) and 88.6% (GPT-5-mini) of responses, compared to 33.2% for human respondents. In the human data, 55.1% of responses indicate that participants are "hardly" or "not at all" interested, versus 5.2% and 7.6% for synthetic respondents, respectively. The option "Not at all interested" was selected by 20.1% of humans but only by 0.1% of model-generated responses. Because 99.99% of model answers to this question match an answer label exactly, the observed pattern arises from the model's response generation process rather than from parsing artifacts.

Concentration explains the political-interest exception documented in Section 3.4. For this outcome, persona-level attributes continue to determine the synthetic response ($R^2$ = 0.44 for Gemini and 0.41 for GPT-5-mini), yet within a constrained and shifted range of values (relative variance = 0.23 for Gemini and 0.13 for GPT-5-mini). Consequently, little persona-driven variation remains in absolute terms. Concentration and above-human spread can coexist: on several 0–10 item questions, models select a narrow subset of values more frequently than humans while distributing those selected values across personas more broadly than humans do.

| | | **Level shift (human SD)** | | **Share of most frequent option** | | |
|---|---|---|---|---|---|---|
| **Outcome** | **Positive shift means** | **Gemini** | **GPT-5-mini** | **Human** | **Gemini** | **GPT-5-mini** |
| Government should reduce income differences (gincdif) | more support for redistribution | +0.48 | +0.47 | 0.429 | 0.509 | 0.514 |
| People mostly try to be helpful (pplhlp) | people seen as more helpful | +0.51 | +1.02 | 0.197 | 0.394 | 0.467 |
| Country needs loyalty towards its leaders (loylead) | more agreement with loyalty to leaders | -0.81 | -0.09 | 0.305 | 0.516 | 0.428 |
| People mostly try to be fair (pplfair) | people seen as fairer | +0.17 | +1.06 | 0.195 | 0.353 | 0.430 |
| Worry about climate change (wrclmch) | more worried | +0.43 | +0.14 | 0.444 | 0.548 | 0.498 |
| Most people can be trusted (ppltrst) | more trusting | -0.56 | +0.49 | 0.188 | 0.433 | 0.332 |
| Immigrants make country a better place (imwbcnt) | immigrants seen as making country better | -0.27 | +0.73 | 0.269 | 0.330 | 0.279 |
| Confidence in own political participation (cptppola) | more confident | +0.62 | +0.53 | 0.330 | 0.587 | 0.523 |
| System allows people like you influence (psppipla) | more perceived influence | +0.38 | +0.47 | 0.345 | 0.507 | 0.566 |
| Interest in politics (polintr) | more interested | +0.79 | +0.64 | 0.350 | 0.797 | 0.886 |
| Personal responsibility for climate change (ccrdprs) | more personal responsibility | +0.66 | +0.54 | 0.173 | 0.470 | 0.253 |
| Life satisfaction (stflife) | more satisfied | -0.09 | +0.02 | 0.273 | 0.446 | 0.424 |
| Children should learn obedience (lrnobed) | more agreement with obedience | -0.41 | -0.58 | 0.395 | 0.393 | 0.442 |
| Allow immigrants of different ethnic group (imdfetn) | more open to immigration | +0.19 | +0.56 | 0.413 | 0.435 | 0.638 |
| How religious (rlgdgr) | more religious | -0.25 | -0.17 | 0.167 | 0.250 | 0.165 |
| **Mean** | **(mean absolute shift)** | **0.44** | **0.50** | **0.298** | **0.465** | **0.456** |

| | | Level shift (human SD) | Share of most frequent option |
|---|---|---|---|

Table 4. Level shift and concentration of synthetic answers by outcome (N = 14,000).

## 3.6 Decomposing the predictable component

### *3.6.1 Purpose and method*

Synthetic responses contain a substantially larger persona-predictable component than human responses. The next question concerns which aspects of the persona account for this predictable variation. Because persona attributes are correlated and can substitute for one another in prediction, individual regression coefficients do not yield a stable decomposition of explained variance. We therefore used the Shapley decomposition of the full ridge-regression fit (Israeli, 2007; Owen & Prieur, 2017). For each outcome, response source, and persona attribute, the Shapley value represents the attribute's average marginal contribution to model fit across alternative subsets of predictors. This procedure allocates the total fitted $R^2$ among all 31 persona attributes, such that attribute-level contributions sum to the full-model fit for each outcome and source.

Shapley values were estimated separately for each outcome and source using Monte Carlo permutation decomposition of the penalized ridge model fitted to the full sample of 14,000 personas. For each sampled attribute ordering, an attribute's contribution was defined as the increase in fitted $R^2$ obtained when that attribute was added to the predictors already present; these marginal contributions were then averaged across sampled orderings.

The fitted $R^2$ of the full-sample model is marginally higher than the out-of-fold $R^2$ used in the primary analysis. The maximum difference across the 45 outcome–source combinations is 0.026, and differences are larger for humans than for the two models. Consequently, Shapley results were employed to decompose the predictable component across attributes rather than serving as an alternative estimate of out-of-sample predictability. To maintain comparability with out-of-fold results, each attribute's Shapley share is multiplied by the out-of-fold AEV for the corresponding outcome and source:

$$\mathrm{AbsoluteShapley(o,s,j)} = \mathrm{ShapleyShare(o,s,j)} \times \mathrm{AEV(o,s)}$$

This rescaling preserves the relative allocation of the decomposition while ensuring that absolute contributions sum to the out-of-fold AEV reported in Section 3.3. Attributes are also aggregated into three predefined blocks: country; nine socio-demographic and lived-context attributes; and 21 personal values. Block results are computed in a fixed order. First, the absolute contributions of the attributes in a block are summed within each outcome and source. Second, each block total is divided by the AEV of that outcome, which gives the block's share of that outcome's predictable component. Third, block shares and block totals are averaged across the 15 outcomes, so every outcome has equal weight. The block shares reported in Figure 4A and Section 8.4 are means of outcome-specific shares. They are not shares of the pooled contributions, and they are not the ratios of the mean block totals in Table 5. Because outcomes differ in AEV, the two summaries differ by up to about five percentage points; they lead to the same conclusions.

The masking targets were derived from the full-sample decomposition presented in Table 6. After masking, predictive structure was re-estimated using a separate cross-validated Shapley decomposition on the 1,000 masking personas (Section 6.11). Because the two decompositions differ in value function and sample composition, their concentration statistics should not be compared directly. This decomposition characterizes the statistical organization of the predictable component rather than identifying an internal causal process within the LLM.

### 3.6.2 Composition of the predictable component

For humans, predictability is distributed relatively evenly across the three information blocks: country accounts for 32.3%, socio-demographic and lived-context attributes for 30.6%, and personal values for 37.1% (Figure 4A). Gemini reallocates the predictable component away from country toward personal values (country 13.8%, socio-demographic and lived-context attributes 33.4%, personal values 52.9%). GPT-5-mini exhibits the strongest reliance on personal values (country 8.9%, socio-demographic and lived-context attributes 21.2%, personal values 69.9%).

Country accounts for more than three times the share of the human predictable component relative to that of GPT-5-mini (32.3% versus 8.9%). Personal values account for approximately one-third of human predictability, slightly more than half of Gemini predictability, and approximately seven-tenths of GPT-5-mini predictability (37.1%, 52.9%, and 69.9%, respectively).

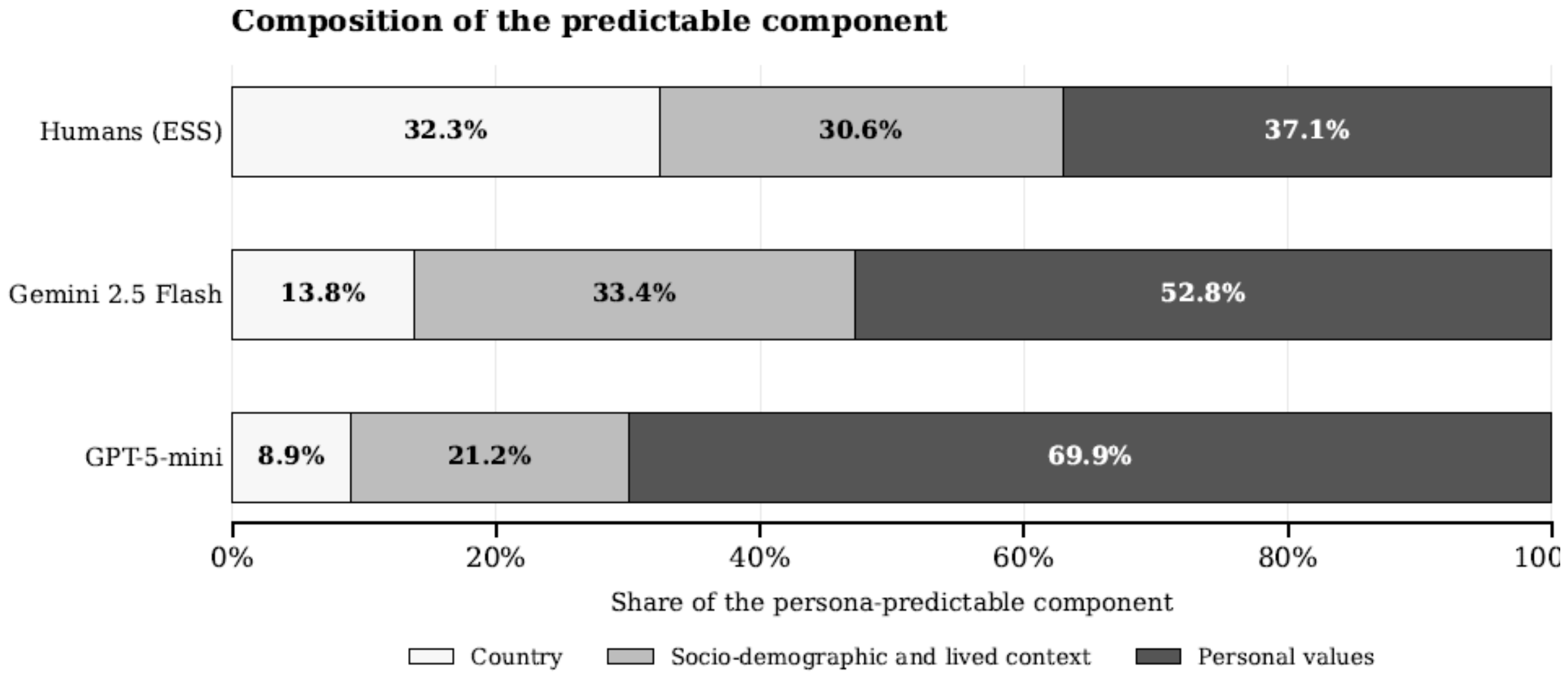

Figure 4A. Composition of the predictable component. Note. Block shares are computed within each outcome and then averaged across the 15 outcomes, with each outcome weighted equally (Section 3.6.1).

### 3.6.3 Absolute contributions

Shares characterize the composition of the predictable component without indicating its absolute magnitude. Because synthetic responses exhibit a substantially larger predictable component than human responses, the absolute differences in predicted variance exceed what the proportional shares alone suggest.

| Information block | Humans | Gemini | GPT-5-mini |
|---|---|---|---|
| Country | 0.060 | 0.087 | 0.037 |
| Socio-demographic and lived context | 0.059 | 0.221 | 0.081 |
| Personal values | 0.072 | 0.310 | 0.351 |
| Total absolute explained variance | 0.192 | 0.617 | 0.468 |

Table 5. Mean absolute contributions by information block, in units of human answer variance. Note. Values are means of outcome-level absolute contributions. Shares in Figure 4A are means of outcome-level shares, so ratios of the block means in this table differ from Figure 4A.

The value-driven component is approximately 4.3 times the human baseline for Gemini and approximately 4.9 times for GPT-5-mini. Block totals sum to total AEV within rounding error (humans 0.191 versus 0.192; Gemini 0.618 versus 0.617). The absolute value-driven component is comparable across both models at 0.310 for Gemini and 0.351 for GPT-5-mini, despite differences in total variance and total AEV. The larger overall predictable component in Gemini is primarily attributable to socio-demographic and contextual information (0.221 versus 0.081). GPT-5-mini does not produce more persona-driven variation than Gemini in absolute terms. Instead, it allocates a larger proportion of the predictable component to personal values, along with a modestly larger absolute amount.

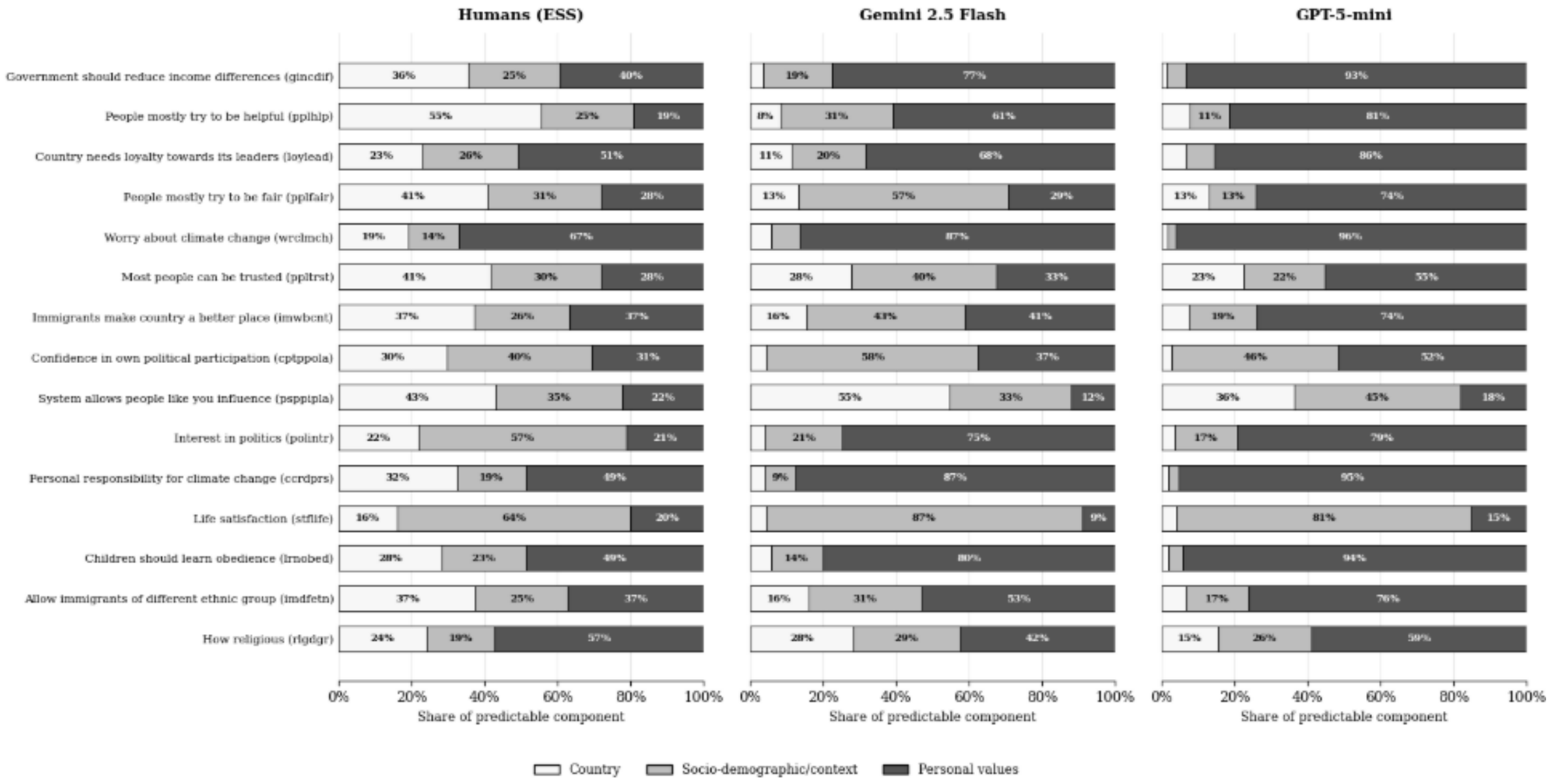

Figure 4B. Outcome-level share of Shapley contributions by source, with country, socio-demographic/context and personal values shown as shares of the outcome-specific predictable component.

### 3.6.4 Attribute-level patterns

At the individual-attribute level, leading predictors vary across outcomes, with neither model relying on a single universal persona attribute. A recurring pattern emerges: where an outcome includes a closely related value item within the persona description, both synthetic models assign that attribute a substantial contribution to prediction.

For the outcome on personal responsibility for climate change (ccrdprs), caring for nature and the environment accounts for 53.5% of the Gemini predictable component and 62.4% of the GPT-5-mini component, compared to 24.9% for humans, in whom country is the leading predictor (32.3%). For support for reducing income inequality (gincdif), belief in equal treatment and equal opportunities accounts for 47.6% (Gemini) and 67.0% (GPT-5-mini), compared to 14.1% for humans, in whom country remains the leading predictor (35.6%). For willingness to allow immigrants from a different ethnic group (imdfetn), country is the leading predictor for humans (37.5%) and also for Gemini (16.0%), whereas GPT-5-mini exhibits the largest contributions from equal treatment and equal opportunities (18.9%) and understanding different people (14.4%).

Political self-efficacy shows a less value-dominated pattern. Years of education is the leading predictor for Gemini and GPT-5-mini (23.2%, 24.3%); for humans it ranks second (14.8%), after country (29.6%) .

The pattern therefore varies across questions, but it has a recognisable logic. When a persona value is semantically close to the question, the models give that cue a large weight; when such value does not exist, contextual and socio-demographic cues stay important. We call this semantic anchoring. In lens-model terms, the difference lies in cue utilisation. The models rely heavily on attributes whose meaning is close to the question (for example, personal responsibility for climate change or support for redistribution), while human responses are often better predicted by cues with no obvious semantic link, such as country. Table 6 lists the three leading GPT-5-mini attributes for each outcome; these are the attributes withheld in the masking experiment. The corresponding rankings for humans and Gemini are given in Table A4.

| Outcome | Top 1 attribute | Top 2 attribute | Top 3 attribute |
|---|---|---|---|
| **Government should reduce income differences**<br>*gincdif* | **Equal treatment and equal opportunities**<br>*ipeqopta*<br>67.0% of predictable component \| Absolute 0.230 | **Helping and caring for others**<br>*iphlppla*<br>6.4% of predictable component \| Absolute 0.022 | **Understanding different people**<br>*ipudrsta*<br>3.8% of predictable component \| Absolute 0.013 |
| **People mostly try to be helpful**<br>*pplhlp* | **Helping and caring for others**<br>*iphlppla*<br>35.0% of predictable component \| Absolute 0.131 | **Understanding different people**<br>*ipudrsta*<br>10.0% of predictable component \| Absolute 0.038 | **Loyalty to close others**<br>*iplylfra*<br>9.1% of predictable component \| Absolute 0.034 |
| **Country needs loyalty towards its leaders**<br>*loylead* | **Following rules and doing what is told**<br>*ipfrulea*<br>28.9% of predictable component \| Absolute 0.241 | **Strong government and safety**<br>*ipstrgva*<br>27.7% of predictable component \| Absolute 0.231 | **Following traditions and customs**<br>*imptrada*<br>6.8% of predictable component \| Absolute 0.057 |
| **People mostly try to be fair**<br>*pplfair* | **Equal treatment and equal opportunities**<br>*ipeqopta*<br>15.7% of predictable component \| Absolute 0.056 | **Understanding different people**<br>*ipudrsta*<br>15.7% of predictable component \| Absolute 0.056 | **Country**<br>*cntry*<br>13.1% of predictable component \| Absolute 0.047 |
| **Worry about climate change**<br>*wrclmch* | **Caring for nature and the environment**<br>*impenva*<br>69.8% of predictable component \| Absolute 0.253 | **Understanding different people**<br>*ipudrsta*<br>3.4% of predictable component \| Absolute 0.012 | **Equal treatment and equal opportunities**<br>*ipeqopta*<br>3.3% of predictable component \| Absolute 0.012 |
| **Most people can be trusted**<br>*ppltrst* | **Country**<br>*cntry*<br>22.6% of predictable component \| Absolute 0.123 | **Understanding different people**<br>*ipudrsta*<br>15.9% of predictable component \| Absolute 0.086 | **Helping and caring for others**<br>*iphlppla*<br>8.9% of predictable component \| Absolute 0.048 |

| Outcome | Top 1 attribute | Top 2 attribute | Top 3 attribute |
|---|---|---|---|
| **Immigrants make country a better place**<br>*imwbcnt* | **Understanding different people**<br>*ipudrsta*<br>16.6% of predictable component \| Absolute 0.095 | **Equal treatment and equal opportunities**<br>*ipeqopta*<br>12.7% of predictable component \| Absolute 0.072 | **Helping and caring for others**<br>*iphlppla*<br>7.4% of predictable component \| Absolute 0.042 |
| **Confidence in own political participation**<br>*cptppola* | **Years of education**<br>*eduyrs*<br>24.3% of predictable component \| Absolute 0.059 | **Freedom and making own decisions**<br>*impfreea*<br>14.7% of predictable component \| Absolute 0.036 | **Feeling about household income**<br>*hincfel*<br>8.7% of predictable component \| Absolute 0.021 |
| **Political system allows people like you influence**<br>*psppipla* | **Country**<br>*cntry*<br>36.4% of predictable component \| Absolute 0.070 | **Feeling about household income**<br>*hincfel*<br>18.4% of predictable component \| Absolute 0.035 | **Years of education**<br>*eduyrs*<br>13.4% of predictable component \| Absolute 0.026 |
| **Interest in politics**<br>*polintr* | **Equal treatment and equal opportunities**<br>*ipeqopta*<br>13.3% of predictable component \| Absolute 0.007 | **Strong government and safety**<br>*ipstrgva*<br>12.4% of predictable component \| Absolute 0.007 | **Freedom and making own decisions**<br>*impfreea*<br>7.6% of predictable component \| Absolute 0.004 |
| **Personal responsibility to reduce climate change**<br>*ccrdprs* | **Caring for nature and the environment**<br>*impenva*<br>62.4% of predictable component \| Absolute 0.402 | **Helping and caring for others**<br>*iphlppla*<br>5.5% of predictable component \| Absolute 0.036 | **Understanding different people**<br>*ipudrsta*<br>4.1% of predictable component \| Absolute 0.027 |
| **Life satisfaction**<br>*stflife* | **Feeling about household income**<br>*hincfel*<br>36.4% of predictable component \| Absolute 0.150 | **Self-rated health**<br>*health*<br>30.9% of predictable component \| Absolute 0.127 | **Main activity**<br>*mnactic*<br>7.8% of predictable component \| Absolute 0.032 |
| **Children should learn obedience and respect for authority**<br>*lrnobed* | **Following rules and doing what is told**<br>*ipfrulea*<br>67.3% of predictable component \| Absolute 0.818 | **Behaving properly**<br>*ipbhprpa*<br>7.2% of predictable component \| Absolute 0.088 | **Strong government and safety**<br>*ipstrgva*<br>4.1% of predictable component \| Absolute 0.050 |
| **Allow immigrants of a different ethnic group**<br>*imdfetn* | **Equal treatment and equal opportunities**<br>*ipeqopta*<br>18.9% of predictable component \| Absolute 0.053 | **Understanding different people**<br>*ipudrsta*<br>14.4% of predictable component \| Absolute 0.041 | **Country**<br>*cntry*<br>6.7% of predictable component \| Absolute 0.019 |
| **Religiosity**<br>*rlgdgr* | **Following traditions and customs**<br>*imptrada*<br>31.1% of predictable component \| Absolute 0.185 | **Country**<br>*cntry*<br>15.4% of predictable component \| Absolute 0.092 | **Age**<br>*agea*<br>7.9% of predictable component \| Absolute 0.047 |

| Outcome | Top 1 attribute | Top 2 attribute | Top 3 attribute |
|---|---|---|---|

Table 6. Top three persona attributes by outcome for GPT-5-mini (attribute label and ESS code, share of the outcome-specific predictable component, absolute contribution in matched human-variance units).

### 3.6.5 Concentration of predictor importance

One potential explanation for over-determination is that synthetic responses are driven primarily by a single cue. The Shapley decomposition does not support this hypothesis. The mean contribution of the leading attribute equals 36.2% for humans, 33.6% for Gemini, and 36.4% for GPT-5-mini. The mean cumulative contribution of the top three attributes is 60.0%, 54.4%, and 58.0%. The effective number of contributing attributes is 5.97, 7.83, and 7.16:

$$N_eff = 1 / \Sigma_j\ Share_j^2$$

Top-1, Top-3 and effective-N are computed per outcome from the 31 attribute shares and then averaged across outcomes. An effective number of six means that the predictable component is spread as if six attributes contributed equally. A lower number means that fewer attributes carry most of it. On these measures, human predictability is about as concentrated as, or more concentrated than, synthetic predictability, and under Top-3 share and effective-N it is more concentrated.

The models differ mainly in the cues they prioritise. Synthetic predictability moves away from country towards personal values, particularly values semantically related to the outcome. Synthetic answers remain multivariate, yet value-based information carries a much larger share of the prediction than it does for real respondents.

One further check bears on the form of the rule. In the gradient-boosting robustness analysis (Table A5), an additive specification without interactions recovers 92% of the out-of-fold $R^2$ of the specification with interactions for humans, 93% for Gemini and 94% for GPT-5-mini. The models show a broadly conventional pattern of additive main effects. What separates them from humans is which cues carry weight, and how much, which is the comparison a cue-utilisation reading is designed to make.

### 3.6.6 From association to intervention

The Shapley decomposition is an observational analysis. Correlated predictors may share information, and a high contribution does not establish that the LLM uses an attribute as a causal decision rule. Consequently, the masking experiment (Section 6) provides a separate test of this question. For each outcome, the three leading GPT-5-mini attributes identified by the decomposition were withheld simultaneously from the persona prompt by either deleting their corresponding rows or marking their values as not recorded. Shapley analysis identifies the attributes most strongly associated with the predictable component. The masking experiment tests whether synthetic responses change when those attributes are no longer available to the model.

## 3.7 Agreement with the matched human

The final question addressed in this section concerns how well synthetic responses represent their corresponding individuals. As a reference point, out-of-fold predictions fitted using human answers and persona-level attributes achieve a correlation of 0.436 with matched human responses. Synthetic model responses correlate with matched human responses at 0.279 (Gemini) and 0.259 (GPT-5-mini), corresponding to 64% and 59%, respectively, of the reference value. In variance terms, this equates to approximately 41% and 35% for Gemini and GPT-5-mini, respectively ($0.279^2/0.192$ and $0.259^2/0.192$). The reference correlation of 0.436 reflects the accuracy achieved by a regression model trained on human responses using only persona-level predictors, thereby demonstrating the predictive capacity of persona information when employed as a human-calibrated benchmark.

Taken together, these results describe a sharp asymmetry. The persona constrains most of the synthetic answer, yet that answer reaches less than two-thirds of the correlation with the real person that a human-fitted persona model attains. Two further questions follow. If synthetic answers consist mainly of the persona-driven component, they should span fewer dimensions than human answers (Section 4). And if the residual is small, much of it may be run-to-run noise (Section 5).

# 4. Dimensional Structure of Response Profiles

## 4.1 Research question

Section 3 demonstrated that persona-level attributes account for substantially more response variation in synthetic respondents than in their matched human counterparts. This section examines whether synthetic response profiles similarly occupy a lower-dimensional space across the 15 outcome variables.

Response variance and response dimensionality capture distinct properties. Variance describes how widely answers are distributed across individuals on a single outcome. Dimensionality describes the number of relatively independent patterns required to represent the joint response profile across all outcomes. A synthetic population can exhibit substantial, or even above-human, variance on individual outcomes while remaining low-dimensional if the same underlying response patterns organize multiple outcomes simultaneously. We examine how the 15 outcomes correlate with one another, rather than looking at the variance of each outcome in isolation.

## 4.2 Participation ratio

Effective dimensionality is measured with the participation ratio (PR) of the 15-outcome correlation matrix (Recanatesi et al., 2022). With eigenvalues $\lambda_1, \ldots, \lambda_{15}$:

$$PR = (\Sigma_k \lambda_k)^2 / \Sigma_k \lambda_k^2$$

The participation ratio can be interpreted as the effective number of dimensions contributing variation to the response profile. When variance is distributed evenly across all 15 components, PR approaches 15. Conversely, when a few components account for most of the correlation structure, PR decreases. In the extreme case where all 15 responses move together as one general tendency, PR would approach 1.

The participation ratio is computed from Pearson correlations of the 15 numeric outcome codes, with each outcome standardised before dimensionality is evaluated. This approach is justified because Section 3 demonstrated that total response variance differs across human respondents and the two models (Gemini and GPT-5-mini). A correlation-based PR measures the organisation of the joint response profile while controlling for differences in outcome scale and total variance between sources. The main analysis uses the same 12,442 respondents with valid answers on all 15 outcomes across all three sources; therefore, differences in missing-data patterns cannot influence the comparison.

## 4.3 Main result

Synthetic response profiles have much lower effective dimensionality than human profiles: 9.67 for humans against 5.03 for Gemini and 5.31 for GPT-5-mini. Synthetic profiles therefore retain 52.0% (Gemini) and 54.9% (GPT-5-mini) of the human effective dimensionality. Relative to the theoretical maximum of 15, humans occupy 64.5%, Gemini 33.5%, and GPT-5-mini 35.4%. Both models organise their answers to the 15 questions into roughly five effective dimensions, where humans need nearly ten, and the result holds across two providers and two sampling configurations. In practice, knowing a synthetic respondent's answers to some questions tells you more about its remaining answers than the same knowledge would tell you about a real person.

## 4.4 Interpretation

A lower participation ratio indicates that responses to different questions exhibit stronger covariation and can be represented by fewer relatively independent response patterns. The model still produces distinct answers across questions, which span different substantive domains and employ different response scales. Lower dimensionality therefore indicates that these answers exhibit stronger interdependence among them. This distinction is important because synthetic responses are not uniformly less variable. Gemini exhibits greater variance than humans on six outcomes and GPT-5-mini on two, with several of the largest positive differences occurring in authority and social-trust outcomes. Lower dimensionality consequently cannot be attributed to a general narrowing of every response distribution.

Total variance and effective dimensionality are distinct properties at the model level. Gemini preserves 0.872 of human response variance with a participation ratio of 5.03, whereas GPT-5-mini preserves 0.615 of human response variance with a slightly higher participation ratio of 5.31. Consequently, Gemini maintains a larger absolute magnitude of variation while organising it across marginally fewer dimensions; GPT-5-mini compresses total variation more substantially and distributes the remaining variation across slightly more dimensions.

The results complement Section 3. Synthetic answers contain a larger persona-predictable component and a more tightly organised joint profile. The Shapley decomposition suggests one source of this structure. Personal values account for 52.9% of the predictable component in Gemini and 69.9% in GPT-5-mini, against 37.1% in humans. If the same few value cues are consulted for many different questions, the answers to those questions will move together, and dimensionality falls. Over-correlation of synthetic judgements across domains has also been reported for silicon samples of professional philosophers (Shi & Haupt, 2026). A rule that runs on half the dimensions of the thing it imitates is, operationally, a heuristic. This is an interpretation drawn from the combined evidence. The participation ratio alone does not identify which attributes produce the lower-dimensional structure; the causal role of the leading attributes is examined through masking in Section 6.

## 4.5 Robustness of dimensionality

Two robustness checks were conducted to examine whether the results depend on analytic choices (Table 7). Using ESS post-stratification weights, PR equals 9.85 for humans, 5.14 for Gemini, and 5.38 for GPT-5-mini. Excluding 629 personas containing at least one item non-response label (leaving 11,954 complete cases), PR is 9.61, 5.05, and 5.41 for humans, Gemini, and GPT-5-mini, respectively. Consequently, the lower dimensionality of synthetic responses is not attributable to ESS post-stratification weights or to the handling of non-substantive persona labels.

| Analysis | Complete-case N | Humans | Gemini | GPT-5-mini |
|---|---|---|---|---|
| Main, unweighted | 12,442 | 9.67 | 5.03 | 5.31 |

| Analysis | Complete-case N | Humans | Gemini | GPT-5-mini |
|---|---|---|---|---|
| ESS post-stratification weight | 12,442 | 9.85 | 5.14 | 5.38 |
| Exclude personas with item non-response labels | 11,954 | 9.61 | 5.05 | 5.41 |

Table 7. Robustness of response-profile participation ratios.

## 4.6 Reliability sensitivity

A separate objection concerns the predictability gap itself. Human responses may appear less predictable because survey data contain measurement error and temporary response variation (Section 1.5). This concern is distinct from dimensionality. Reliability affects the estimated relationship between persona-level attributes and observed responses, whereas the participation ratio characterizes the correlation structure across outcomes. We examine this objection using a symmetric sensitivity analysis rather than treating observed human responses as free of measurement error.

For each outcome o, observed predictability is divided by an outcome reliability coefficient and capped at 1:

$$R^2_o,\text{corrected} = \min(R^2_o,\text{observed} \,/\, \rho_o, 1)$$

Human reliability is not directly observed in the present dataset. Corrected human predictability was therefore evaluated across a grid of assumed item-reliability values ($\rho_H$), with no single assumed value presented as an empirical property of the ESS outcomes. For GPT-5-mini, reliability was estimated empirically and separately for each outcome as intraclass correlation coefficient ICC(1) derived from three independent repeated generations of the same personas and questions (Section 5):

$$R^2_\text{GPT},o,\text{corrected} = \min(R^2_\text{GPT},o,\text{observed} \,/\, \text{ICC}(1)_o, 1)$$

The overall corrected estimate for GPT-5-mini represents the mean of the 15 outcome-level corrected values. The resulting corrected full-persona $R^2$ equals 0.88. This differs from dividing the mean observed $R^2$ by the mean ICC (0.731 / 0.825 = 0.886), because the correction is applied to each outcome before averaging. No GPT-5-mini outcome reaches the cap of 1. Gemini was not reliability-corrected because it was elicited at temperature = 0 and was excluded from the test–retest experiment.

At an assumed human reliability of 0.70, the corrected mean human $R^2$ equals 0.274 versus 0.880 for GPT-5-mini, corresponding to a ratio of 3.21. Under this condition, GPT-5-mini exceeds the corrected human estimate on all 15 outcomes. At an assumed reliability of 0.50, the corrected mean human $R^2$ equals 0.383, and GPT-5-mini remains above the human estimate on all 15 outcomes (Table 8).

The more informative quantity concerns the level of human reliability at which measurement error alone would suffice to close the gap. The median assumed human reliability at which corrected human predictability equals corrected GPT-5-mini predictability is 0.222. Measurement error alone would therefore account for closing the gap only if more than three-quarters of the variation in a typical human answer were attributable to noise. This break-even threshold (Table A6) ranges from 0.111 for support for reducing income differences to 0.328 for interest in politics. Measurement reliability thus reduces the estimated magnitude of the predictability gap but does not eliminate it under conventional assumptions regarding measurement reliability.

The analysis functions as a sensitivity test. Human reliability values are specified rather than estimated from repeated ESS measurements, and GPT-5-mini reliability is derived from repeated model generations instead of being obtained through psychometric instrumentation. The two concepts are comparable as attenuation corrections but are not identical measurements.

| Assumed human reliability | Corrected human mean $R^2$ | GPT-5-mini / human |
|---|---|---|
| 1.0 | 0.192 | 4.59 |
| 0.9 | 0.213 | 4.13 |
| 0.8 | 0.240 | 3.67 |
| 0.7 | 0.274 | 3.21 |
| 0.6 | 0.320 | 2.75 |
| 0.5 | 0.383 | 2.30 |

Table 8. Reliability sensitivity of the predictability gap.
Note. Human estimates are hypothetical corrections of observed mean $R^2$ = 0.192. GPT-5-mini uses the mean outcome-specific reliability correction from the independent test-retest experiment. At assumed human reliability 0.70 and 0.50, GPT-5-mini exceeds the corrected human estimate for all 15 outcomes. Because the correction is monotonic in the assumed reliability, GPT-5-mini also exceeds the corrected human estimate on all 15 outcomes at every assumed reliability between 0.5 and 1.

### 4.7 What the results establish

Synthetic response profiles exhibit substantially lower effective dimensionality than matched human response profiles. This pattern is observed for both Gemini and GPT-5-mini and remains robust to weighting procedures as well as the exclusion of personas containing item non-response labels. Lower dimensionality does not imply uniformly lower variance; rather, it reflects stronger organization of responses across questions into fewer independent patterns.

Together with Section 3, the evidence demonstrates three interrelated features of synthetic responses: a larger persona-predictable component, a smaller residual component, and a lower-dimensional joint structure. Although human measurement unreliability can account for part of the difference in predictability, it would have to be severe across all outcomes to eliminate the gap entirely. These findings are consistent with a response process that preserves and amplifies a structured persona-related component while containing less of the additional variation present in human response profiles. The participation ratio does not establish whether the missing variation represents stable individual differences, omitted variables, temporary response instability, or measurement error. The next section examines one aspect of this question directly: how much of the synthetic residual is attributable to run-to-run variation.

## 5. Test–Retest Stability and Run-to-Run Variation

### 5.1 Research question

Sections 3 and 4 demonstrate that GPT-5-mini responses exhibit a large persona-predictable component, a small residual component, and a lower-dimensional joint structure. This finding does not imply that the model operates as a fully deterministic system. Repeated prompting of the same persona with identical questions may yield different responses. If missing individual-level variation could be recovered simply through repeated model sampling, then over-determination would reflect data collection methodology instead of a structural characteristic of the generated response.

The test–retest experiment therefore examines three aspects of GPT-5-mini stability across independent model calls: (1) how much variation is introduced between successive prompts, (2) the proportion of total variance attributable to repeated generation within the same persona, and (3) whether this variation is comparable to the individual-level differences observed among human respondents. The key methodological distinction lies between stable between-persona variation, where different personas consistently produce distinct answers, and within-persona variation, where a single persona generates different responses across calls.

## 5.2 Design

The experiment utilizes 1,000 personas randomly drawn from the main analytical sample and stratified by country distribution. Each persona responds to all 15 outcome variables across three repetitions, yielding a total of 45,000 responses. Every repetition constitutes a separate GPT-5-mini call with no conversation history and no access to previous responses. Persona-level attributes, question wording, and available answer options remain constant throughout the experiment. The design therefore measures variability generated by repeated model invocations while observed respondent information remains fixed.

The experiment was conducted exclusively for GPT-5-mini, whose API uses a fixed default temperature of 1; thus, within-persona variance reflects call-to-call variation under this sampling configuration. Gemini 2.5 Flash was not incorporated into the analysis because it was elicited at temperature = 0 and would be expected to yield minimal sampling variation upon repeated calls. Any remaining differences observed in such a scenario could reflect service-side nondeterminism or changes in the served model.

## 5.3 Variance decomposition

For persona i, outcome o and repetition r, the repeated synthetic answer is represented as:

$$Y_ior = \mu_o + u_io + \varepsilon_ior$$

where μ_o denotes the overall mean for outcome o, u_io represents the stable difference between personas, and ε_ior captures variation between repeated calls for the same persona and outcome. For each outcome, total synthetic variance is decomposed into between-persona and within-persona components, both expressed in units of the matched human response variance. Within-persona variance therefore indicates how much human-scale variation is generated solely through repeated model invocations of the same question. Test–retest reliability was measured using the one-way intraclass correlation coefficient (Shrout & Fleiss, 1979):

$$ICC_o = \sigma^2_between,o \, / \, (\sigma^2_between,o + \sigma^2_within,o)$$

A higher ICC indicates that response differences are stable across personas, whereas a lower ICC indicates a larger contribution from run-to-run variation. An ICC of 1.0 would indicate that the model produces identical responses for the same persona across repeated calls. An ICC of 0.0 would indicate that variation attributable to differences between personas is no greater than variation arising from differences between repeated calls. Additionally, we report pairwise exact agreement between repetitions and the proportion of persona–outcome combinations yielding identical answers across all three repetitions.

## 5.4 Overall stability

GPT-5-mini is highly but not perfectly stable. Mean test–retest ICC(1) across the 15 outcomes is 0.825. On average, 82.5% of the variance in GPT-5-mini responses therefore reflects stable differences between personas, and the remaining 17.5% reflects within-persona variation across independent generations. This 17.5% is the mean of the 15 outcome-level within shares. On average, 70.5% of response pairs are exactly identical, and all three answers are identical for 57.9% of persona–outcome combinations. The complementary 29.5% of answer pairs that differ between identical calls serves as the noise baseline for answer changes in the masking experiment (Section 6.9).

Mean total GPT-5-mini variance in the test–retest sample equals 0.616 in human-variance units, closely matching the 0.615 observed in the main analytical sample. Mean within-persona variance equals 0.098 in the same units. Both figures characterize variation in model-generated responses, expressed relative to the variance of matched human responses to the same outcome.

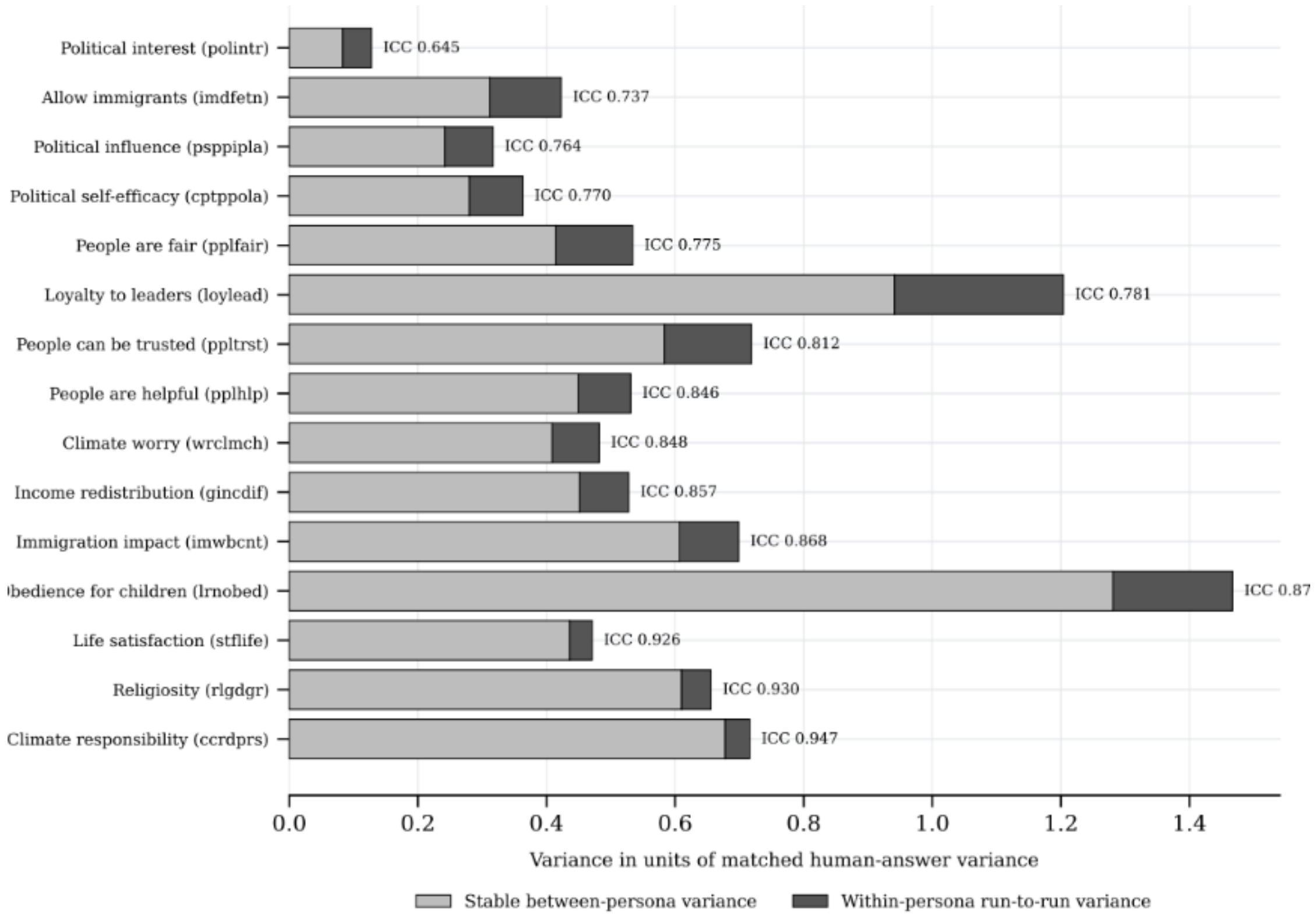

Figure 5. Test–retest variance decomposition: stable between-persona and within-persona run-to-run components of GPT-5-mini response variance, outcomes sorted by ICC.

## 5.5 Differences across outcomes

Stability exhibits substantial variation across outcomes (Table 9). Intraclass correlation coefficients range from 0.645 for political interest to 0.947 for personal responsibility to reduce climate change. Outcomes characterized by high stability include personal responsibility for climate change (0.947), religiosity (0.930), and life satisfaction (0.926). Outcomes characterized by lower stability include political interest (0.645), willingness to allow immigrants from a different ethnic group (0.737), perceived political influence (0.764), and confidence in political participation (0.770).

Within-persona variance ranges from 0.035 human-variance units for life satisfaction to 0.263 for loyalty towards leaders. Expressed as a proportion of total model variance, run-to-run variation accounts for 5.3% for personal climate responsibility, 7.0% for religiosity, 7.5% for life satisfaction, and 35.5% for political interest. Exact agreement depends in part on the number and utilization of response categories; consequently, it should be interpreted alongside ICC values and within-persona variance rather than examined in isolation.

| Outcome | Within-persona variance | Total model variance | Within share | ICC(1) | Pairwise exact agreement | All 3 answers identical |
|---|---|---|---|---|---|---|
| Climate responsibility (ccrdprs) | 0.038 | 0.716 | 5.3% | 0.947 | 62.2% | 46.3% |
| Political self-efficacy (cptppola) | 0.083 | 0.363 | 23.0% | 0.770 | 80.4% | 70.6% |
| Income redistribution (gincdif) | 0.076 | 0.528 | 14.3% | 0.857 | 90.0% | 85.2% |
| Allow immigrants (imdfetn) | 0.111 | 0.423 | 26.3% | 0.737 | 81.4% | 72.1% |
| Immigration impact (imwbcnt) | 0.092 | 0.699 | 13.2% | 0.868 | 48.0% | 27.6% |
| Loyalty to leaders (loylead) | 0.263 | 1.204 | 21.9% | 0.781 | 73.6% | 60.9% |
| Obedience for children (lrnobed) | 0.186 | 1.467 | 12.7% | 0.873 | 80.4% | 71.0% |
| Political interest (polintr) | 0.045 | 0.128 | 35.5% | 0.645 | 92.1% | 88.2% |
| People are fair (pplfair) | 0.120 | 0.534 | 22.5% | 0.775 | 48.7% | 29.1% |
| People are helpful (pplhlp) | 0.082 | 0.531 | 15.5% | 0.846 | 54.4% | 35.5% |
| People can be trusted (ppltrst) | 0.136 | 0.719 | 18.9% | 0.812 | 49.3% | 30.4% |
| Political influence (psppipla) | 0.075 | 0.317 | 23.6% | 0.764 | 85.8% | 78.7% |
| Religiosity (rlgdgr) | 0.046 | 0.656 | 7.0% | 0.930 | 51.9% | 33.5% |
| Life satisfaction (stflife) | 0.035 | 0.471 | 7.5% | 0.926 | 72.5% | 59.0% |
| Climate worry (wrclmch) | 0.073 | 0.482 | 15.2% | 0.848 | 86.5% | 79.8% |
| **Mean** | **0.098** | **0.616** | **17.5%** | **0.825** | **70.5%** | **57.9%** |

Table 9. GPT-5-mini test–retest stability and variance components by outcome.
**Note.** *N = 1,000 personas per outcome and three independent calls per persona. Within-persona and total model variance are expressed in units of matched human-answer variance. Within share is within-persona variance divided by total model variance. ICC(1) is the one-way intraclass correlation. Exact agreement is scale-dependent and should be interpreted together with ICC and variance components.*

## 5.6 Political interest

Political interest again exhibits a distinctive pattern that warrants further examination. This outcome demonstrates the lowest ICC value (0.645) and the largest within-persona share of model variance (35.5%), while simultaneously exhibiting the highest pairwise exact agreement rate (92.1%) and the highest proportion of personas yielding identical answers across all three repetitions (88.2%). The explanation concerns concentration effects. GPT-5-mini assigns the response label "Quite interested" to 88.6% of personas (Section 3.5), so most personas receive the same modal answer in every repetition. Only a small minority of personas shift between categories, and these shifts generate a relatively large share of variance within an already narrow distribution with total variance of only 0.128 human-variance units. High exact agreement results from one response category dominating the distribution; ICC is lower because the limited remaining variance contains a relatively large run-to-run component; and absolute within-persona variance remains small at 0.045 human-variance units.

## 5.7 Relationship to the main-sample residual

The test–retest results may be compared with the residual variance identified in Section 3. In the main analysis, GPT-5-mini residual variance unexplained by persona-level attributes equals 0.147 human-variance units. In the test–retest experiment, mean within-persona variance equals 0.098 human-variance units. Consequently, the run-to-run component constitutes approximately two-thirds of the mean residual (0.098 / 0.147 ≈ 0.67), and the difference between the two means equals 0.049.

At the outcome level, within-persona variance ranges from 54.5% to 74.9% of the corresponding GPT-5-mini residual, with a mean outcome-level ratio of 65.2% (Figure 6). This differs slightly from the ratio of means (0.098 / 0.147 ≈ 0.67) because each ratio is computed per outcome before aggregation across outcomes. Consequently, this comparison suggests that a substantial portion of the GPT-5-mini residual is compatible with variation between repeated model calls.

The remaining 0.049 may include stable persona-related structure not captured by the ridge specification, nonlinear or interaction effects, sampling differences between the main and test–retest samples, outcome-specific variance differences between the two samples, and other systematic model variation outside the 31-attribute model. Consequently, this remainder should not be interpreted directly as stable individual-level variation. The residual and the test–retest component were estimated in different samples using distinct decomposition procedures; therefore, their subtraction constitutes descriptive variance accounting rather than an exact identity. Nevertheless, the comparison provides a useful approximation of scale: most of the model variation unexplained by the persona regression is approximately equal in magnitude to run-to-run instability.

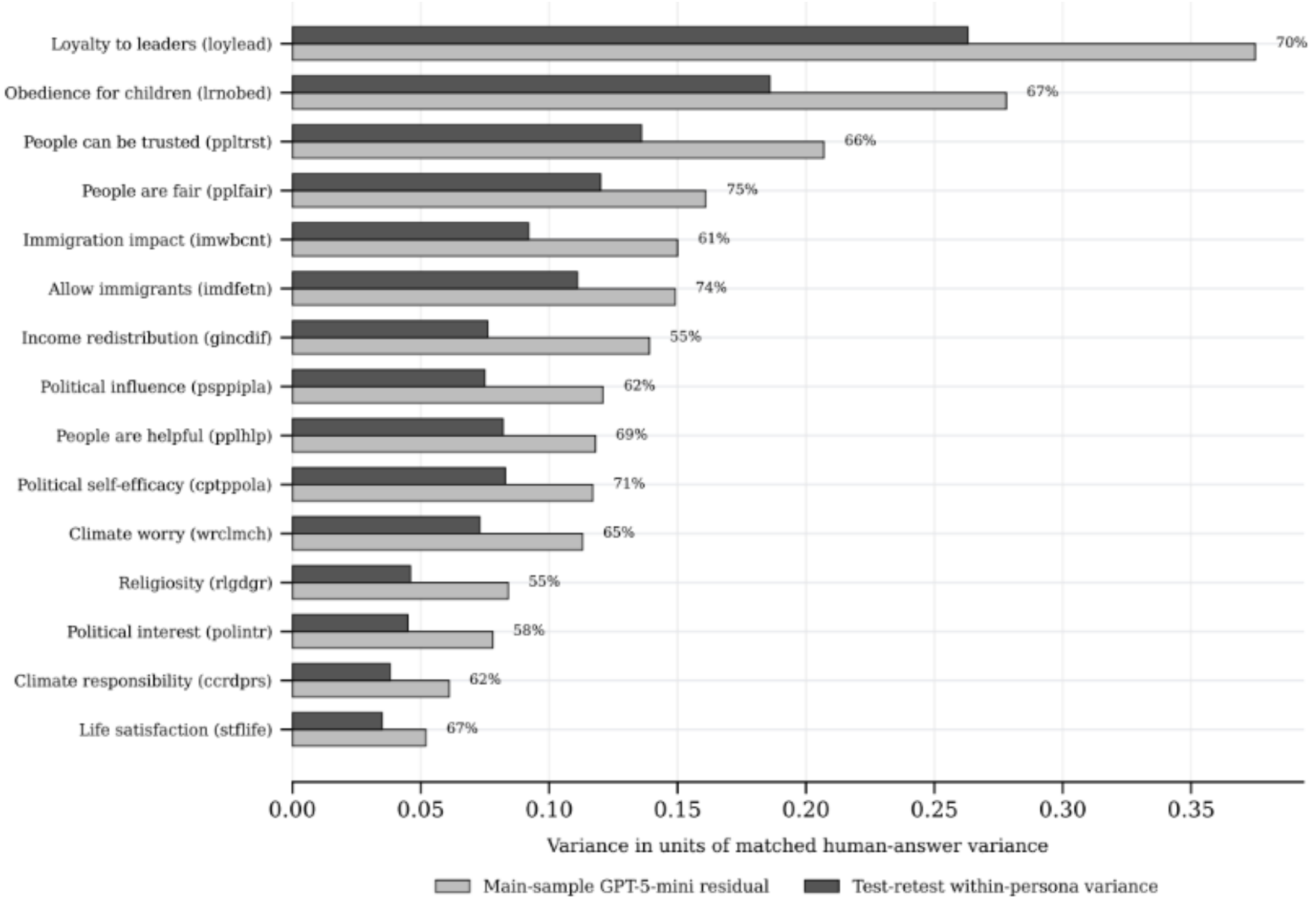

Figure 6. Residual accounting by outcome. GPT-5-mini residual variance from the main sample and within-persona variance from test-retest, both in human units.

## 5.8 Does repeated generation restore human variation?

The central comparison concerns the human residual variance. Mean within-persona GPT-5-mini variance equals 0.098 versus the human residual variance of 0.808, corresponding to a ratio of approximately 12.0%. Consequently, run-to-run variation accounts for only 12.0% of the residual variation observed in human responses. This metric contextualizes the variation generated by repeatedly querying the model with identical inputs against the variation among real persons that the persona leaves unexplained. At the outcome level, the ratio ranges from 4.5% for life satisfaction to 31.5% for loyalty towards leaders.

Repeated generation introduces some variation, but its scale is substantially lower than the residual variation in human responses that the persona does not predict. Averaging repeated generations would diminish the run-to-run component additionally; however, this approach would yield a more stable estimate of the model's conditional response without recovering the missing human-level variation. Consequently, introducing additional stochasticity through multiple model invocations cannot be considered equivalent to sampling individuals from the human conditional distribution.

The results distinguish two sources of variation. Model randomness captures variation in repeated outputs for the same persona. Individual heterogeneity refers to stable differences among persons that cannot be explained by observed persona-level attributes. The experiment provides evidence on model randomness but does not recover individual heterogeneity.

## 5.9 What the results establish

GPT-5-mini responses exhibit high but imperfect stability across independent model calls. Approximately 82.5% of total synthetic variance captures stable differences between personas, whereas 17.5% captures variation between repeated invocations. The run-to-run component accounts for a substantial portion of the model residual yet remains substantially smaller than residual variation observed among human respondents. Consequently, repeated elicitation does not recover the individual-level layer missing from the synthetic response profile.

These findings refine what over-determination means. The model's answers do vary across independent calls. But the variation is small, it does not persist, and it is far too limited to reproduce human-level individual differences; most stable synthetic differentiation is still organised by persona attributes. Adding randomness through repeated LLM calls increases response variation, but does not reproduce the heterogeneity observed in human populations. What remains open is which persona information carries this organisation, and what the model does when that information is removed.

# 6. Masking Dominant Persona Information

## 6.1 Research question

The Shapley decomposition in Section 3.6 identifies, for each outcome, the three persona attributes that contribute most to the GPT-5-mini predictable component. That finding is observational. The masking experiment intervenes instead: it withholds the values of these attributes before GPT-5-mini answers the question. In the terms of Section 1.6, it asks what a cue-based judge does when its strongest cues are missing.

The experiment addresses four research questions. First, to what extent is persona-explained variation lost when the three most predictive attribute values are unavailable? Second, does GPT-5-mini recover any portion of the lost structure from remaining persona information beyond what can be accounted for statistically by those remaining attributes alone? Third, does predictive importance redistribute across the remaining attributes or instead re-concentrate on a few of them relative to purely statistical redistribution? Fourth, is the effect driven primarily by information removal or by physically shortening the prompt length? Masking constitutes an intervention on input variables. This procedure does not reveal the model's internal causal computations.

## 6.2 Selection of the masked attributes

The masking targets are specified separately for each of the 15 outcome variables as the three highest-ranked attributes in the final GPT-5-mini Shapley decomposition (Table 6). Consequently, the identified targets vary across questions. Personal responsibility to reduce climate change (ccrdprs), for example, is masked on caring for nature and the environment, helping and caring for others, and understanding different people. Life satisfaction (stflife) is masked on household-income situation, self-rated health, and main activity. Political interest (polintr) is masked on equal treatment and opportunities, preference for strong government and safety, and freedom to make one's own decisions. These targets were selected prior to the analysis of masked responses. The three values are withheld simultaneously, and the remaining 28 persona-level attributes remain unchanged throughout the experiment.

## 6.3 Experimental conditions

The same 1,000 personas responded to identical 15 outcome variables under three experimental conditions.

In the FULL condition, serving as the experimental baseline, GPT-5-mini receives the complete persona containing all 31 attributes. FULL responses were newly elicited within the same analytical run as the two masking conditions; consequently, all three conditions share the same personas, model snapshot, collection period, question wording, response options, and prompt template structure.

In the DELETE TOP-3 condition, the three selected attribute rows are removed entirely from the persona prompt. Neither the attribute labels nor their values are displayed to the model. For example, if the selected attributes for an outcome include household-income situation ("Coping on present income"), self-rated health ("Good"), and main activity ("Paid work"), none of these rows appears in the prompt. The model receives a physically shorter persona containing 28 attributes rather than the full set. This condition removes substantive information while simultaneously shortening the prompt length and altering the spacing between remaining rows.

In the NOT RECORDED TOP-3 condition, the same three attribute rows retain their original positions, while their values are replaced with the neutral label "Not Recorded". In the illustrative example, the prompt would contain "Feeling about household income: Not Recorded", "Self-rated health: Not Recorded", and "Main activity: Not Recorded". The model can identify which attributes lack specified values but cannot access their actual values. This procedure preserves attribute names, their original positions, the order of all remaining attributes, and the overall prompt structure and approximate length. The label "Not Recorded" functions as a researcher-facing placeholder for missing information; it does not imply that the respondent refused to provide the data or characterized it as unknown.

## 6.4 Why two masking conditions are required

Both masking conditions withhold the values of the same three outcome-specific attributes across all observations. The two conditions differ exclusively in how absence is represented to the model. This design addresses a potential confound related to prompt length and structural organization. If the masking effect were primarily driven by physical shortening of the prompt, DELETE TOP-3 would produce materially different results from NOT RECORDED TOP-3; for example, it would manifest as larger changes in response variance, greater loss of persona-explained variation, or a more dispersed predictive structure. Conversely, if the main effect stems from loss of substantive information, the two conditions should yield similar results despite their distinct prompt structures.

The design cannot completely disentangle prompt length from the various consequences associated with different representations of missing information. Instead, these two conditions bracket this concern: DELETE TOP-3 represents complete physical removal of the attributes, while NOT RECORDED TOP-3 preserves the attribute slots and removes their content only. Interpretation of agreement between the two conditions is taken as evidence that information removal, rather than prompt shortening, drives the principal effect observed in this experiment.

## 6.5 Sample, elicitation and audit

The experiment utilizes 1,000 personas randomly drawn from the main analytical sample and stratified by country distribution. With 15 outcome variables and three experimental conditions, the design produces a total of 45,000 responses (15,000 responses per condition). Every persona–outcome combination yields exactly one valid response across all three conditions. Each response was generated through a separate API invocation without conversation history or access to responses from other conditions. Attribute ordering conforms to that used in the main study, and question wording along with response options remain identical across conditions. All 45,000 responses passed parsing procedures and permissible-range audits; there were no unresolved output truncations or incomplete persona–outcome–condition combinations.

Two robustness checks support the comparability of the masking experiment with Stage 1. First, 39 of the 1,000 personas carry at least one item non-response label (Don't Know, Refusal, or No Answer). Excluding these cases alters AEV retention by less than one percentage point across conditions (DELETE: 53.7% to 54.1%; NOT RECORDED: 51.1% to 51.7%). Second, to assess snapshot drift, FULL responses from the masking run were compared with Stage 1 GPT-5-mini answers for the same personas; exact agreement occurs in 69.9% of persona–outcome pairs versus 70.5% agreement between independent repeated calls in Stage 2 (test–retest sample). The latter figure serves as a reference level rather than a matched comparison because it comes from the test–retest sample rather than the masking personas. The mean level difference is 0.023 human standard deviations, and the variance ratio is 0.98. Consequently, the model behaved consistently with Stage 1, and FULL responses are comparable with the main study estimates.

## 6.6 Prediction models

A separate ridge regression model was estimated for each outcome and condition. FULL responses were predicted from all 31 persona-level attributes, whereas DELETE TOP-3 and NOT RECORDED TOP-3 responses were predicted from the same 28 remaining attributes. The withheld attributes were not reintroduced into the masked regressions as missing-value indicators; consequently, each masked regression measures how much response structure remains recoverable from information actually available to the model. All conditions employed a ridge penalty of $\lambda = 1$, used the same five country-stratified cross-validation folds, maintained identical coding of the remaining attributes, and generated out-of-fold predictions. The attribute coding followed the main study design: categorical indicators with one reference category dropped per variable; age grouped into seven bands; education into six bands; household size capped at a maximum value of six; and, unlike the main study, non-substantive values retained as their own category, giving 177 design columns under FULL and 138–162 under the masked conditions. These cross-validation folds remained identical to those used for the post-masking Shapley analysis described in Section 6.11.

## 6.7 Outcome measures

As in Section 3, AEV is the primary outcome:

$$\mathrm{AEV(o,c)} = \mathrm{R^2(o,c)} \times \mathrm{Var(Y(o,c))} \,/\, \mathrm{Var(Y_H(o))}$$

This metric estimates the persona-explained component of synthetic responses expressed in units of matched human response variance. Out-of-fold $R^2$ serves as a secondary outcome and is not interpreted independently of response variance, since masking may leave a condition proportionally predictable while substantially reducing the magnitude of variation in its responses. Consequently, we report relative response variance, out-of-fold $R^2$ values, AEV metrics, retention rates relative to the FULL condition, modal concentration measures, and the share of individual answers changed relative to the FULL baseline.

## 6.8 Main masking result

Withholding the three dominant attribute values removes about half of the persona-explained variation (Table 10).

| Condition | Relative response variance | OOF $R^2$ | Absolute explained variance |
|---|---|---|---|
| FULL | 0.613 | 0.661 | 0.427 |
| DELETE TOP-3 | 0.428 | 0.495 | 0.230 |
| NOT RECORDED TOP-3 | 0.420 | 0.486 | 0.219 |

Table 10. Main masking result: means across 15 outcomes.

Retention represents the mean across all 15 outcomes within each masked condition, divided by the corresponding mean under the FULL baseline, calculated from unrounded outcome-level estimates. Relative to the FULL condition, DELETE TOP-3 retains 69.8% of response variance, 74.9% of out-of-fold $R^2$ values, and 53.7% of AEV. NOT RECORDED TOP-3 retains 68.5% of response variance, 73.5% of out-of-fold $R^2$ values, and 51.1% of AEV. All reported values describe GPT-5-mini responses within the masking sample. Relative response variance characterizes the spread of model-generated answers relative to the spread of matched human answers, computed per outcome and averaged across outcomes.

Respondent-level uncertainty was estimated using 2,000 paired persona bootstrap samples stratified within country, with identical resampled personas applied across all three conditions. The bootstrap procedure was applied to frozen cross-fitted predictions; consequently, it quantifies sampling uncertainty conditional on the fixed predictive modeling framework. The contrast between absolute and proportional predictability remains stable under this design. DELETE TOP-3 retains 53.7% of AEV (95% CI [51.2%, 56.2%]) compared with 74.9% of out-of-fold $R^2$ (95% CI [72.7%, 77.1%]). NOT RECORDED TOP-3 retains 51.1% of AEV (95% CI [48.6%, 53.7%]) compared with 73.5% of out-of-fold $R^2$ (95% CI [71.2%, 75.6%]).

The gap between these two retention figures has a straightforward source. Masking also contracts the variance in model-generated responses. Approximately three-quarters of proportional predictability remains after masking; however, this applies to responses that exhibit substantially less variation than under the FULL condition.

The two masking implementations produce substantively similar results, although they are not numerically identical. DELETE TOP-3 retains 2.6 percentage points more AEV than NOT RECORDED TOP-3 (95% CI [0.5, 4.6]); the corresponding difference in $R^2$ retention is 1.5 percentage points (95% CI [−0.8, 3.8]). The AEV difference aligns with a small prompt-length effect operating in the expected direction: deleting rows physically shortens the prompt and may slightly increase the relative importance of remaining information. Conversely, the phrase "Not Recorded" may itself signal that information has been withheld, and the current design cannot distinguish between these two secondary mechanisms. Under either interpretation, the difference remains substantially smaller than the loss of approximately 46–49 percentage points of AEV relative to the FULL condition and therefore cannot account for the main result.

These findings demonstrate both a strong dependence on selected attributes and continued, but weaker, organisation around remaining information. The three selected Top-3 attributes are behaviourally consequential, whereas the remaining 28 attributes continue to organise model responses, though substantially less in absolute terms.

### *Matched-predictor control*

The retained structure may merely reflect the analyst fitting fewer predictors. To evaluate this hypothesis, FULL responses were additionally predicted using the same 28 attributes available in the DELETE TOP-3 condition. Under this FULL-28 benchmark specification, however, the model received information about all 31 attributes during response generation, while prediction was conducted using only the 28 surviving attributes.

Mean AEV equals 0.184 for FULL-28 and 0.230 for DELETE-28. After masking, the same 28 attributes therefore explain a larger absolute amount of response variation despite reduced total response variance (0.230 of 0.428 after masking versus 0.184 of 0.613 before masking). Once the strongest cues are unavailable, the model relies more heavily on the attributes it still observes than when the full persona was provided. The bootstrap mean difference is:

$$\Delta\text{AEV} = \text{AEV(DELETE-28)} - \text{AEV(FULL-28)} = 0.0459, 95\%\ \text{CI}\ [0.0298, 0.0623]$$

The ratio of mean AEV for DELETE-28 to FULL-28 equals 1.25 (95% CI [1.15, 1.38]). Consequently, the increase in the absolute explanatory contribution of the surviving attributes is unlikely to reflect sampling variation within the persona panel. This control provides evidence of genuine reorganisation after masking: the remaining attributes carry greater absolute predictive weight once Top-3 information is withheld compared with their weight in responses generated from the complete persona.

The same comparison can be expressed as a recovery share. The Top-3 attributes contribute FULL (0.427) minus FULL-28 (0.184), equaling 0.243 of absolute explained variance that the surviving 28 attributes do not account for statistically. After masking, the surviving attributes account for an additional 0.046 in DELETE TOP-3 and 0.035 in NOT RECORDED TOP-3, corresponding to differences between their respective AEV values (0.230 and 0.219) and the FULL-28 baseline (0.184). The surviving attributes therefore recover approximately 19% in DELETE TOP-3 and 14% in NOT RECORDED TOP-3 of the withheld contribution from remaining persona information:

Recovery = (AEV_masked-28 − AEV_FULL-28) / (AEV_FULL − AEV_FULL-28)

### *Sequential masking: ordered cue substitution*

An earlier sequential masking experiment, which removed the leading Shapley attributes one at a time, shows how this reorganisation begins. When the original Top-1 attribute was withheld, the original Top-2 attribute became the leading predictor in 13 of 15 outcomes (Table A8). This is cue substitution, observed item by item: when the strongest cue is unavailable, GPT-5-mini typically moves to the next-ranked one in the pre-specified overall Shapley ranking, much as a lexicographic judge would. It is the most direct evidence in the paper for an ordered cue hierarchy. The sequential run used an earlier ranking, so the headline masking estimates come from the final two-arm experiment; the substitution pattern itself is the mechanistic result.

## 6.9 Individual answer changes and modal concentration

Masking alters the substantive content of model responses rather than merely adjusting the regression estimates fitted to them. Relative to the FULL baseline, 52.9% of DELETE TOP-3 answers change versus 54.3% of NOT RECORDED TOP-3 answers, exceeding a noise baseline derived from identical repeated calls (Section 5.4): approximately 29.5% of answer pairs differ under this control condition. The mean absolute answer change equals 0.523 human standard deviations for DELETE TOP-3 and 0.538 for NOT RECORDED TOP-3.

Mean modal concentration rises from 45.1% under FULL to 55.0% under DELETE TOP-3 and 54.8% under NOT RECORDED TOP-3. One might expect missing information to show up as more dispersed, less committed answers. On average the opposite happens: answers move towards a narrower and more common pattern. The tendency is not universal, and some outcomes become more dispersed after masking (Section 6.15).

## 6.10 Deleting rows versus marking values as not recorded

The two masking conditions produce substantively similar macro-level results. Variance retention, $R^2$ retention, AEV retention, the share of answers changed from FULL and the mean modal share are close in the two conditions (Table 11). Across all 15 outcomes, the mean absolute difference between conditions equals 0.024 in AEV, 0.029 in $R^2$, and 0.026 in relative response variance.

| Metric | DELETE TOP-3 | NOT RECORDED TOP-3 |
|---|---|---|
| Variance retention | 69.8% | 68.5% |
| $R^2$ retention | 74.9% | 73.5% |

| Metric | DELETE TOP-3 | NOT RECORDED TOP-3 |
|---|---|---|
| AEV retention | 53.7% | 51.1% |
| Answers changed from FULL | 52.9% | 54.3% |
| Mean modal share | 55.0% | 54.8% |

Table 11. Masking macro model results

Preserving the attribute rows with "Not Recorded" does not materially restore the lost response structure, nor does deleting the rows produce a substantially larger effect. Consequently, prompt shortening cannot account for the main masking effect. As noted in Section 6.4, these two conditions bracket the length confound rather than isolate it as an independent mechanism. Although some outcome-level differences persist, they do not alter the overall conclusion that information removal, rather than physical prompt shortening, drives the principal masking result.

## 6.11 Post-masking Shapley decomposition

To examine how predictive structure reorganises, Shapley contributions were re-estimated separately for each outcome and condition, rather than applying the FULL-condition Shapley shares to masked responses. The cooperative game includes 31 attributes under FULL and the same 28 remaining attributes under both masked conditions. The value function employs five-fold cross-validation using out-of-fold ridge regression $R^2$ with identical cross-validation folds and ridge penalty parameters as in the primary masking analysis. The decomposition uses 48 randomly sampled attribute orderings together with their reverses, yielding 96 antithetic paths for each outcome and condition. The mean feature-level Monte Carlo standard error equals 0.00157, with a maximum of 0.01237. This uncertainty reflects the inherent variation in permutation averaging and is not a respondent-level bootstrap.

Cross-validated Shapley contributions may assume negative values because adding a weak predictor can degrade out-of-fold prediction performance. Raw signed contributions are released in the replication files. Concentration statistics utilize only positive Shapley mass; specifically, negative contributions are set to zero, and remaining positive contributions are renormalized to sum to 100%. Because this cross-validated decomposition differs from the full-sample decomposition described in Section 3.6 in both design and sample composition, these two sets of concentration statistics should not be compared directly.

## 6.12 Concentration after masking: statistical spreading, model re-anchoring

After masking, predictive contribution appears to spread out.

| Condition | Leading-attribute share | Top-3 cumulative share | Effective-N |
|---|---|---|---|
| FULL | 42.8% | 66.4% | 5.20 |
| DELETE TOP-3 | 26.1% | 53.5% | 8.15 |
| NOT RECORDED TOP-3 | 27.0% | 54.2% | 7.80 |

Table 12. Concentration of positive Shapley contribution by condition (means across outcomes).

Under the FULL condition, the leading attribute accounts on average for 42.8% of positive predictive contribution. After masking, the leading remaining attribute accounts for approximately 26–27%, the cumulative share of the three leading remaining attributes decreases by approximately 12–13 percentage points, and effective-N increases from 5.20 to approximately eight.

Most of this increase is attributable to a mechanical effect. Removing the three largest contributions from any distribution inherently raises effective-N. Two baselines demonstrate how much spreading would be expected without any change in model behaviour. A mechanical baseline removes the Top-3 Shapley contributions from the FULL decomposition and renormalises the remaining contributions to 100%; this yields a leading share of 21.2% and an effective-N of 9.14. A statistical baseline decomposes the FULL responses using only the 28 remaining attributes; in this design, the analyst withholds the Top-3 while the model retains access to all information (same cross-validation folds, ridge penalty parameters, and 96 antithetic paths). This baseline yields a leading share of 21.5% and an effective-N of 8.91.

The observed masking conditions exhibit greater concentration than either baseline. Effective-N equals 8.15 for DELETE TOP-3 and 7.80 for NOT RECORDED TOP-3, corresponding to differences of 0.76 and 1.11 below the statistical baseline, respectively. The leading share is 26.1% for DELETE and 27.0% for NOT RECORDED versus 21.5% in the statistical baseline. When the model's strongest predictive cues are withheld, GPT-5-mini does not distribute its response across the remaining attributes as evenly as statistical redistribution alone would predict. Instead, it re-anchors its answers on a subset of the available attributes.

The mean reflects two contrasting outcome-level patterns. Effective-N declines by more than one unit below the statistical baseline in both masking conditions for six outcomes, whereas it increases by more than one unit above the baseline for two outcomes (Section 6.15). The effective number is computed as 1 divided by the sum of squared positive, renormalised Shapley shares s (i.e., $1 / \Sigma s^2$), and the audit reproduces the FULL Shapley values from Section 6.11 exactly (maximum difference < 1e-15). Baseline values represent means across the 15 outcome variables.

## 6.13 Consistency of redistribution across the two masking conditions

The patterns of predictive redistribution are substantially similar whether attribute rows are deleted or their values are marked as "Not Recorded". The DELETE TOP-3 condition yields marginally lower concentration and a higher effective-N relative to the NOT RECORDED condition. This pattern aligns with the expectation that prompt shortening produces a small additional dispersive effect, yet the difference remains minor relative to the overall change from FULL. At the outcome level, the two conditions agree closely: the difference in effective-N from the statistical baseline correlates at $r = 0.94$ across all 15 outcomes and maintains the same sign for 14 of them. Consequently, the re-anchoring pattern observed after masking does not depend on how unavailable attributes are represented.

## 6.14 Redistribution across information blocks

Under the FULL condition, positive Shapley contributions are distributed among three attribute blocks: country (9.4%), socio-demographic and life-situation characteristics (21.9%), and personal values (68.7%). After masking, personal values remain the largest component; however, their relative share declines while the socio-demographic and contextual share increases.

| Condition | Country | Socio-demographic/context | Personal values |
|---|---|---|---|
| FULL | 9.4% | 21.9% | 68.7% |
| DELETE TOP-3 | 4.9% | 33.5% | 61.6% |
| NOT RECORDED TOP-3 | 3.9% | 32.6% | 63.5% |

Table 13. Relative composition of persona-explained variation by information block.

Because the total predictable component contracts after masking, a larger share does not necessarily mean a larger absolute contribution. The absolute contributions clarify what happens.

| Information block | FULL | DELETE TOP-3 | NOT RECORDED TOP-3 |
|---|---|---|---|
| Country | 0.037 | 0.010 | 0.009 |
| Socio-demographic/context | 0.079 | 0.074 | 0.074 |
| Personal values | 0.312 | 0.145 | 0.136 |

Table 14. Absolute explained variance by information block, in units of matched human answer variance.

Block shares represent means of outcome-level positive contributions, and absolute contributions sum to the AEV for each condition within rounding (FULL 0.427, DELETE 0.230, NOT RECORDED 0.219). Masking does not generate additional socio-demographic signal in absolute terms. The absolute value-driven component decreases by more than half; conversely, the country-level contribution contracts sharply while the pre-existing socio-demographic and contextual component remains almost fully preserved. Socio-demographic information occupies a larger relative share because it persists within a smaller total AEV. Consequently, the aggregate response to masking combines three elements: loss of value-driven structure, preservation of contextual structure, and re-concentration on a subset of remaining attributes (Section 6.12).

## 6.15 Outcome-specific heterogeneity

Aggregate masking results obscure substantial heterogeneity across outcomes. The outcome-level analyses covering AEV retention, variance retention, modal concentration, leading surviving attributes, and the distribution of positive Shapley contributions provide descriptive characterisations. These patterns are exploratory in nature and do not constitute statistically defined categories. Full outcome-level results are provided in Table A7. Two contrasting response patterns are described hereafter.

The first pattern is fallback homogenisation. For certain outcomes, masking substantially reduces response variance and shifts answers toward a narrower, more concentrated distribution. Loyalty towards leaders retains 10.3% of FULL AEV under DELETE TOP-3 and 8.6% under NOT RECORDED TOP-3; its variance retention amounts to approximately one-quarter, and modal concentration increases by more than 22 percentage points. Life satisfaction retains 9.8% and 12.0% of FULL AEV under DELETE TOP-3 and NOT RECORDED TOP-3, respectively; its variance retention is approximately one-fifth, and modal concentration increases by more than 26 percentage points. In these cases, the retained $R^2$ describes predictability within a much narrower fallback distribution rather than reconstruction of the original response structure.

The second pattern is expansion after masking. Three outcomes exhibit greater absolute persona-explained variation after masking than under the FULL condition. For allowing immigrants from a different ethnic group, AEV retention equals 125.5% under DELETE TOP-3 and 118.3% under NOT RECORDED TOP-3, with openness to new and different experiences emerging as the leading surviving predictor in both conditions. For perceived impact of immigration, AEV retention equals 114.6% and 110.7%, again with openness to new and different experiences as the leading predictor. For political interest, AEV retention equals 156.9% and 152.5%, whereas education becomes the leading predictor in both conditions, and modal concentration decreases by approximately seven percentage points. Because FULL AEV is small for political interest (0.028), a large percentage retention corresponds to a small absolute change (AEV of 0.045 and 0.043 after masking; Table A7). Masking produces a new persona-conditioned organisational structure centred on the surviving attributes.

The exact leading surviving attribute is not identical across the two masking conditions for every outcome, though broader patterns remain generally consistent. The same leading replacement appears in both conditions for several substantively important outcomes: education for political interest; openness to new and different experiences for both immigration-related outcomes; place of residence for social trust; main activity for perceived political influence; and helping and caring for others for climate worry. Where the exact leading attribute differs between conditions, the broader pattern nevertheless holds: contracted outcomes remain contracted, expansion outcomes remain expanded, and outcomes characterised by concentrated re-anchoring continue to rely on a restricted subset of attributes. The mechanism is therefore more stable at the level of overall response strategy than at the level of individual attribute replacement.

## 6.16 Interpretation

The masking experiment demonstrates both signal loss and reorganisation around the remaining persona-level information. Relative to the complete persona, withholding the three dominant attribute values removes approximately half of the absolute persona-explained variation. Relative to the matched FULL-28 benchmark, however, the same 28 surviving attributes explain greater absolute variation after masking (0.230 versus 0.184; difference = 0.0459, 95% CI [0.0298, 0.0623]). This finding supports active reorganisation rather than random sampling variation.

This reorganisation should not be characterised as recovery of the original Top-3 contribution. The intervention may generate an alternative response rule, producing severe signal contraction in certain outcomes and expansion beyond the FULL AEV level in other cases. The earlier sequential experiment points to an initial step in this process, in which the original Top-2 attribute emerges as the leading predictor in 13 of 15 outcomes following removal of the Top-1 attribute. The simultaneous Top-3 experiment reveals the final state: significant absolute variation loss, partial retention of proportional predictability, and increased reliance on remaining attributes. Post-masking Shapley results indicate that substantial apparent spreading results from mechanical effects; relative to the statistical baseline, masked responses maintain greater concentration (leading share 26.1–27.0% versus 21.5%; effective-N 7.80–8.15 versus 8.91). The evidence therefore supports re-anchoring on a subset of surviving attributes rather than uniform redistribution across remaining predictors.

For the paper's central argument, over-determination persists even when the dominant attributes are removed. Synthetic answers reorganise around the persona information that remains available. When the dominant information is withheld, that organisation weakens in absolute terms, but it is not replaced by human-like individual variation. GPT-5-mini builds a new, outcome-specific structure around what remains.

Seen as judgement under limited information, this behaviour may be seen as lawful. Incomplete information produces substitution. People also substitute: nobody refuses to judge because one cue is missing. Judgements made with limited information often rely on a small number of cues. The model re-anchors and answers as firmly as before, so nothing in its output signals that information was missing. In that sense the failure is that it does not represent its own ignorance. Our design records answers, so this is a behavioural description of the output.

## 6.17 Limitations of the masking experiment

The final masking experiment was conducted exclusively for GPT-5-mini. The Top-3 targets were identified through a predictive Shapley decomposition; consequently, their removal demonstrates behavioural dependence on the corresponding input information but does not reveal the model's internal causal computations. Surviving attributes may be correlated or conceptually overlapping, which enables substitution via redundant semantic information. The two masking arms bracket the prompt-length concern without perfectly isolating physical shortening from every consequence of representing missing information differently. The respondent-level bootstrap captures sampling uncertainty over personas conditional on frozen cross-fitted predictions. The sequential experiment used an earlier ranking and is treated as a check on the replication mechanism rather than a source of the basis for the main estimate. Concentration measures utilise positive, renormalised cross-validated Shapley contributions, with signed contributions are released in the replication files. The outcome-level patterns show how the effects of masking vary across questions, without implying distinct or latent types of outcomes.

# 7. Proximity to the Human Conditional Mean

## 7.1 Research question

The preceding sections demonstrate that synthetic responses contain a larger persona-predictable component, occupy a lower-dimensional response space, and exhibit limited run-to-run variation. These findings are consistent with the theoretical account set forth in Section 1.6, though they do not test it directly. The present section addresses this gap through direct empirical evaluation. It examines whether synthetic responses resemble individual draws from the human conditional distribution, or alternatively, whether they approximate the expected human response given the observed persona-level attributes.

For each respondent–outcome pair, we estimate the expected human response conditional on the same 31 persona-level attributes used to construct the synthetic persona. We then quantify the distance between the matched human response and this conditional expectation, as well as the distance between the synthetic response and this expectation. A human respondent provides one observed response drawn from the conditional distribution around the expected value. If the model behaves more like a statistical predictor than an individual draw from the population, its predicted answer should lie closer to that expectation, particularly once systematic model-level shifts have been removed.

## 7.2 Estimating the human conditional mean

The human conditional mean is estimated separately for each outcome using the same ridge specification and persona-level attributes employed in Section 3. Predictions are generated out-of-fold to ensure that the estimate for respondent i derives from a model that was not fitted using data from that respondent:

$$\mu_{io} = E(Y_{io}^H \mid X_i)$$

where $Y_{io}^H$ denotes respondent i's observed human response to outcome o, $X_i$ is a vector containing the 31 persona-level attributes for respondent i, and $\mu_{io}$ represents the cross-validated human prediction. Responses from all three sources (human respondents, Gemini, and GPT-5-mini) are evaluated against this common conditional expectation. This quantity constitutes a model-based expectation conditioned on the observed persona information. It captures the expected response based on the information included in the persona. It does not represent the respondent's latent true attitude.

## 7.3 Raw and centered comparisons

Synthetic responses may deviate from the estimated human conditional mean for two reasons. The model may fail to follow the persona-related response pattern observed in human data, or it may adopt that pattern while placing its entire response distribution at a systematically different level. The raw comparison preserves both differences. The centered comparison removes the average difference between model and human answers separately for each outcome:

$$Y_io\text{^}(M,c) = Y_io\text{^}M - (\text{mean of } Y_o\text{^}M - \text{mean of } Y_o\text{^}H)$$

Centering preserves differences between personas, correlations, and the shape of the model's response gradient while removing only the outcome-specific mean shift. Consequently, raw distance evaluates overall calibration to the human conditional mean; centered distance, by contrast, evaluates structural tracking after accounting for average level bias.

## 7.4 Distance from the conditional mean

Distance is measured as the mean squared deviation from the estimated human conditional mean, expressed in units of matched human response variance:

$$d_o\text{^}S = E[(Y_io\text{^}S - \mu_io)^2]$$

The mean distance between human responses and the conditional mean equals 0.808. Because the conditional mean is estimated from observed human data, this value corresponds to the human residual variance reported in Section 3.3 and serves as the human reference point for this analysis. In the raw comparison, Gemini's distance equals 0.829 and GPT-5-mini's distance equals 0.773. Before level adjustment, Gemini is therefore slightly farther from the estimated human conditional mean than matched human respondents; conversely, GPT-5-mini is slightly closer. Raw distance alone does not support a general claim that model responses approximate the conditional mean more closely: Gemini exhibits distance smaller than the matched human in 36.0% of observations, and GPT-5-mini in 36.3% of observations. These low proportions reflect substantial model-specific level shifts. Raw distance captures two distinct sources of disagreement with the human conditional mean: a model-specific shift in the average response level, and differences in how the model follows the persona-related gradient structure. Consequently, a model may be relatively close to the conditional mean on average while still being farther away than matched humans in many individual cases. The centered comparison separates these effects by removing the average level shift.

## 7.5 Results after removing level shifts

Centering reduces model distance from 0.829 to 0.587 for Gemini and from 0.773 to 0.434 for GPT-5-mini; human distance remains at 0.808 because the matched human responses define the reference level. Relative to human distance, centered model distance is reduced by 27.4% for Gemini and 46.3% for GPT-5-mini. After centering, Gemini exhibits smaller distance than the matched human in 56.9% of observations, and GPT-5-mini in 61.9%. Centered model distance is lower than human distance for 11 of 15 outcomes for Gemini and 13 of 15 outcomes for GPT-5-mini (Figure 7; Tables 15A and 15B).

Exceptions are identified by mean-centered distance. Using this criterion, Gemini remains farther from the conditional mean than matched human respondents for loyalty towards leaders, obedience, helpfulness, and social trust, while GPT-5-mini exhibits such deviations only for loyalty towards leaders and obedience. The proportion of observations in which the model is closer provides a slightly different perspective for Gemini's case. For helpfulness, Gemini is closer in 59.2% of cases despite a higher mean distance; this indicates a minority of large deviations. In contrast, for immigration impact, Gemini exhibits lower mean distance but is closer in only 49.5% of observations, which is close to parity with the matched human. Mean distance serves as the primary criterion because it is the metric reported in Table 15A and Table 15B and Figure 7. Authority outcomes represent cases in which synthetic responses exhibit above-human variance alongside strong value-based Shapley contributions. Centering removes average level shifts but does not eliminate amplified differentiation between personas.

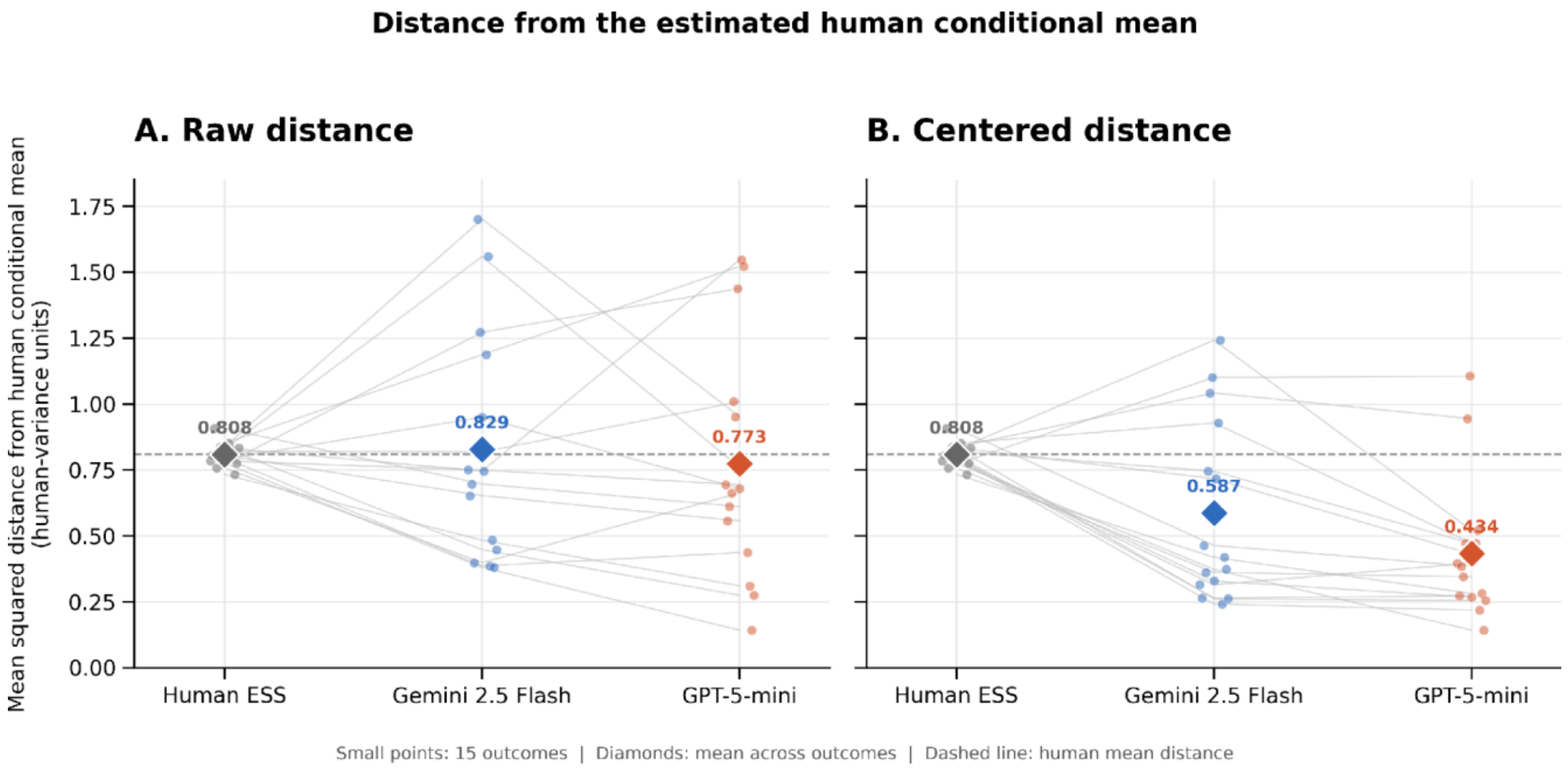

Figure 7. Distance from the estimated human conditional mean. Two panels show raw and outcome-centered distances for Human ESS, Gemini 2.5 Flash and GPT-5-mini. Large markers show means across 15 outcomes; smaller outcome-level points are connected within outcomes.

## 7.6 Correlation with the conditional mean

A second perspective employs correlation rather than distance as the evaluation criterion. The mean correlation between human responses and the estimated conditional mean equals 0.436. Using this same conditional mean, the mean correlation is 0.586 for Gemini and 0.549 for GPT-5-mini. Consequently, both models follow the predictable ordering of human responses more closely than individual humans do.

The conditional mean is fitted to human answers, so the human benchmark here is directly linked to the predictability analysis in Section 3. For out-of-fold ridge predictions, the correlation between observed answers and their own fitted values approximates the square root of out-of-fold $R^2$, and this holds outcome by outcome in our data. The value of 0.436 restates human predictability on a correlation scale. Synthetic answers follow the conditional mean more closely across personas. This is consistent with, but distinct from, their smaller residual variance in Section 3.3.

Closer tracking of the conditional mean is not the same as predicting the matched respondents more accurately. Respondent-level model–human correlations remain 0.279 for Gemini and 0.259 for GPT-5-mini (Section 3.7). The two findings answer different questions. Correlation with the conditional mean measures alignment with the predictable human gradient. Correlation with the matched respondent measures individual-level recovery. Synthetic answers align more closely with the estimated human gradient while remaining weak substitutes for the answers of specific individuals.

Alignment varies by outcome variable. For Gemini, correlation with the conditional mean ranges from 0.282 for political interest to 0.817 for life satisfaction; for GPT-5-mini, it ranges from 0.273 to 0.837 across the same two outcomes. Political interest represents the boundary case: because responses concentrate on "Quite interested," there is limited capacity to reproduce variation along the human conditional gradient.

## 7.7 Strength of the conditional-mean gradient

Regressing each response source against the human conditional mean reveals the extent to which persona-level differences are expressed. The mean slope equals 0.940 for human responses, 1.151 for Gemini, and 0.906 for GPT-5-mini. A slope of 1.0 indicates that persona-related differences are expressed at the same scale as in matched human data. Slopes exceeding 1.0 reflect amplified variation between personas, whereas slopes below 1.0 reflect muted variation. On average, Gemini exhibits slight amplification of the human conditional gradient, while GPT-5-mini reproduces it at approximately the human scale.

The mean obscures substantial heterogeneity across outcomes. Gemini exhibits pronounced amplification of the gradient for social trust (slope = 2.005), life satisfaction (1.551), immigration impact (1.521), fairness (1.440), and obedience (1.438). GPT-5-mini amplifies the gradient most notably for obedience, loyalty towards leaders, and religiosity, but attenuates it across several political outcomes. Political interest is strongly attenuated in both models (slopes = 0.278 and 0.206), respectively. The narrow distribution of responses to political-interest questions therefore constrains both level calibration and the model's capacity to reproduce conditional differences between personas.

Consequently, a model may exhibit proximity to the conditional mean in squared-deviation units while simultaneously expressing the persona-related gradient either too weakly or too strongly. Distance measures how far each answer lies from the person-specific conditional mean; correlation quantifies alignment with the predictable human response pattern; and slope captures the proportional scaling of persona-level variation. Each metric therefore addresses distinct aspects of model behaviour.

## 7.8 Level bias

Centering yields a substantial improvement because synthetic responses exhibit significant outcome-specific level shifts. Mean absolute level shifts equal 0.44 for Gemini and 0.50 for GPT-5-mini, expressed in human standard deviation units (Table 4).

Under the original ESS coding scheme (Tables 15A and 15B), the largest level shifts include Gemini political interest (−0.789), Gemini loyalty towards leaders (+0.812), and GPT-5-mini fairness (+1.056), helpfulness (+1.024), and immigration impact (+0.733). The substantive direction of these shifts is documented in Table 4.

Life satisfaction exhibits minimal average level bias (Gemini −0.090 SD; GPT-5-mini +0.016 SD) and yields the clearest centered result. Human response distance equals 0.776 versus 0.373 for Gemini and 0.143 for GPT-5-mini, implying that centered distance for GPT-5-mini is approximately 82% lower than matched human distance. Level calibration and structural tracking should therefore be reported separately; poor raw calibration can coexist with strong tracking of persona-related differences in the human gradient.

## 7.9 Outcome-level exceptions

Centering does not render synthetic responses universally closer to the conditional mean. Authority-related outcomes represent important exceptions to this pattern. For loyalty towards leaders, centered distance equals 0.837 for humans versus 1.042 for Gemini and 0.944 for GPT-5-mini. For obedience, it equals 0.772, 1.101, and 1.106 respectively. These outcomes reflect amplified differences linked to semantically related personal values; removing the mean shift does not correct this amplification. Social trust functions as a Gemini-specific exception (human distance = 0.828,

Gemini = 1.242, GPT-5-mini = 0.521). The models differ in both level bias and the strength of persona-related gradients. These results vary across outcomes, so the averages do not describe every item equally well.

**Panel A. Gemini 2.5 Flash**

| Outcome | Human distance | Raw model distance | Centered model distance | Model closer after centering | Human corr. with μ | Model corr. with μ | Human slope | Model slope | Level bias (human SD) |
|---|---|---|---|---|---|---|---|---|---|
| Climate responsibility (ccrdprs) | 0.784 | 0.750 | 0.315 | 67.2% | 0.465 | 0.662 | 0.946 | 1.008 | +0.660 |
| Political self-efficacy (cptppola) | 0.805 | 0.653 | 0.265 | 68.6% | 0.442 | 0.567 | 0.947 | 0.738 | +0.623 |
| Income redistribution (gincdif) | 0.908 | 0.697 | 0.464 | 64.0% | 0.307 | 0.468 | 0.880 | 1.036 | -0.483 |
| Allow immigrants (imdfetn) | 0.757 | 0.399 | 0.362 | 62.8% | 0.493 | 0.663 | 0.957 | 1.033 | -0.194 |
| Immigration impact (imwbcnt) | 0.820 | 0.818 | 0.745 | 49.5% | 0.425 | 0.639 | 0.936 | 1.521 | -0.271 |
| Loyalty to leaders (loylead) | 0.837 | 1.701 | 1.042 | 42.2% | 0.405 | 0.440 | 0.931 | 1.148 | +0.812 |
| Obedience for children (lrnobed) | 0.772 | 1.272 | 1.101 | 35.6% | 0.478 | 0.573 | 0.957 | 1.438 | +0.412 |
| Political interest (polintr) | 0.792 | 0.950 | 0.329 | 65.5% | 0.457 | 0.282 | 0.949 | 0.278 | -0.789 |
| People are fair (pplfair) | 0.834 | 0.746 | 0.717 | 58.1% | 0.409 | 0.608 | 0.933 | 1.440 | +0.169 |
| People are helpful (pplhlp) | 0.851 | 1.187 | 0.928 | 59.2% | 0.387 | 0.394 | 0.925 | 0.986 | +0.509 |
| People can be trusted (ppltrst) | 0.828 | 1.559 | 1.242 | 27.6% | 0.415 | 0.656 | 0.938 | 2.005 | -0.563 |
| Political influence (psppipla) | 0.796 | 0.386 | 0.242 | 70.9% | 0.452 | 0.643 | 0.948 | 0.858 | +0.380 |
| Religiosity (rlgdgr) | 0.733 | 0.484 | 0.419 | 60.3% | 0.517 | 0.735 | 0.965 | 1.275 | -0.254 |
| Life satisfaction (stflife) | 0.776 | 0.381 | 0.373 | 55.9% | 0.474 | 0.817 | 0.952 | 1.551 | -0.090 |
| Climate worry (wrclmch) | 0.833 | 0.448 | 0.262 | 66.6% | 0.410 | 0.636 | 0.930 | 0.957 | +0.431 |

| Outcome | Human distance | Raw model distance | Centered model distance | Model closer after centering | Human corr. with μ | Model corr. with μ | Human slope | Model slope | Level bias (human SD) |
|---|---|---|---|---|---|---|---|---|---|
| **Mean** | **0.808** | **0.829** | **0.587** | **56.9%** | **0.436** | **0.586** | **0.940** | **1.151** | **0.443** |

Table 15A. Proximity to the estimated human conditional mean by outcome and model. Panel A: Gemini 2.5 Flash. Note. Distances are mean squared deviations from the out-of-fold human conditional mean, in human-variance units. Centering removes the model–human mean difference for each outcome. "Model closer" is the share of matched cases where the centered model answer is closer to the conditional mean than the observed human answer. Level bias uses raw ESS coding; substantive direction depends on item coding. The Mean row reports the mean absolute level bias.

**Panel B. GPT-5-mini**

| Outcome | Human distance | Raw model distance | Centered model distance | Model closer after centering | Human corr. with μ | Model corr. with μ | Human slope | Model slope | Level bias (human SD) |
|---|---|---|---|---|---|---|---|---|---|
| Climate responsibility (ccrdprs) | 0.784 | 0.694 | 0.397 | 61.1% | 0.465 | 0.652 | 0.946 | 1.099 | +0.544 |
| Political self-efficacy (cptppola) | 0.805 | 0.557 | 0.273 | 68.2% | 0.442 | 0.544 | 0.947 | 0.698 | +0.534 |
| Income redistribution (gincdif) | 0.908 | 0.611 | 0.385 | 66.9% | 0.307 | 0.454 | 0.880 | 0.905 | -0.474 |
| Allow immigrants (imdfetn) | 0.757 | 0.662 | 0.345 | 63.9% | 0.493 | 0.518 | 0.957 | 0.660 | -0.564 |
| Immigration impact (imwbcnt) | 0.820 | 1.009 | 0.472 | 57.5% | 0.425 | 0.588 | 0.936 | 1.099 | +0.733 |
| Loyalty to leaders (loylead) | 0.837 | 0.952 | 0.944 | 42.9% | 0.405 | 0.474 | 0.931 | 1.198 | +0.087 |
| Obedience for children (lrnobed) | 0.772 | 1.438 | 1.106 | 36.2% | 0.478 | 0.522 | 0.957 | 1.278 | +0.576 |
| Political interest (polintr) | 0.792 | 0.680 | 0.268 | 67.9% | 0.457 | 0.273 | 0.949 | 0.206 | -0.642 |
| People are fair (pplfair) | 0.834 | 1.547 | 0.431 | 63.4% | 0.409 | 0.437 | 0.933 | 0.716 | +1.056 |
| People are helpful (pplhlp) | 0.851 | 1.522 | 0.473 | 63.8% | 0.387 | 0.332 | 0.925 | 0.557 | +1.024 |
| People can be trusted (ppltrst) | 0.828 | 0.760 | 0.521 | 59.3% | 0.415 | 0.555 | 0.938 | 1.087 | +0.489 |
| Political influence (psppipla) | 0.796 | 0.437 | 0.219 | 70.9% | 0.452 | 0.602 | 0.948 | 0.705 | +0.468 |

| Outcome | Human distance | Raw model distance | Centered model distance | Model closer after centering | Human corr. with μ | Model corr. with μ | Human slope | Model slope | Level bias (human SD) |
|---|---|---|---|---|---|---|---|---|---|
| Religiosity (rlgdgr) | 0.733 | 0.311 | 0.282 | 66.6% | 0.517 | 0.776 | 0.965 | 1.194 | -0.171 |
| Life satisfaction (stflife) | 0.776 | 0.143 | 0.143 | 73.8% | 0.474 | 0.837 | 0.952 | 1.142 | +0.016 |
| Climate worry (wrclmch) | 0.833 | 0.275 | 0.255 | 65.5% | 0.410 | 0.675 | 0.930 | 1.047 | +0.139 |
| **Mean** | **0.808** | **0.773** | **0.434** | **61.9%** | **0.436** | **0.549** | **0.940** | **0.906** | **0.501** |

**Table 15B. Proximity to the estimated human conditional mean by outcome: GPT-5-mini. Note.** *Distances are mean squared deviations from the out-of-fold human conditional mean, in human-variance units. Centering removes the model–human mean difference for each outcome. "Model closer" is the share of matched cases where the centered model answer is closer to the conditional mean than the observed human answer. Level bias uses raw ESS coding; substantive direction depends on item coding. The Mean row reports the mean absolute level bias.*

## 7.10 Interpretation

The centered results support a conditional-mean interpretation of synthetic responding. Synthetic responses track predictable differences among human respondents more closely than observed human answers do; they correlate more strongly with the human conditional mean, exhibit greater proximity to it after centering, and are more frequently closer to it than matched human respondents. Simultaneously, these responses do not reproduce the full human conditional distribution. Their joint response structure is lower-dimensional, residual variance is substantially smaller, repeated model invocations contribute limited variation that depends on configuration, and respondent-level agreement remains low.

Behaviourally, the synthetic response can be represented as:

$$Y_{io}^{M} \approx g_{mo} \cdot f_o(X_i) + \delta_{mo} + \varepsilon_{io}^{M}$$

where $f_o(X_i)$ denotes the persona-predictable human response pattern for outcome o, $g_{mo}$ represents a model- and outcome-specific scaling factor that permits the model to attenuate or amplify the gradient, $\delta_{mo}$ is the model- and outcome-specific level shift, and $\varepsilon_{io}^{M}$ represents the relatively small call-specific residual. The $g_{mo}$ term is required because the outcome-level slopes demonstrate that models do not consistently express predictable differences at the human scale. Because model responses correlate only 0.59 (Gemini) and 0.55 (GPT-5-mini) with the human conditional mean, $g_{mo}$ should be interpreted as partial alignment: the model's own gradient only partially coincides with $f_o$. A more general form is $Y \approx f_M(X) + \delta + \varepsilon$, where $f_M$ is partly aligned to $f_o$.

This representation explains how a model can reproduce rankings and systematic contrasts, approximate the centered human conditional mean, and show systematic level differences, all while failing to recover the matched individual answer. This fits Dawes's (1979) observation that a consistently applied rule captures the predictable component well; its weakness at the individual level is what our data add. Here the averages are 0.586 and 0.549 against 0.436 for humans, and the individual correlations are 0.279 and 0.259.

## 7.11 Limits of the conditional-mean analysis

The human conditional mean is estimated from the same 31 persona-level attributes employed in the synthetic response generation; it does not condition on unobserved respondent characteristics. The ridge regression specification may omit nonlinearities and interactions, which means this conditional mean represents the predictable component under our main analytical framework. Centering removes only the average model–human difference for each outcome; it does not correct differences in variance, category concentration, slope, or nonlinear response shape. A lower centered distance therefore does not establish that an LLM internally represents a conditional probability distribution. Instead, these findings provide behavioural evidence that model responses approximate a transformed conditional expectation more closely than individual draws from the human response distribution.

Synthetic responses reproduce the persona-explained component of human responses more strongly than they recover individual-level differences among matched respondents. The conditional-mean result connects the paper's primary findings: a substantially larger persona-predictable component, stronger reliance on personal values, lower joint response dimensionality, limited run-to-run variation, and low respondent-level agreement. The discussion examines how the limited representation of individual variation affects the use of synthetic respondents.

# 8. Discussion, Implications and Limitations

## 8.1 Main finding

Across two model families, synthetic answers are far more structured by the observed persona than the answers of the matched humans. In units of human answer variance, the persona accounts for 0.192 of human variance, against 0.617 for Gemini and 0.468 for GPT-5-mini. As a share of each source's own variance, out-of-fold $R^2$ is 0.192 for humans, 0.695 for Gemini, and 0.731 for GPT-5-mini. Synthetic profiles also occupy a lower-dimensional space: the participation ratio falls from 9.67 for humans to 5.03 for Gemini and 5.31 for GPT-5-mini. None of this means synthetic answers are uniformly less variable; on some outcomes they vary more than human answers. The general pattern is different. Whatever variation synthetic answers have is organised, to an unusual degree, by the information in the persona.

## 8.2 Over-determination rather than simple variance compression

Variance compression and persona over-determination are interrelated yet distinct properties. GPT-5-mini exhibits the higher $R^2$ alongside the lower AEV because its total response variance is more compressed. Gemini preserves greater total variance, though a substantially larger absolute amount of that variation remains predictable from the persona. On several outcomes, particularly authority and social-trust items, synthetic variance equals or exceeds human variance, with most of this additional variation still attributable to the persona. The central finding is that the variation LLM-generated responses produce is disproportionately determined by the observed persona attributes.

## 8.3 What synthetic respondents reproduce

Synthetic responses reproduce the predictable structure of human attitudes more successfully than they recover individual human differences. After outcome-specific level shifts are removed, both models exhibit closer average proximity to the estimated human conditional mean relative to matched human responses. Synthetic responses also correlate more strongly with this conditional mean than observed human responses (0.436 for humans, 0.586 for Gemini, and 0.549 for GPT-5-mini). Correlations with matched individuals, however, remain low at 0.279 (Gemini) and 0.259 (GPT-5-mini).

This combination constitutes the core of the validation problem raised in the introduction. An LLM can reproduce predictable gradients, rankings, and group contrasts while remaining a weak substitute for any specific respondent. Agreement at the aggregate level and fidelity at the individual level represent distinct properties; evidence for the former does not constitute evidence for the latter.

## 8.4 Role of persona information

Human predictability is distributed relatively evenly across country, socio-demographic and life-situation characteristics, and personal values. Synthetic predictability shifts toward personal values: 37.1% in humans versus 52.9% in Gemini and 69.9% in GPT-5-mini. The clearest patterns involve strong conceptual alignment between a persona value and the outcome question, such as environmental concern and climate-related outcomes, equality and redistribution, and rule-following and obedience. This pattern is consistent with semantic anchoring to persona-level values, although the decomposition remains observational.

Measures of concentration of attribute importance (Top-1, Top-3 and effective-N) do not indicate greater overall concentration relative to human responses. The observed difference pertains to the type of information organising synthetic answers, particularly the enhanced role of personal values in predicting model-generated responses.

The masking experiment provides interventional evidence of behavioural dependence on these attributes. Withholding the three leading attributes removes approximately half of the absolute persona-explained variation. The surviving attributes recover only about 14–19% of the withheld contribution (Section 6.8), and, on average, the model re-anchors on a few of them (Section 6.12).

## 8.5 How the synthetic answer is made: a cue-based reading

The results so far can be reported as statistics: AEV, participation ratios, Shapley shares, retention rates. They can also be read as a description of how the model makes a judgement about a person, and that reading is the more useful one. Across the analyses, synthetic responses are consistent with a low-dimensional cue-based rule. The answers remain organised around persona information, even as particular inputs are removed or replaced. In this sense, low-dimensional describes the response structure and the substitutability of persona inputs. It does not describe the number of attributes used: Shapley importance is slightly more dispersed for the models than for humans (Section 3.6.5). Synthetic answers occupy about half as many effective dimensions as humans, and a withheld cue is replaced rather than compensated for. Synthetic responses occupy about half as many effective dimensions as human responses. Their structure also reflects the meaning of the survey item: models often give greatest weight to the persona attribute most closely related to the question, while country is frequently more important for human answers. The masking experiment shows how this structure adapts. After the leading attribute is removed, the next attribute in the overall Shapley ranking becomes dominant in 13 of the 15 outcomes. Removing the top three produces a similar response, with importance concentrating on a small set of remaining attributes beyond what statistical redistribution alone would predict. This is the vocabulary of the lens model (Brunswik, 1956), and our matched design supplies both of its sides: cue validities from the human data and cue utilisation from the model's answers. It also clarifies what does not differ. Both humans and models combine cues in a largely additive way (Section 3.6.5); the difference is in which cues carry weight and how much. The result is consistent with Dawes's (1979) observation that simple, consistently applied rules can perform well despite imperfect cue weights. In our data, this advantage appears in the stronger alignment with the predictable component of human responses. And the masking results add the point we consider most consequential. When information is missing, the model substitutes, and its answers become, if anything, more concentrated. People substitute as well, but missing information usually leaves some trace of doubt; in the model's output it leaves none. We describe all of this as behaviour consistent with a cue-based rule. The design does not observe the model's internal computation, and we do not claim to have identified a specific heuristic.

## 8.6 Where the synthetic person comes from

A synthetic answer can be thought of as the product of several interacting sources: what the model learned about people during pre-training, the persona and prompt through which that knowledge is accessed, the cue-based inference described above, and post-training alignment, which shapes what the model is willing to express. These sources are not independent, and the present design cannot separate them. We therefore treat the decomposition as a future research agenda.

One of these sources leaves a visible trace in our data. The level shifts in Section 3.5 are not random. Synthetic respondents are almost uniformly interested in politics: "Quite interested" accounts for 79.7% of Gemini answers and 88.6% of GPT-5-mini answers, against 33.2% of human answers. They are also more worried about climate change (+0.43 and +0.14 human SD) and more likely to feel personally responsible for it (+0.66 and +0.54). The common direction, towards engagement, efficacy, pro-environmental attitudes, and lower authoritarianism, is consistent with evidence that preference-tuned models shift survey answers towards socially desirable positions (Salecha et al., 2024; Santurkar et al., 2023) and reduce output diversity (Kirk et al., 2024). We cannot attribute these shifts to alignment with the present design. They do, however, show that the persona is not the only thing speaking in a synthetic answer. Separating learned knowledge, prompt, cue use, and alignment, for example by comparing base and instruction-tuned versions of the same model, is the natural next step. Evidence from the Pew American Trends Panel suggests the comparison is informative: base models produced response distributions closer to human data and better preserved demographic structure than their post-trained counterparts (Grief-Albert et al., 2026).

## 8.7 Model-specific bias and calibration

Strong structural alignment with the persona does not guarantee correct calibration to human response levels. Mean absolute level shifts equal 0.44 human standard deviations for Gemini and 0.50 for GPT-5-mini (Table 4). Political interest represents the clearest example of this phenomenon: "Quite interested" is selected in 79.7% of Gemini responses, 88.6% of GPT-5-mini responses, and 33.2% of human responses. Once the outcome-specific level shift is removed, both models move substantially closer to the human conditional mean. Synthetic responses can be characterised as a transformed predictable component rather than direct reproductions of human conditional expectations; models may attenuate or amplify the human gradient and introduce an outcome-specific level shift. Post-hoc calibration can correct part of this at the aggregate level. Leng et al. (2026) learn non-negative weights that sum to one over outputs obtained with different personas, prompts and model versions, so that the weighted aggregate matches a human benchmark; the calibrated aggregate therefore remains within the convex hull of the outputs being combined. Such calibration realigns aggregates; it does not by construction supply the individual-level variation documented here.

## 8.8 Stochasticity does not recreate individuality

GPT-5-mini exhibits high but imperfect stability across repeated model invocations, with a mean test–retest ICC of 0.825 under the API's default temperature setting of 1. Run-to-run variation accounts for 17.5% of total GPT-5-mini variance and averages 0.098 in human-variance units; this is substantially smaller than the human residual variance of 0.808. Repeated generation consequently introduces stochasticity but does not recover the considerably larger individual-level component observed among matched human respondents. Consequently, sampling variability between model invocations is not equivalent to stable heterogeneity among individuals who share the same observed persona attributes. This agrees with evidence that instruction-tuned models collapse to a single output, returning the same answer for the same persona and question on more than half of the items of a public-opinion benchmark (Jang et al., 2026).

## 8.9 Robustness of the main conclusion

Several checks address alternative explanations. The predictability ordering remains unchanged after applying ESS post-stratification weights and after excluding personas containing item non-response labels; the dimensionality ordering is likewise stable under both conditions. A gradient-boosting specification with interactions changes the model-to-human predictability ratios only marginally (3.54 and 3.70, against 3.63 and 3.82 for ridge; Table A5) and does not replace the linear model employed for the main decomposition. Reliability correction narrows but does not eliminate the predictability gap: at an assumed human reliability of 0.70, corrected $R^2$ equals 0.274 for humans and 0.880 for GPT-5-mini, whereas the median human reliability required to eliminate outcome-level differences is 0.222. In the masking experiment, DELETE and NOT RECORDED arms yield comparable results, bracketing the prompt-length confound; excluding the 39 personas with non-response labels changes retention by less than one percentage point; and FULL responses show no drift relative to Stage 1.

These checks address alternative explanations based on weighting schemes, missing information handling, functional form specification, measurement attenuation, and prompt-length concerns. They do not establish that the reported values remain invariant to all prompts, models, or decoding configurations.

## 8.10 Implications for synthetic research

The results explain more precisely the capabilities and limitations of synthetic responses. Synthetic respondents may be useful when the objective is to recover structured population patterns strongly associated with observed persona-level attributes. Potentially suitable applications include rapid directional exploration, concept or question-wording pretesting, comparison of broad population segments, and hypothesis generation for subsequent human research projects.

Strong aggregate or conditional-pattern recovery should not, however, be interpreted as validation of person-level digital twins. Applications requiring individual substitution, realistic joint distributions, or inference about within-segment heterogeneity face more demanding standards. For such applications, matching averages is insufficient; synthetic populations must also reproduce conditional variance, covariance, and dimensional structure, stable individual differences, item non-response patterns, and both contradictory responses and weakly structured attitudes. Put differently, validation should inquire whether synthetic respondents produce answers with similar conditional uncertainty to human respondents.

## 8.11 Implications for digital-twin design

Current persona prompting appears better suited to producing profile-consistent responses than to simulating individual-level draws from a human conditional distribution. Adding additional persona attributes may increase predictability without recovering the missing individual layer; however, greater information availability can render synthetic answers still more determined by the prompt specification itself.

Related evidence from a very different design points the same way. In parsimony and dose-response analyses of the Twin-2K-500 digital twins (Toubia et al., 2025), where twins were conditioned on up to 500 of each person's own prior survey answers rather than on about thirty attributes, the twins were more predictable from a few profile variables than the people they represented (Florès, 2026). Adding the person's own answers did not improve accuracy: 0.765 with demographics alone against 0.757 with five hundred of the person's own answers. The claim is bounded to accuracy; for correlation, a gain cannot be ruled out, as the subsample analysed was too small to detect it. The instruments and metrics differ, so the two results cannot be pooled. Together they suggest that over-determination does not shrink as more individual information is supplied, at least as measured by accuracy. The published mega-study points the same way at larger scale: twins built from each person's answers to more than 500 questions were only modestly more accurate than the base model, and correlated weakly with their humans (average r = 0.20; Peng et al., 2026). In the same study, full-persona twins were not significantly more accurate than twins given demographics only (0.748 against 0.746). Adding variables adds candidates for anchoring. The form of the information may matter as well as its amount: agents built from two-hour interviews predicted held-out General Social Survey items more accurately than agents given demographics only (Park et al., 2024).

Future systems may therefore need to separate two tasks: estimation of a conditional mean and generation of calibrated individual deviations around that mean. Those deviations should not be introduced as arbitrary noise; instead, they should reproduce empirically observed human residual variance, covariance across outcomes, stability over time, and heterogeneity across individuals. Evaluation should extend beyond aggregate alignment to include person-level, distributional, and mechanism-oriented diagnostics.

## 8.12 Limitations

The study has several boundaries and limitations. It is based on a single survey round (ESS Round 11), covering 28 countries and 15 outcome variables; findings may differ for other surveys, domains, or response formats. All prompts were written in English, and country entered as a persona attribute rather than through the language of the national ESS questionnaire. Eliciting answers in the questionnaire language of a non-English market is the most direct next test of this choice. Persona attributes were presented in a fixed order, with personal values positioned immediately before the outcome question; prompt ordering was not experimentally varied. The order of answer options was likewise fixed; because order and labelling effects could shape LLM answers (Dominguez-Olmedo, Hardt & Mendler-Dünner, 2024), this is a further boundary of the prompt architecture. The persona was framed as a consumer, although several outcomes concern political and civic attitudes. The conditional mean was estimated from the same 31 observed attributes using a single ridge specification; it is therefore an approximate rather than complete expectation conditional on all information about the respondent. Test–retest stability is configuration-specific: GPT-5-mini was evaluated at the API's fixed default temperature of 1, whereas Gemini was elicited at temperature = 0 and was not included in repeated generations. Finally, models were accessed through provider APIs. The results describe the documented model versions and collection period and may not generalise to later releases or other deployment configurations.

## 8.13 Conclusion

Synthetic responses closely reproduce the persona-explained component of human responding, with personal values playing an especially strong role. The answers are highly predictable, occupy a lower-dimensional space, and approximate the centered human conditional mean, yet they remain only weakly aligned with the individuals they were built from. They behave like dthe output of a consistent rule applied to a few key cues.

The central methodological challenge is now to reproduce the stable, structured individual-level variation that remains after conditioning on observed persona characteristics. Until that layer has been modelled and validated, synthetic responses should be described as simulations of persona-conditioned expectations, not as substitutes for individual respondents.

The question reaches beyond survey research, though this paper makes no claim there. Systems that recommend, anticipate, or purchase on someone's behalf also act on a representation of "a person like me", built from limited information. How that representation is constructed, which cues it uses, how few, and what it does when cues are missing, is the same question we have asked of synthetic respondents. Synthetic respondents simply make it unusually easy to study, because a real person stands behind every synthetic one.

### *Use of generative AI*

Large language models were used for language editing and stylistic refinement of the manuscript, as the authors are not native English speakers. The tools were not used to generate data, conduct analyses, or determine the study's findings or conclusions. All substantive content and revisions were reviewed by the authors.

### *Replication package*

The replication package contains the analysis code, the frozen result files, the raw and parsed model responses, and the identifiers of the sampled respondents (country and idno) https://doi.org/10.5281/zenodo.24286641. ESS microdata are not redistributed: the package does not contain persona attributes, human answers or weights. Users download ESS11e03 from the ESS Data Portal and prepare one input file as described in Input/README.md; run_all.py then reproduces all results and compares them with the frozen files.

### *Competing interests.*

K. Suchecki works as a researcher at a consulting and technology company, the views and opinions expressed in this paper are his own and do not reflect those of his employer. L. Florès is the founder of FlashInsight, a company that offers synthetic samples. The study was conducted independently of any commercial product, and neither organisation had a role in its design, analysis or interpretation.

# Appendix

## A1. Persona attribute coding

| Attribute | Block | Coding | Categories | Reference | Columns |
|---|---|---|---|---|---|
| Country (cntry) | Country | categorical | 28 countries (Austria … United Kingdom) | Austria | 27 |
| Age (agea) | Socio-demographic/context | 7 bands | 15–24 \| 25–34 \| 35–44 \| 45–54 \| 55–64 \| 65–74 \| 75–100 | 15–24 | 6 |
| Gender (gndr) | Socio-demographic/context | categorical | Male \| Female | Male | 1 |
| Place of residence (domicil) | Socio-demographic/context | categorical | Big city \| Suburbs \| Town or small city \| Country village \| Farm or countryside home | Big city | 4 |
| Years of education (eduyrs) | Socio-demographic/context | 6 bands | 0–8 \| 9–10 \| 11–12 \| 13–14 \| 15–16 \| 17–30 | 0–8 | 5 |
| Main activity (mnactic) | Socio-demographic/context | categorical | Paid work \| Education \| Unemployed, looking \| Unemployed, not looking \| Sick or disabled \| Retired \| Community or military service \| Housework \| Other | Paid work | 8 |
| Household size (hhmmb) | Socio-demographic/context | capped at 6 | 1 \| 2 \| 3 \| 4 \| 5 \| 6+ | 1 | 5 |
| Children ever in household (chldhhe) | Socio-demographic/context | categorical | Yes \| No \| Not applicable (routing) | Yes | 2 |
| Household income feeling (hincfel) | Socio-demographic/context | categorical | Living comfortably \| Coping \| Difficult \| Very difficult | Living comfortably | 3 |
| Self-rated health (health) | Socio-demographic/context | categorical | Very good \| Good \| Fair \| Bad \| Very bad | Very good | 4 |
| Trying new things (impdiffa) | Personal values | categorical | Very much like me \| Like me \| Somewhat like me \| A little like me \| Not like me \| Not like me at all | Very much like me | 5 |
| Caring for nature (impenva) | Personal values | categorical | Very much like me \| Like me \| Somewhat like me \| A little like me \| Not like me \| Not like me at all | Very much like me | 5 |
| Freedom (impfreea) | Personal values | categorical | Very much like me \| Like me \| Somewhat like me \| A little like me \| Not like me \| Not like me at all | Very much like me | 5 |
| Fun and pleasure (impfuna) | Personal values | categorical | Very much like me \| Like me \| Somewhat like me \| A little like me \| Not like me \| Not like me at all | Very much like me | 5 |
| Being rich (impricha) | Personal values | categorical | Very much like me \| Like me \| Somewhat like me \| A little like me \| Not like me \| Not like me at all | Very much like me | 5 |
| Secure surroundings (impsafea) | Personal values | categorical | Very much like me \| Like me \| Somewhat like me \| A little like me \| Not like me \| Not like me at all | Very much like me | 5 |

| Attribute | Block | Coding | Categories | Reference | Columns |
|---|---|---|---|---|---|
| Tradition (imptrada) | Personal values | categorical | Very much like me \| Like me \| Somewhat like me \| A little like me \| Not like me \| Not like me at all | Very much like me | 5 |
| Adventure (ipadvnta) | Personal values | categorical | Very much like me \| Like me \| Somewhat like me \| A little like me \| Not like me \| Not like me at all | Very much like me | 5 |
| Behaving properly (ipbhprpa) | Personal values | categorical | Very much like me \| Like me \| Somewhat like me \| A little like me \| Not like me \| Not like me at all | Very much like me | 5 |
| Creativity (ipcrtiva) | Personal values | categorical | Very much like me \| Like me \| Somewhat like me \| A little like me \| Not like me \| Not like me at all | Very much like me | 5 |
| Equal treatment (ipeqopta) | Personal values | categorical | Very much like me \| Like me \| Somewhat like me \| A little like me \| Not like me \| Not like me at all | Very much like me | 5 |
| Following rules (ipfrulea) | Personal values | categorical | Very much like me \| Like me \| Somewhat like me \| A little like me \| Not like me \| Not like me at all | Very much like me | 5 |
| Good time (ipgdtima) | Personal values | categorical | Very much like me \| Like me \| Somewhat like me \| A little like me \| Not like me \| Not like me at all | Very much like me | 5 |
| Helping others (iphlppla) | Personal values | categorical | Very much like me \| Like me \| Somewhat like me \| A little like me \| Not like me \| Not like me at all | Very much like me | 5 |
| Loyalty to close others (iplylfra) | Personal values | categorical | Very much like me \| Like me \| Somewhat like me \| A little like me \| Not like me \| Not like me at all | Very much like me | 5 |
| Modesty (ipmodsta) | Personal values | categorical | Very much like me \| Like me \| Somewhat like me \| A little like me \| Not like me \| Not like me at all | Very much like me | 5 |
| Respect (iprspota) | Personal values | categorical | Very much like me \| Like me \| Somewhat like me \| A little like me \| Not like me \| Not like me at all | Very much like me | 5 |
| Showing abilities (ipshabta) | Personal values | categorical | Very much like me \| Like me \| Somewhat like me \| A little like me \| Not like me \| Not like me at all | Very much like me | 5 |
| Strong government (ipstrgva) | Personal values | categorical | Very much like me \| Like me \| Somewhat like me \| A little like me \| Not like me \| Not like me at all | Very much like me | 5 |
| Success (ipsucesa) | Personal values | categorical | Very much like me \| Like me \| Somewhat like me \| A little like me \| Not like me \| Not like me at all | Very much like me | 5 |
| Understanding others (ipudrsta) | Personal values | categorical | Very much like me \| Like me \| Somewhat like me \| A little like me \| Not like me \| Not like me at all | Very much like me | 5 |
| **Total** | | | | | **170** |

Table A1. Coding of the 31 persona attributes in the main-study ridge models (170 design columns). Note. One-hot coding with the first category as reference. Non-substantive values (Don't know, Refusal, No answer) are set to missing and pooled with the reference category; “Not applicable” for chldhhe forms its own category. Eighteen respondents with more than 30 years of education fall outside the education bands and are pooled with the reference band. The masking analysis (Section 6.6) uses its own encoder in which non-substantive values form a separate category (177 columns under FULL; 138–162 under the masked conditions).

## A2. Question wording and parser

The full prompt for one persona and one question is shown in Box 1, and the wording and answer options of all 15 outcomes are listed in Table 2, with the three adaptations described in Section 2.7. Table A2 reports missing answers by outcome and source.

| Outcome | Gemini: unmatched | GPT-5-mini: unmatched | Human: non-substantive |
|---|---|---|---|
| Climate responsibility (ccrdprs) | 1 | 0 | 332 |
| Political self-efficacy (cptppola) | 0 | 0 | 192 |
| Income redistribution (gincdif) | 0 | 1 | 175 |
| Allow immigrants (imdfetn) | 0 | 0 | 244 |
| Immigration impact (imwbcnt) | 0 | 0 | 333 |
| Loyalty to leaders (loylead) | 0 | 0 | 309 |
| Obedience for children (lrnobed) | 0 | 0 | 101 |
| Political interest (polintr) | 2 | 0 | 18 |
| People are fair (pplfair) | 0 | 0 | 52 |
| People are helpful (pplhlp) | 0 | 0 | 19 |
| People can be trusted (ppltrst) | 0 | 0 | 23 |
| Political influence (psppipla) | 0 | 0 | 191 |
| Religiosity (rlgdgr) | 9 | 0 | 75 |
| Life satisfaction (stflife) | 1 | 0 | 77 |
| Climate worry (wrclmch) | 5 | 0 | 225 |
| **Total (of 210,000)** | **18** | **1** | **2366** |

Table A2. Missing answers by outcome and source in the main study (N = 14,000 per outcome). Note. Model columns count answers the rule-based parser could not map to an answer option; these are set to missing and are not imputed or re-elicited. The human column counts ESS non-substantive answers (Don't know, Refusal, No answer) and is not a parser result. A breakdown by parser step will be released with the replication package.

## A3. Outcome-level predictability

| Outcome | Human $R^2$ = AEV | Human resid. | Gemini rel. var. | Gemini $R^2$ | Gemini AEV | Gemini resid. | GPT rel. var. | GPT $R^2$ | GPT AEV | GPT resid. |
|---|---|---|---|---|---|---|---|---|---|---|
| Climate responsibility (ccrdprs) | 0.216 | 0.784 | 0.587 | 0.800 | 0.469 | 0.117 | 0.705 | 0.913 | 0.644 | 0.061 |
| Political self-efficacy (cptppola) | 0.195 | 0.805 | 0.370 | 0.668 | 0.248 | 0.123 | 0.359 | 0.674 | 0.242 | 0.117 |
| Income redistribution (gincdif) | 0.092 | 0.908 | 0.594 | 0.618 | 0.367 | 0.226 | 0.482 | 0.712 | 0.343 | 0.139 |
| Allow immigrants (imdfetn) | 0.243 | 0.757 | 0.645 | 0.703 | 0.453 | 0.192 | 0.431 | 0.656 | 0.283 | 0.149 |
| Immigration impact (imwbcnt) | 0.180 | 0.820 | 1.171 | 0.736 | 0.862 | 0.310 | 0.721 | 0.792 | 0.571 | 0.150 |
| Loyalty to leaders (loylead) | 0.163 | 0.837 | 1.293 | 0.556 | 0.719 | 0.575 | 1.209 | 0.690 | 0.834 | 0.375 |
| Obedience for children (lrnobed) | 0.228 | 0.772 | 1.568 | 0.778 | 1.220 | 0.348 | 1.494 | 0.814 | 1.217 | 0.278 |
| Political interest (polintr) | 0.208 | 0.792 | 0.226 | 0.442 | 0.100 | 0.126 | 0.132 | 0.410 | 0.054 | 0.078 |
| People are fair (pplfair) | 0.166 | 0.834 | 1.081 | 0.746 | 0.806 | 0.274 | 0.516 | 0.688 | 0.355 | 0.161 |
| People are helpful (pplhlp) | 0.149 | 0.851 | 1.099 | 0.640 | 0.703 | 0.396 | 0.494 | 0.760 | 0.375 | 0.118 |
| People can be trusted (ppltrst) | 0.172 | 0.828 | 1.832 | 0.662 | 1.213 | 0.620 | 0.751 | 0.724 | 0.544 | 0.207 |
| Political influence (psppipla) | 0.204 | 0.796 | 0.406 | 0.694 | 0.282 | 0.124 | 0.313 | 0.613 | 0.192 | 0.121 |
| Religiosity (rlgdgr) | 0.267 | 0.733 | 0.864 | 0.815 | 0.704 | 0.160 | 0.680 | 0.877 | 0.597 | 0.084 |
| Life satisfaction (stflife) | 0.224 | 0.776 | 0.898 | 0.920 | 0.826 | 0.072 | 0.463 | 0.887 | 0.411 | 0.052 |
| Climate worry (wrclmch) | 0.167 | 0.833 | 0.446 | 0.650 | 0.290 | 0.156 | 0.475 | 0.762 | 0.362 | 0.113 |
| **Mean** | **0.192** | **0.808** | **0.872** | **0.695** | **0.617** | **0.255** | **0.615** | **0.731** | **0.468** | **0.147** |

Table A3. Outcome-level predictability by source (values underlying Figure 3). Note. Out-of-fold ridge $R^2$ (five country-stratified folds, $\lambda = 1$). Relative variance, AEV and residual variance are in units of matched human answer variance; for

| Outcome | Human $R^2$ = AEV | Human resid. | Gemini rel. var. | Gemini $R^2$ | Gemini AEV | Gemini resid. | GPT rel. var. | GPT $R^2$ | GPT AEV | GPT resid. |
|---|---|---|---|---|---|---|---|---|---|---|
| humans relative variance is 1, so AEV equals $R^2$ and residual variance equals $1 - R^2$. Mean row computed from unrounded values. | | | | | | | | | | |

## A4. Leading persona attributes for humans and Gemini

| Source | Outcome | Top 1 | Top 2 | Top 3 |
|---|---|---|---|---|
| Humans | Climate responsibility (ccrdprs) | Country (cntry) 32.3% \| 0.070 | Caring for nature (impenva) 24.9% \| 0.054 | Years of education (eduyrs) 6.3% \| 0.014 |
| Humans | Political self-efficacy (cptppola) | Country (cntry) 29.6% \| 0.058 | Years of education (eduyrs) 14.8% \| 0.029 | Household income feeling (hincfel) 7.4% \| 0.014 |
| Humans | Income redistribution (gincdif) | Country (cntry) 35.6% \| 0.033 | Equal treatment (ipeqopta) 14.1% \| 0.013 | Household income feeling (hincfel) 11.6% \| 0.011 |
| Humans | Allow immigrants (imdfetn) | Country (cntry) 37.5% \| 0.091 | Years of education (eduyrs) 8.2% \| 0.020 | Household income feeling (hincfel) 7.0% \| 0.017 |
| Humans | Immigration impact (imwbcnt) | Country (cntry) 37.2% \| 0.067 | Years of education (eduyrs) 8.6% \| 0.016 | Household income feeling (hincfel) 7.5% \| 0.013 |
| Humans | Loyalty to leaders (loylead) | Country (cntry) 23.0% \| 0.037 | Following rules (ipfrulea) 13.9% \| 0.023 | Tradition (imptrada) 10.6% \| 0.017 |
| Humans | Obedience for children (lrnobed) | Country (cntry) 28.2% \| 0.064 | Following rules (ipfrulea) 10.4% \| 0.024 | Behaving properly (ipbhprpa) 9.1% \| 0.021 |
| Humans | Political interest (polintr) | Country (cntry) 21.9% \| 0.046 | Years of education (eduyrs) 18.5% \| 0.039 | Gender (gndr) 10.6% \| 0.022 |
| Humans | People are fair (pplfair) | Country (cntry) 40.8% \| 0.068 | Household income feeling (hincfel) 12.9% \| 0.021 | Years of education (eduyrs) 9.2% \| 0.015 |
| Humans | People are helpful (pplhlp) | Country (cntry) 55.4% \| 0.083 | Household income feeling (hincfel) 10.9% \| 0.016 | Years of education (eduyrs) 4.5% \| 0.007 |
| Humans | People can be trusted (ppltrst) | Country (cntry) 41.5% \| 0.071 | Household income feeling (hincfel) 11.9% \| 0.021 | Years of education (eduyrs) 8.3% \| 0.014 |
| Humans | Political influence (psppipla) | Country (cntry) 43.0% \| 0.087 | Household income feeling (hincfel) 13.1% \| 0.027 | Years of education (eduyrs) 11.7% \| 0.024 |
| Humans | Religiosity (rlgdgr) | Tradition (imptrada) 40.6% \| 0.108 | Country (cntry) 24.0% \| 0.064 | Gender (gndr) 4.4% \| 0.012 |
| Humans | Life satisfaction (stflife) | Household income feeling (hincfel) 29.0% \| 0.065 | Self-rated health (health) 22.4% \| 0.050 | Country (cntry) 16.0% \| 0.036 |
| Humans | Climate worry (wrclmch) | Caring for nature (impenva) 47.5% \| 0.079 | Country (cntry) 19.0% \| 0.032 | Gender (gndr) 4.3% \| 0.007 |
| Gemini | Climate responsibility (ccrdprs) | Caring for nature (impenva) 53.5% \| 0.251 | Equal treatment (ipeqopta) 5.7% \| 0.027 | Understanding others (ipudrsta) 4.5% \| 0.021 |
| Gemini | Political self-efficacy (cptppola) | Years of education (eduyrs) 23.2% \| 0.057 | Household income feeling (hincfel) 13.7% \| 0.034 | Freedom (impfreea) 8.4% \| 0.021 |
| Gemini | Income redistribution (gincdif) | Equal treatment (ipeqopta) 47.6% \| 0.175 | Household income feeling (hincfel) 10.9% \| 0.040 | Helping others (iphlppla) 5.8% \| 0.021 |
| Gemini | Allow immigrants (imdfetn) | Country (cntry) 16.0% \| 0.073 | Equal treatment (ipeqopta) 11.0% \| 0.050 | Household income feeling (hincfel) 7.9% \| 0.036 |
| Gemini | Immigration impact (imwbcnt) | Household income feeling (hincfel) 17.0% \| 0.147 | Country (cntry) 15.5% \| 0.134 | Equal treatment (ipeqopta) 6.5% \| 0.056 |
| Gemini | Loyalty to leaders (loylead) | Following rules (ipfrulea) 18.4% \| 0.132 | Strong government (ipstrgva) 12.0% \| 0.086 | Country (cntry) 11.5% \| 0.082 |
| Gemini | Obedience for children (lrnobed) | Following rules (ipfrulea) 39.1% \| 0.477 | Tradition (imptrada) 10.4% \| 0.127 | Behaving properly (ipbhprpa) 6.7% \| 0.081 |
| Gemini | Political interest (polintr) | Strong government (ipstrgva) 19.9% \| 0.020 | Equal treatment (ipeqopta) 7.6% \| 0.008 | Freedom (impfreea) 6.5% \| 0.007 |

| Source | Outcome | Top 1 | Top 2 | Top 3 |
|---|---|---|---|---|
| Gemini | People are fair (pplfair) | Household income feeling (hincfel) 34.7% \| 0.280 | Country (cntry) 13.3% \| 0.107 | Self-rated health (health) 12.2% \| 0.098 |
| Gemini | People are helpful (pplhlp) | Helping others (iphlppla) 20.1% \| 0.141 | Household income feeling (hincfel) 13.6% \| 0.096 | Country (cntry) 8.2% \| 0.057 |
| Gemini | People can be trusted (ppltrst) | Country (cntry) 27.6% \| 0.335 | Household income feeling (hincfel) 21.3% \| 0.259 | Self-rated health (health) 6.5% \| 0.079 |
| Gemini | Political influence (psppipla) | Country (cntry) 54.6% \| 0.154 | Household income feeling (hincfel) 18.7% \| 0.053 | Years of education (eduyrs) 4.8% \| 0.014 |
| Gemini | Religiosity (rlgdgr) | Country (cntry) 28.2% \| 0.199 | Tradition (imptrada) 20.1% \| 0.141 | Age (agea) 9.0% \| 0.063 |
| Gemini | Life satisfaction (stflife) | Household income feeling (hincfel) 56.1% \| 0.463 | Self-rated health (health) 16.7% \| 0.138 | Main activity (mnactic) 9.2% \| 0.076 |
| Gemini | Climate worry (wrclmch) | Caring for nature (impenva) 47.6% \| 0.138 | Equal treatment (ipeqopta) 8.4% \| 0.024 | Country (cntry) 5.8% \| 0.017 |

Table A4. Top three persona attributes by outcome for humans and Gemini 2.5 Flash. Note. Full-sample Monte Carlo Shapley decomposition (N = 14,000). Each cell gives the share of the outcome-specific predictable component and the absolute contribution in matched human-variance units. The corresponding GPT-5-mini rankings are in Table 6.

## A5. Nonlinear robustness check

| Outcome | Human ridge | Human GBM add. | Human GBM full | Gemini ridge | Gemini GBM add. | Gemini GBM full | GPT ridge | GPT GBM add. | GPT GBM full |
|---|---|---|---|---|---|---|---|---|---|
| Climate responsibility (ccrdprs) | 0.216 | 0.217 | 0.236 | 0.800 | 0.795 | 0.837 | 0.913 | 0.909 | 0.927 |
| Political self-efficacy (cptppola) | 0.195 | 0.196 | 0.209 | 0.668 | 0.660 | 0.693 | 0.674 | 0.670 | 0.705 |
| Income redistribution (gincdif) | 0.092 | 0.095 | 0.107 | 0.618 | 0.615 | 0.710 | 0.712 | 0.713 | 0.793 |
| Allow immigrants (imdfetn) | 0.243 | 0.241 | 0.263 | 0.703 | 0.679 | 0.718 | 0.656 | 0.641 | 0.662 |
| Immigration impact (imwbcnt) | 0.180 | 0.179 | 0.211 | 0.736 | 0.710 | 0.753 | 0.792 | 0.768 | 0.812 |
| Loyalty to leaders (loylead) | 0.163 | 0.158 | 0.184 | 0.556 | 0.541 | 0.640 | 0.690 | 0.679 | 0.729 |
| Obedience for children (lrnobed) | 0.228 | 0.219 | 0.257 | 0.778 | 0.767 | 0.807 | 0.814 | 0.812 | 0.850 |
| Political interest (polintr) | 0.208 | 0.207 | 0.220 | 0.442 | 0.428 | 0.513 | 0.410 | 0.401 | 0.514 |
| People are fair (pplfair) | 0.166 | 0.165 | 0.179 | 0.746 | 0.730 | 0.766 | 0.688 | 0.678 | 0.716 |
| People are helpful (pplhlp) | 0.149 | 0.150 | 0.160 | 0.640 | 0.620 | 0.692 | 0.760 | 0.750 | 0.802 |
| People can be trusted (ppltrst) | 0.172 | 0.173 | 0.182 | 0.662 | 0.649 | 0.713 | 0.724 | 0.706 | 0.760 |
| Political influence (psppipla) | 0.204 | 0.206 | 0.218 | 0.694 | 0.706 | 0.777 | 0.613 | 0.626 | 0.691 |
| Religiosity (rlgdgr) | 0.267 | 0.267 | 0.283 | 0.815 | 0.806 | 0.856 | 0.877 | 0.868 | 0.902 |
| Life satisfaction (stflife) | 0.224 | 0.220 | 0.229 | 0.920 | 0.915 | 0.934 | 0.887 | 0.879 | 0.903 |
| Climate worry (wrclmch) | 0.167 | 0.167 | 0.186 | 0.650 | 0.633 | 0.663 | 0.762 | 0.759 | 0.790 |
| Mean | 0.192 | 0.191 | 0.208 | 0.695 | 0.684 | 0.738 | 0.731 | 0.724 | 0.770 |
| Ratio to humans | | | | 3.63 | | 3.54 | 3.82 | | 3.70 |

Table A5. Out-of-fold $R^2$ from ridge and gradient boosting, by outcome and source. Note. Ridge values as in Table A3. Gradient boosting: scikit-learn HistGradientBoostingRegressor, max_iter = 300, learning_rate = 0.05, max_leaf_nodes = 31, min_samples_leaf = 40, seed = 42. “Additive” constrains the model to no interactions (interaction_cst); “full” allows interactions. GBM folds (five) are assigned cyclically and are not stratified by country, unlike the ridge folds. Ratio row: mean $R^2$ of each model divided by the human mean for the same specification.

## A6. Reliability break-even by outcome

| Outcome | Human $R^2$ | GPT $R^2$ | GPT ICC(1) | GPT $R^2$ corrected | Human $R^2$ corr. ($\rho$ = 0.7) | Human $R^2$ corr. ($\rho$ = 0.5) | Break-even human $\rho$ |
|---|---|---|---|---|---|---|---|
| Climate responsibility (ccrdprs) | 0.216 | 0.913 | 0.947 | 0.964 | 0.308 | 0.431 | 0.224 |
| Political self-efficacy (cptppola) | 0.195 | 0.674 | 0.770 | 0.875 | 0.278 | 0.389 | 0.222 |
| Income redistribution (gincdif) | 0.092 | 0.712 | 0.857 | 0.831 | 0.132 | 0.184 | 0.111 |
| Allow immigrants (imdfetn) | 0.243 | 0.656 | 0.737 | 0.890 | 0.347 | 0.486 | 0.273 |
| Immigration impact (imwbcnt) | 0.180 | 0.792 | 0.868 | 0.912 | 0.256 | 0.359 | 0.197 |
| Loyalty to leaders (loylead) | 0.163 | 0.690 | 0.781 | 0.883 | 0.233 | 0.326 | 0.184 |
| Obedience for children (lrnobed) | 0.228 | 0.814 | 0.873 | 0.933 | 0.326 | 0.456 | 0.245 |
| Political interest (polintr) | 0.208 | 0.410 | 0.645 | 0.635 | 0.297 | 0.416 | 0.328 |
| People are fair (pplfair) | 0.166 | 0.688 | 0.775 | 0.888 | 0.237 | 0.332 | 0.187 |
| People are helpful (pplhlp) | 0.149 | 0.760 | 0.846 | 0.899 | 0.213 | 0.298 | 0.166 |
| People can be trusted (ppltrst) | 0.172 | 0.724 | 0.812 | 0.892 | 0.245 | 0.343 | 0.192 |
| Political influence (psppipla) | 0.204 | 0.613 | 0.764 | 0.802 | 0.291 | 0.407 | 0.254 |
| Religiosity (rlgdgr) | 0.267 | 0.877 | 0.930 | 0.943 | 0.382 | 0.534 | 0.283 |
| Life satisfaction (stflife) | 0.224 | 0.887 | 0.926 | 0.958 | 0.321 | 0.449 | 0.234 |
| Climate worry (wrclmch) | 0.167 | 0.762 | 0.848 | 0.899 | 0.239 | 0.334 | 0.186 |
| **Mean (break-even: median)** | **0.192** | **0.731** | **0.825** | **0.880** | **0.274** | **0.383** | **0.222** |

Table A6. Reliability correction and break-even reliability by outcome. Note. Corrected $R^2 = \min(R^2 / \rho, 1)$. GPT-5-mini $\rho$ is the outcome-level ICC(1) from the test–retest experiment (Section 5). Human $\rho$ is assumed. Break-even $\rho$ is the assumed human reliability at which corrected human $R^2$ equals corrected GPT-5-mini $R^2$.

## A7. Masking results by outcome

| Outcome | Cond. | Masked Top-3 | Rel. var. | Var. ret. | $R^2$ | $R^2$ ret. | AEV | AEV ret. | Modal share | Change d vs FULL | Leading attribute |
|---|---|---|---|---|---|---|---|---|---|---|---|
| Climate responsibility (ccrdprs) | FULL | impenva, iphlppla, ipudrsta | 0.669 | | 0.882 | | 0.590 | | 23.7% | | impenva |
| | DELETE | | 0.377 | 56.3% | 0.654 | 74.2% | 0.246 | 41.8% | 38.5% | 73.9% | ipeqopta |
| | NOT REC. | | 0.364 | 54.4% | 0.628 | 71.3% | 0.229 | 38.8% | 27.5% | 81.6% | ipeqopta |
| Political self-efficacy (cptppola) | FULL | eduyrs, impfreea, hincfel | 0.393 | | 0.590 | | 0.232 | | 48.6% | | eduyrs |
| | DELETE | | 0.303 | 77.1% | 0.492 | 83.5% | 0.149 | 64.4% | 57.9% | 34.0% | ipcrtiva |
| | NOT REC. | | 0.298 | 75.8% | 0.492 | 83.4% | 0.147 | 63.2% | 54.4% | 34.3% | ipshabta |
| Income redistribution (gincdif) | FULL | ipeqopta, iphlppla, ipudrsta | 0.444 | | 0.678 | | 0.301 | | 53.9% | | ipeqopta |
| | DELETE | | 0.308 | 69.3% | 0.342 | 50.5% | 0.105 | 35.0% | 80.0% | 40.2% | hincfel |
| | NOT REC. | | 0.365 | 82.2% | 0.412 | 60.7% | 0.150 | 49.9% | 82.0% | 43.1% | impricha |
| Allow immigrants (imdfetn) | FULL | ipeqopta, ipudrsta, cntry | 0.423 | | 0.532 | | 0.225 | | 65.9% | | ipeqopta |
| | DELETE | | 0.482 | 114.1% | 0.585 | 110.0% | 0.282 | 125.5% | 60.1% | 39.7% | impdiffa |
| | NOT REC. | | 0.475 | 112.3% | 0.561 | 105.3% | 0.266 | 118.3% | 61.3% | 39.3% | impdiffa |
| Immigration impact (imwbcnt) | FULL | ipudrsta, ipeqopta, iphlppla | 0.683 | | 0.725 | | 0.495 | | 28.5% | | ipudrsta |
| | DELETE | | 0.784 | 114.8% | 0.723 | 99.8% | 0.567 | 114.6% | 29.3% | 80.4% | impdiffa |

| Outcome | Cond. | Masked Top-3 | Rel. var. | Var. ret. | R² | R² ret. | AEV | AEV ret. | Modal share | Change d vs FULL | Leading attribute |
|---|---|---|---|---|---|---|---|---|---|---|---|
| | NOT REC. | | 0.758 | 110.9% | 0.723 | 99.8% | 0.548 | 110.7% | 30.7% | 78.6% | impdiffa |
| Loyalty to leaders (loylead) | FULL | ipfrulea, ipstrgva, imptrada | 1.194 | | 0.600 | | 0.717 | | 43.5% | | ipstrgva |
| | DELETE | | 0.290 | 24.3% | 0.256 | 42.6% | 0.074 | 10.3% | 65.8% | 59.2% | impfreea |
| | NOT REC. | | 0.279 | 23.4% | 0.220 | 36.6% | 0.061 | 8.6% | 68.6% | 60.6% | ipeqopta |
| Obedience for children (lrnobed) | FULL | ipfrulea, ipbhprpa, ipstrgva | 1.462 | | 0.754 | | 1.102 | | 42.1% | | ipfrulea |
| | DELETE | | 1.022 | 69.9% | 0.516 | 68.5% | 0.528 | 47.9% | 48.9% | 57.1% | imptrada |
| | NOT REC. | | 0.938 | 64.2% | 0.472 | 62.7% | 0.443 | 40.2% | 53.4% | 57.3% | impfreea |
| Political interest (polintr) | FULL | ipeqopta, ipstrgva, impfreea | 0.124 | | 0.229 | | 0.028 | | 88.9% | | ipstrgva |
| | DELETE | | 0.171 | 138.2% | 0.260 | 113.5% | 0.045 | 156.9% | 81.5% | 16.3% | eduyrs |
| | NOT REC. | | 0.167 | 135.2% | 0.259 | 112.8% | 0.043 | 152.5% | 81.8% | 16.9% | eduyrs |
| People are fair (pplfair) | FULL | ipeqopta, ipudrsta, cntry | 0.549 | | 0.607 | | 0.333 | | 42.5% | | iphlppla |
| | DELETE | | 0.436 | 79.3% | 0.457 | 75.4% | 0.199 | 59.8% | 54.5% | 58.8% | iphlppla |
| | NOT REC. | | 0.394 | 71.7% | 0.461 | 76.0% | 0.182 | 54.5% | 55.1% | 59.8% | iphlppla |
| People are helpful (pplhlp) | FULL | iphlppla, ipudrsta, iplylfra | 0.490 | | 0.687 | | 0.336 | | 44.4% | | iphlppla |
| | DELETE | | 0.451 | 92.2% | 0.540 | 78.6% | 0.244 | 72.4% | 48.2% | 59.5% | cntry |
| | NOT REC. | | 0.486 | 99.2% | 0.553 | 80.5% | 0.269 | 79.9% | 42.5% | 62.9% | domicil |
| People can be trusted (ppltrst) | FULL | cntry, ipudrsta, iphlppla | 0.810 | | 0.689 | | 0.558 | | 30.0% | | cntry |
| | DELETE | | 0.729 | 90.0% | 0.571 | 82.8% | 0.416 | 74.5% | 31.2% | 68.2% | domicil |
| | NOT REC. | | 0.727 | 89.8% | 0.550 | 79.8% | 0.400 | 71.7% | 31.2% | 68.9% | domicil |
| Political influence (psppipla) | FULL | cntry, hincfel, eduyrs | 0.312 | | 0.510 | | 0.159 | | 55.2% | | cntry |
| | DELETE | | 0.200 | 64.0% | 0.343 | 67.2% | 0.069 | 43.0% | 74.4% | 35.5% | mnactic |
| | NOT REC. | | 0.229 | 73.3% | 0.374 | 73.4% | 0.086 | 53.8% | 69.2% | 35.4% | mnactic |
| Religiosity (rlgdgr) | FULL | imptrada, cntry, agea | 0.641 | | 0.833 | | 0.534 | | 16.0% | | imptrada |
| | DELETE | | 0.417 | 65.1% | 0.754 | 90.5% | 0.314 | 58.9% | 23.5% | 76.2% | ipfrulea |
| | NOT REC. | | 0.414 | 64.6% | 0.729 | 87.5% | 0.302 | 56.5% | 24.7% | 77.8% | ipfrulea |
| Life satisfaction (stflife) | FULL | hincfel, health, mnactic | 0.523 | | 0.870 | | 0.455 | | 41.9% | | hincfel |
| | DELETE | | 0.092 | 17.7% | 0.485 | 55.7% | 0.045 | 9.8% | 70.5% | 54.5% | impfreea |
| | NOT REC. | | 0.107 | 20.5% | 0.512 | 58.9% | 0.055 | 12.0% | 68.1% | 55.1% | cntry |
| Climate worry (wrclmch) | FULL | impenva, ipudrsta, ipeqopta | 0.475 | | 0.728 | | 0.346 | | 51.0% | | impenva |
| | DELETE | | 0.356 | 74.9% | 0.450 | 61.8% | 0.160 | 46.3% | 60.9% | 39.6% | iphlppla |
| | NOT REC. | | 0.294 | 61.9% | 0.338 | 46.4% | 0.099 | 28.7% | 71.4% | 42.5% | iphlppla |

Table A7. Masking results by outcome and condition (GPT-5-mini, N = 1,000). Note. Relative variance and AEV in matched human-variance units. Retention is the outcome-level ratio to FULL (the macro retention in Section 6.8 is a ratio of means). Leading attribute is the top attribute in the cross-validated post-masking Shapley decomposition (Section 6.11); under FULL

| Outcome | Cond. | Masked Top-3 | Rel. var. | Var. ret. | $R^2$ | $R^2$ ret. | AEV | AEV ret. | Modal share | Changed vs FULL | Leading attribute |
|---|---|---|---|---|---|---|---|---|---|---|---|
| it can differ from the Table 6 ranking used to select the masked attributes, because the two decompositions differ in sample and value function. | | | | | | | | | | | |

## A8. Sequential masking

| Outcome | Removed 1st | Removed 2nd | Removed 3rd | Leading after 1st removed | Original Top-2 leads | $R^2$ ret. step 1 | $R^2$ ret. step 2 | $R^2$ ret. step 3 |
|---|---|---|---|---|---|---|---|---|
| Climate responsibility (ccrdprs) | impenva | iphlppla | iplylfra | iphlppla | Yes | 87.3% | 77.1% | 76.6% |
| Political self-efficacy (cptppola) | eduyrs | impfreea | hincfel | impfreea | Yes | 109.6% | 96.1% | 99.3% |
| Income redistribution (gincdif) | ipeqopta | iphlppla | ipudrsta | iphlppla | Yes | 58.0% | 45.6% | 50.3% |
| Allow immigrants (imdfetn) | ipeqopta | ipudrsta | cntry | ipudrsta | Yes | 104.8% | 100.0% | 102.0% |
| Immigration impact (imwbcnt) | ipudrsta | ipeqopta | iphlppla | ipeqopta | Yes | 96.9% | 94.3% | 93.4% |
| Loyalty to leaders (loylead) | ipfrulea | ipstrgva | imptrada | ipstrgva | Yes | 86.7% | 50.4% | 41.3% |
| Obedience for children (lrnobed) | ipfrulea | ipbhprpa | ipstrgva | ipstrgva | No | 73.2% | 68.7% | 67.0% |
| Political interest (polintr) | ipeqopta | ipstrgva | iphlppla | ipstrgva | Yes | 101.3% | 108.2% | 94.8% |
| People are fair (pplfair) | ipudrsta | ipeqopta | iphlppla | ipeqopta | Yes | 97.5% | 88.1% | 85.7% |
| People are helpful (pplhlp) | iphlppla | ipudrsta | iplylfra | cntry | No | 86.3% | 84.3% | 82.1% |
| People can be trusted (ppltrst) | cntry | ipudrsta | iphlppla | ipudrsta | Yes | 87.6% | 82.4% | 68.8% |
| Political influence (psppipla) | cntry | hincfel | eduyrs | hincfel | Yes | 59.3% | 46.5% | 44.1% |
| Religiosity (rlgdgr) | imptrada | cntry | agea | cntry | Yes | 85.8% | 80.2% | 82.4% |
| Life satisfaction (stflife) | hincfel | health | mnactic | health | Yes | 83.0% | 50.4% | 73.3% |
| Climate worry (wrclmch) | impenva | iphlppla | iplylfra | iphlppla | Yes | 74.4% | 65.3% | 68.4% |
| **Total** | | | | | **13 of 15** | | | |

Table A8. Earlier sequential masking experiment (GPT-5-mini, N = 1,000). Note. Attributes were removed cumulatively in the order of the earlier (pre-N = 14,000) GPT-5-mini ranking; $R^2$ retention is relative to the full persona at each step. The two exceptions are lrnobed, where the original Top-3 attribute takes over, and pplhlp, where country takes over. This run is complementary mechanistic evidence; headline masking estimates come from the final two-arm experiment.